\documentclass[twocolumn]{aastex63}

\usepackage{bm}
\usepackage{amsmath}
\usepackage{comment}

\newcommand{\prot}{P_\mathrm{rot}}
\newcommand{\tauc}{\tau_\mathrm{c}}
\newcommand{\ro}{\mathrm{Ro}}
\newcommand{\amp}{R_\mathrm{per}}
\newcommand{\teff}{T_\mathrm{eff}}
\newcommand{\kepler}{\textit{Kepler}}
\newcommand{\sigmakep}{\sigma_\mathrm{Kep}}
\newcommand{\pbreak}{P_\mathrm{break}}
\newcommand{\robreak}{\ro_\mathrm{break}}
\newcommand{\rbreak}{R_\mathrm{break}}
\newcommand{\betasat}{{\beta_\mathrm{sat}}}
\newcommand{\betaunsat}{{\beta_\mathrm{unsat}}}
\newcommand{\lx}{L_X/L_\mathrm{bol}}

\shorttitle{Understanding the McQuillan Sample}
\shortauthors{Masuda}
\graphicspath{{./}{figures/}{figures_model_all/}{figures_pdet_all/}}

\begin{document}

\title{
On the Evolution of Rotational Modulation Amplitude in Solar-mass Main-Sequence Stars
}

\correspondingauthor{Kento Masuda}
\email{kmasuda@ess.sci.osaka-u.ac.jp}

\author[0000-0003-1298-9699]{Kento Masuda}
\affiliation{Department of Earth and Space Science, Osaka University, Osaka 560-0043, Japan}



\begin{abstract}

We investigate the relation between rotation periods $P_\mathrm{rot}$ and photometric modulation amplitudes $R_\mathrm{per}$ for $\approx 4,000$ Sun-like main-sequence stars observed by \textit{Kepler}, using $P_\mathrm{rot}$ and $R_\mathrm{per}$ from \citet{2014ApJS..211...24M}, effective temperature $T_\mathrm{eff}$ from LAMOST DR6, and parallax data from \textit{Gaia} EDR3.  
As has been suggested in previous works, we find that $P_\mathrm{rot}$ scaled by the convective turnover time $\tau_\mathrm{c}$, or the Rossby number $\mathrm{Ro}=P_\mathrm{rot}/\tau_\mathrm{c}$, serves as a good predictor of $R_\mathrm{per}$: $R_\mathrm{per}$ plateaus around $1\%$ in relative flux for $0.2 \lesssim \ro/\mathrm{Ro}_\odot \lesssim 0.4$, and decays steeply with increasing $\ro$ for $0.4 \lesssim \mathrm{Ro}/\mathrm{Ro}_\odot \lesssim 0.8$, where $\ro_\odot$ denotes $\ro$ of the Sun. In the latter regime we find $\mathrm{d}\ln R_\mathrm{per}/\mathrm{d}\ln\mathrm{Ro} \sim -4.5$ to $-2.5$, although the value is sensitive to detection bias against weak modulation and may depend on other parameters including $T_\mathrm{eff}$ and surface metallicity. The existing X-ray and Ca \textsc{ii} H\&K flux data also show transitions at $\mathrm{Ro}/\mathrm{Ro}_\odot\sim 0.4$, suggesting that all these transitions share the same physical origin.
We also find that the rapid decrease of $R_\mathrm{per}$ with increasing $\mathrm{Ro}$ causes rotational modulation of fainter \textit{Kepler} stars with $\mathrm{Ro}/\mathrm{Ro}_\odot \gtrsim 0.6$
to be buried under the photometric noise. This effect sets the longest $\prot$ detected in the \citet{2014ApJS..211...24M} sample as a function of $\teff$, and obscures the signature of stalled spin down that has been proposed to set in around $\mathrm{Ro}/\mathrm{Ro}_\odot \sim 1$.
\end{abstract}



\section{Introduction}\label{sec:intro}

Recent studies of rotation of old Sun-like stars 
suggest a change in stellar activity of middle-aged 
main-sequence stars. 
There has been accumulating evidence that the rotation periods of Sun-like stars cease to scale as the square-root of age \citep{1972ApJ...171..565S} in the latter halves of their lives
\citep[e.g.,][]{2015MNRAS.450.1787A,2016Natur.529..181V,2021NatAs...5..707H,2022MNRAS.510.5623M}. A similar transition has also been noted in chromospheric activities \citep{2016ApJ...826L...2M}. 
These may suggest a corresponding change in the mechanism of magnetic field generation that occurs once the rotation period $\prot$ becomes comparable to the convective turnover timescale $\tauc$.

Quasi-periodic brightness modulation of stars in broad-band photometry is well-suited for statistical studies of rotational evolution.
In particular, light curves from the prime \kepler\ mission \citep{2010Sci...327..977B} have been used to derive rotation periods up to months for tens of thousands of Sun-like stars \citep[e.g.,][]{2013A&A...557L..10N, 2013A&A...560A...4R, 2014ApJS..211...24M, 2014A&A...572A..34G,2019ApJS..244...21S,2021ApJS..255...17S}.
Previous investigations of the $\prot$ distribution of \kepler\ stars using the sample of \citet{2014ApJS..211...24M} 
have shown that the observed distribution is truncated roughly around the solar Rossby number $\ro=\prot/\tauc$ \citep{2019ApJ...872..128V} and also exhibits a pile-up 
around slightly shorter $\prot$ \citep{2022arXiv220308920D}. 
While these features are in qualitative agreement with the stalled spin down scenario \citep{2019ApJ...872..128V},
the effects of rotational evolution and detection bias have not been clearly disentangled.
In general, longer-period tail of the $\prot$ distribution is most prone to the detection bias, and so it requires good understanding of the bias to correctly interpret the observed distribution.

In this work, we attempt to better understand the detection bias in the $\prot$ sample constructed by \citet{2014ApJS..211...24M} to aid statistical interpretation of the sample.
To do so, we first investigate the generic relation between spot-modulation amplitudes $\amp$ and rotation periods $\prot$ for Sun-like main-sequence stars in the \citet{2014ApJS..211...24M} sample (as defined in Section~\ref{sec:sample}), and derive a relation that predicts the modulation amplitude $\amp$ given $\prot$ and $\teff$ (Section~\ref{sec:analysis}).
Then we clarify how this dependence is combined with the magnitude-dependent detection threshold for rotational modulation to sculpt the observed distribution of $\prot$ as a function of $\teff$ (Section \ref{sec:detection}).
We also show that the general pattern derived here is consistent with the sample from
\citet{2019ApJS..244...21S, 2021ApJS..255...17S} that include more detections of rotation periods, and that the latter catalog is subject to a different detection function. 
In Section \ref{sec:discussion}, we discuss our finding in connection with coronal and chromospheric activity indicators and the weakened magnetic braking hypothesis, and propose a test to check the veracity of our view on the detection bias further.

\section{The Sample}\label{sec:sample}

\citet{2014ApJS..211...24M} performed a homogeneous search for quasi-periodic brightness modulation associated with stellar rotation in \kepler\ light curves, and 
reported detections of robust rotational modulation for $\approx 34,000$ stars.
All these stars are assigned the rotation period $\prot$ as determined from the auto-correlation analysis, along with the average amplitude of variability within one period in units of parts-per-million (ppm), $\amp$, 
defined as the median of the differences between 95th and 5th percentiles of normalized flux in each rotation period cycle.
The Large Sky Area Multi-Object Fibre Spectroscopic Telescope (LAMOST) project \citep{2012RAA....12.1197C,2014IAUS..306..340W,2015RAA....15.1095L,2016ApJS..225...28R}, on the other hand, provided spectroscopic parameters for $\approx 60,000$ \kepler\ stars in their sixth data release (DR6). We work on the overlap of the two samples, 
for which $\prot$, $\amp$, and $\teff$ have been derived homogeneously.

One concern is the presence of unresolved binaries. 
The contaminating flux from the secondary affects both the inferred modulation amplitude and stellar classification. Tidal interactions with a close-in companion also affect the rotation period, although such close-in companions would occur in $\lesssim 10\%$ for Sun-like stars \citep{2010ApJS..190....1R}.
The analysis may also be complicated by evolved stars whose rotation may have changed due to evolution of internal structure rather than magnetic braking. We use the information on absolute magnitudes made available by {\it Gaia} \citep{2016A&A...595A...1G} to remove such objects as possible.

\begin{figure}
    \epsscale{1.18}
    \plotone{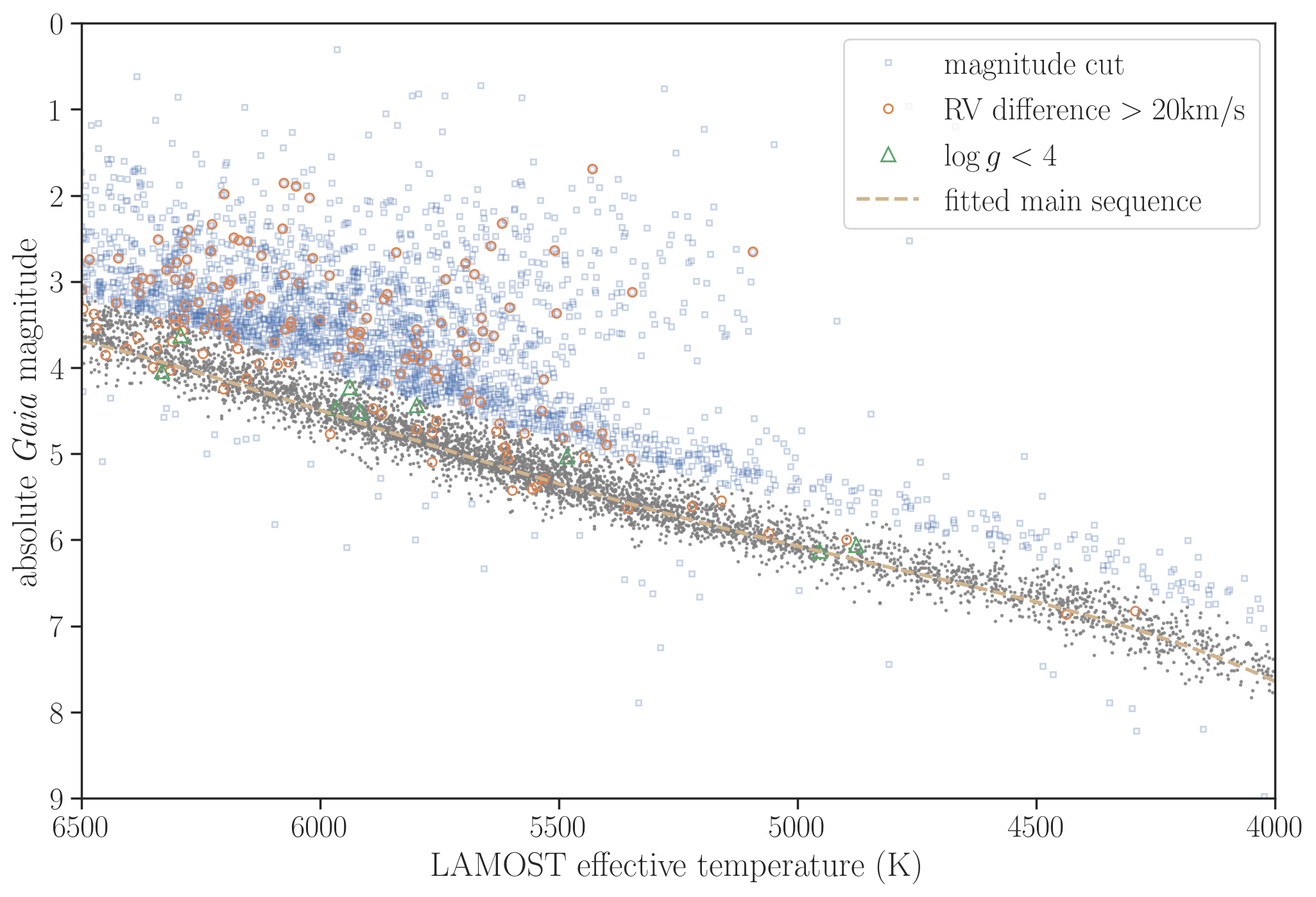}
    \caption{
    Absolute {\it Gaia} magnitudes and LAMOST $\teff$ of our sample stars (gray dots). Stars removed by the cuts described in Section~\ref{sec:sample} are shown with open symbols (see legends). The tan dashed line shows the fitted main sequence (see text).
    }
    \label{fig:hr}
\end{figure}

We start with 8,772 unique stars, for which robust periods are detected in \citet{2014ApJS..211...24M} and LAMOST DR6 data are publicly available.\footnote{We used \kepler\ IDs in ``tcomment" columns in the latter catalog.} 
For LAMOST stars with multi-epoch observations, the mean of $\teff$ was adopted.\footnote{Scatters of $\teff$ from multi-epoch observations are typically smaller than the $\sim 100\,\mathrm{K}$ uncertainty estimated for dwarfs \citep{2016ApJS..225...28R}. The bin size in Section~\ref{sec:analysis} is chosen to be larger than this latter value.}
We then used the cross-match service of the Centre de Donn\'{e}es astronomiques de Strasbourg (CDS) to find closest {\it Gaia} EDR3 sources \citep{2021A&A...649A...1G} within 5~arcsec and with {\tt parallax\_over\_error} greater than 10. We find 8,309 matches. The difference between the $G$-band and \kepler-band magnitudes, latter taken from \citet{2017ApJS..229...30M}, has the mean of $-0.02$ and standard deviation of 0.09, indicating correct matches.
We then placed these stars on the absolute {\it Gaia} magnitude--LAMOST $\teff$ diagram focusing on stars with $4,000\,\mathrm{K}<\teff<6500\,\mathrm{K}$, defined the main sequence by fitting a 5th-order polynomial 
iteratively clipping $1\sigma$ and $3\sigma$ outliers below and above the sequence respectively, and removed stars deviating by more than 0.5~magnitudes at a given $\teff$ from the derived sequence.
This removes bright sources that may be either evolved stars or unresolved binaries, where the threshold of 0.5 is chosen to remove the sequence of equal-brightness binaries that are brighter than single stars by 0.75~magnitudes at a given $\teff$. This cut left us with 5,022 stars. We also removed stars with multiple LAMOST measurements in which maximum and minimum radial velocities differ by more than $20\,\mathrm{km/s}$, above which the distribution of velocity differences exhibits a clear excess from a Gaussian distribution that appears to represent measurement uncertainties. This velocity cut left us with 4,977 stars. 
We also removed stars with LAMOST $\log g<4$ in the remaining sample because they shared the same locations in the $\log g$--$\teff$ plane as the stars removed in the above cuts,
and obtained the final sample of 4,968 stars.
The remaining discussion relies on (subsets of) these 4,968 stars with LAMOST $\teff=4,000$--$6,500\,\mathrm{K}$. 
The selection is visualized in the absolute {\it Gaia} magnitude--$\teff$ plane in Figure~\ref{fig:hr}.

\section{Evolution of Modulation Amplitudes}\label{sec:analysis}

Here we investigate how the photometric modulation amplitude $\amp$ evolves as a function of $\prot$. As has been shown in previous works \citep{2021ApJ...912..127S, 2021A&A...652L...2C}, the relation between $\amp$ and $\prot$ for roughly solar-mass stars is concisely summarized in terms of the Rossby number $\ro=\prot/\tauc$. We revisit such a relation for our sample stars and use it for the discussion of detectability in Section~\ref{sec:detection}.  

\begin{figure}
    \epsscale{1.18}
    \plotone{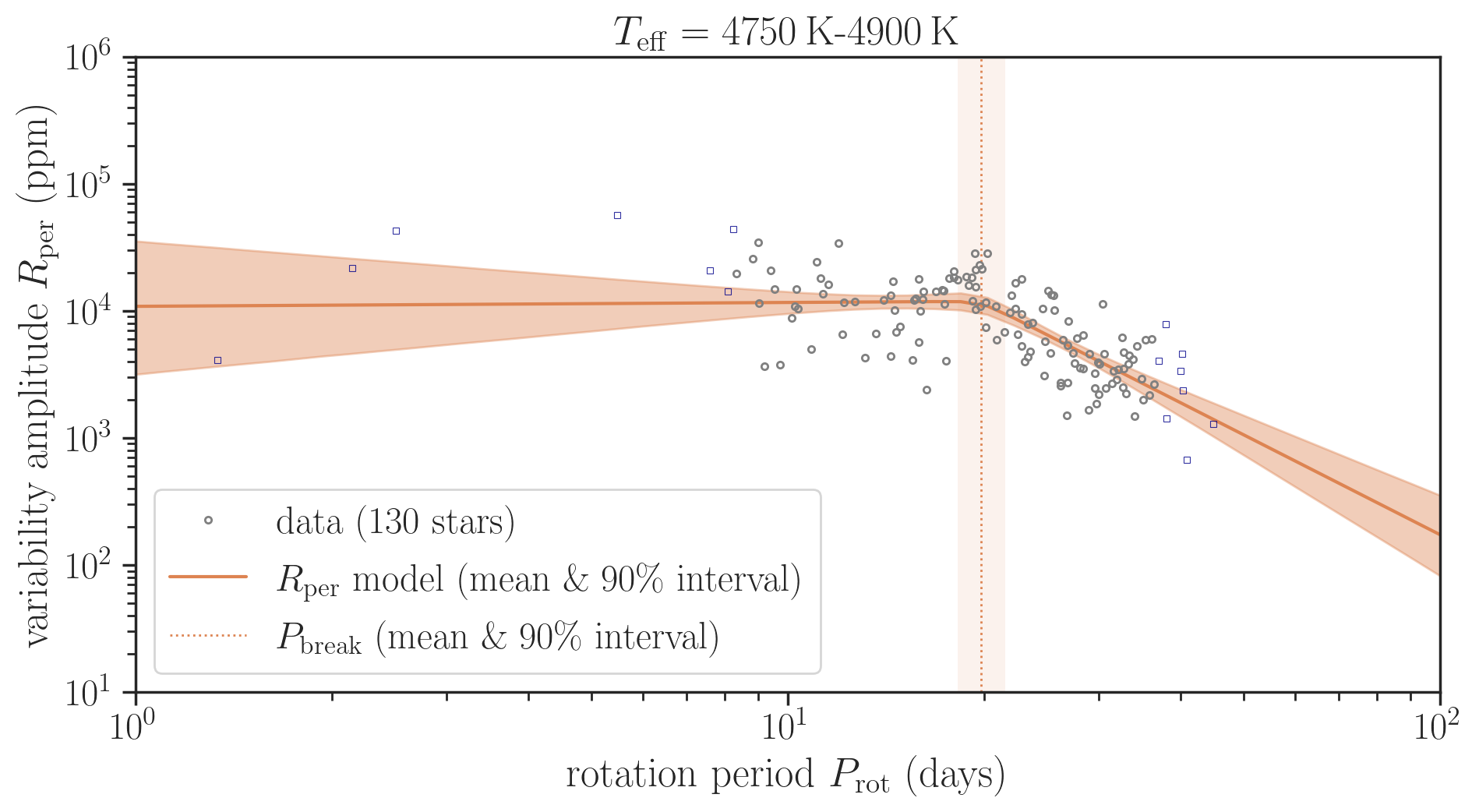}
    \plotone{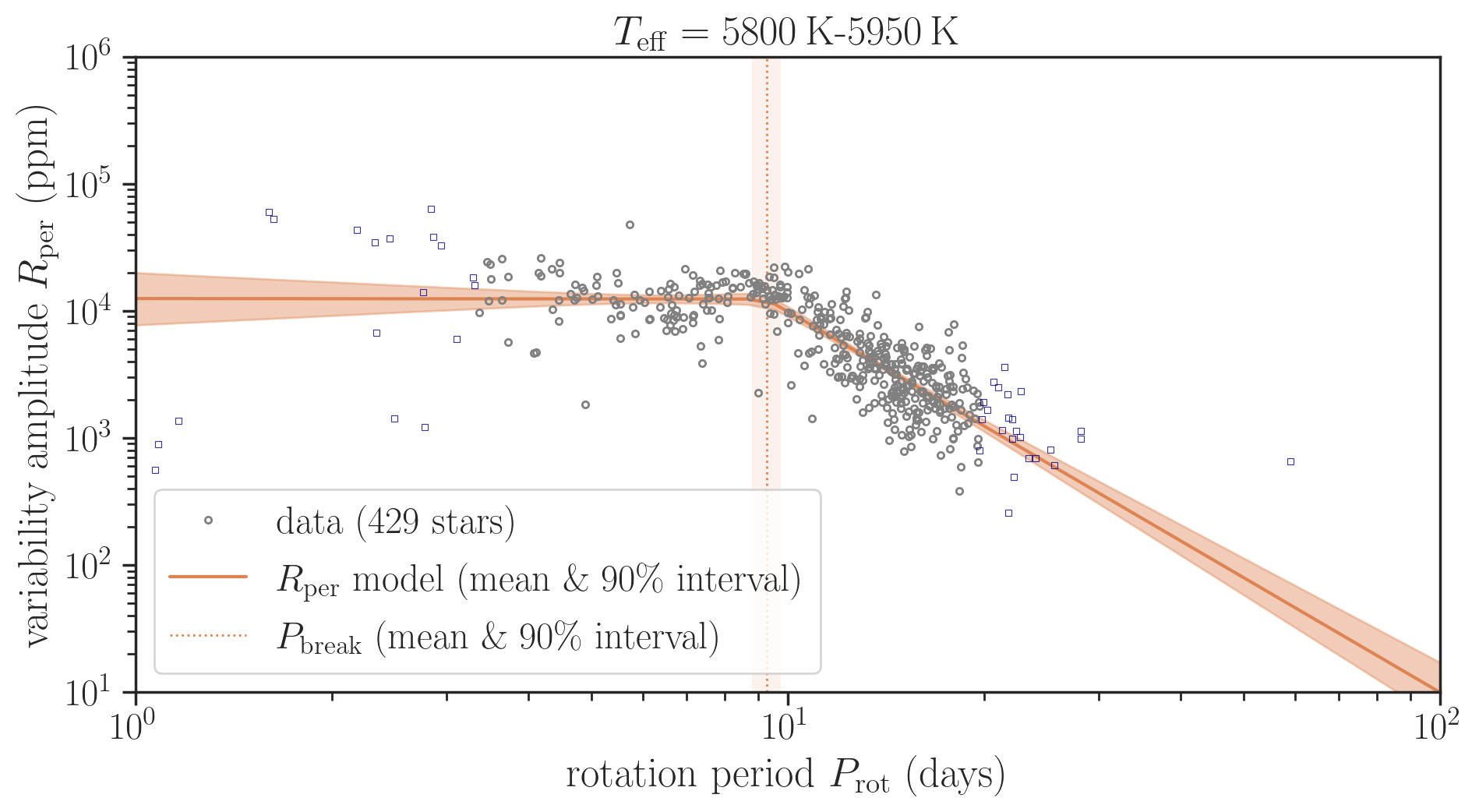}
    \caption{
    Spot-modulation amplitudes $\amp$ and rotation periods $\prot$ for stars with $\teff=4750$--$4900\,\mathrm{K}$ {\it (top)} and with $\teff=5,800$--$5,950\,\mathrm{K}$ {\it (bottom)} in our sample.
    {\it Gray circles}:  Data points. 
    Blue open squares show the ones that were not used for modeling.
    {\it Orange solid line and shade}: Broken power-law model. 
    {\it Vertical orange dotted line and shade}: Inferred location of the break, $\pbreak$. See Section \ref{ssec:amp_prot} for details, and Figure~\ref{fig:r_prot_all} for results in other $\teff$ bins.
    }
    \label{fig:r_prot_example}
\end{figure}

\subsection{Spot-Modulation Amplitude vs Rotation Period}\label{ssec:amp_prot}

In Figure~\ref{fig:r_prot_example} and Figure~\ref{fig:r_prot_all} in Appendix, we show $\amp$ and $\prot$ for our sample stars separated into 150$\,\mathrm{K}$ bins ranging from 4000 to 6400~K.
In Figure~\ref{fig:r_prot_example}, we show two $\teff$ bins separated by $\sim 1,000\,\mathrm{K}$ to illustrate two typical behaviors: 
(i) $\amp$ is roughly constant at shorter $\prot$ and exhibits a power-law decay at longer $\prot$, and 
(ii) the transition period, which we denote by $\pbreak$, is shorter for hotter stars (bottom panel).
The data for other $\teff$ ranges in Figure \ref{fig:r_prot_all} show that the same trend holds continuously over the most $\teff$ range, except for the coolest and hottest stars in the sample (see below).

To quantify this visual trend, we model the data with the following broken power-law function:
\begin{equation}
    \label{eq:plmodel}
    \amp(\prot; \bm{\theta}) = \begin{cases}
    \rbreak\left(\prot \over \pbreak\right)^\betasat \quad &\mathrm{for}\quad \prot < \pbreak\\
    \rbreak\left(\prot \over \pbreak\right)^\betaunsat&\mathrm{for}\quad \prot > \pbreak
    \end{cases}
\end{equation}
and infer $\bm{\theta}\equiv (\rbreak, \pbreak, \betasat, \betaunsat)$ 
for stars in each $\teff$ bin. The subscripts ``sat" and ``unsat" stand for saturated and unsaturated regimes, respectively, following the existing nomenclature --- although both regimes here fall within the so-called ``unsaturated" regimes of other activity indicators (see Section \ref{ssec:comparison}). 
We also model the measurement uncertainties as well as intrinsic scatters around this deterministic relation, assuming that the measured values of $\ln \amp^{\mathrm{obs}}$ and $\ln \prot^{\mathrm{obs}}$ for each star follow independent Gaussian distributions around the model, with common standard deviations $\sigma_{\ln\amp}$ and $\sigma_{\ln\prot}$. 
We also infer these parameters as well as the ``true" value of $\prot$ of each star,
bringing the total number of parameters to be five plus the number of stars in each $\teff$ subsample. 
The likelihood function is therefore
\begin{align}
    \notag 
    &p(\{\ln \amp^{\mathrm{obs},j}\}, \{\ln \prot^{\mathrm{obs},j}\}|\{\prot^j\}, \bm{\theta}, \sigma_{\ln\amp}, \sigma_{\ln\prot})
    \\
    \notag
    &= \prod_j \left[\mathcal{N}(\ln\amp^{\mathrm{obs},j}; \ln\amp(\prot^j; \bm{\theta}), \ln\sigma_{\amp}) \right.\\
    & \left. \qquad \quad \times \ 
    \mathcal{N}(\ln\prot^{\mathrm{obs},j}; \ln\prot^j, \ln\sigma_{\prot})\right]
\end{align}
where $j$ is the label for stars in each subsample, $\{x^j\}$ denotes the set of $x$ in the subsample, and $\mathcal{N}(x; \mu, \sigma)$ is the Gaussian distribution for $x$ with mean $\mu$ and standard deviation $\sigma$.

We consider this broken-power law model as a simple mathematical tool that is useful to quantify how the $\amp$--$\prot$ (or $\ro$) relation depends on $\teff$, and do not claim that this function provides the correct description of the relation (nor do we attempt to identify it). Indeed, we see a hint of more detailed structures than described by this model, as will be discussed below.  

The inference was performed in a Bayesian manner. We adopt independent prior probability density functions (PDFs) 
for $\bm{\theta}$, $\sigma_{\ln\amp}$, $\sigma_{\ln\prot}$, and $\{\prot^{\mathrm{obs},j}\}$ as summarized in Table \ref{tab:priors}, 
and infer the joint posterior PDF for these parameters given the data $(\{\amp^{\mathrm{obs},j}\}, \{\prot^{\mathrm{obs},j}\})$ by drawing samples from the posterior PDF using Hamiltonian Monte Carlo \citep{DUANE1987216, 2017arXiv170102434B} with No-U-Turn sampler \citep{2011arXiv1111.4246H} as implemented in {\tt NumPyro} \citep{bingham2018pyro, phan2019composable}.

In each subsample, we only model the stars whose rotation periods fall between their 5th and 95th percentiles. 
In Figure \ref{fig:r_prot_example} and \ref{fig:r_prot_all}, those stars omitted from fitting are shown as blue open squares.
They are both more sensitive to detection bias against weaker modulation in fainter stars that has not been taken into account in our model: the few rapid rotators tend to be rarer and thus fainter, and the slowest rotators tend to have smaller $\amp$. 
The lack of detection model may be considered as a limitation of our analysis and may introduce systematic errors in the inferred parameters; see Section~\ref{ssec:r_ro_bias} for further discussion.

\begin{deluxetable}{l@{\hspace{2cm}}c}
\caption{Parameters and Priors of the Model.}
\label{tab:priors}
\tablehead{
	\colhead{Parameters} & \colhead{Priors}
} 
\startdata
\multicolumn{2}{l}{{\it (Common Parameters)}}\\
$\ln\rbreak$ (ppm)  & $\mathcal{N}(\ln 10^4, 1)$\\
$\betasat$	     & $\mathcal{N}(0, 5)$\\
$\betaunsat$	 & $\mathcal{N}(0, 5)$\\
$\sigma_{\ln\amp}$  & $\mathcal{N}_\mathrm{half}(1)$\\
\multicolumn{2}{l}{{\it ($\amp$--$\prot$ analysis in Section~\ref{ssec:amp_prot})}}\\
$\ln\pbreak$ (day)	     & $\mathcal{U}(\ln 3, \ln 30)$\\
$\sigma_{\ln\prot}$ (day) & $\mathcal{N}_\mathrm{half}(0.1)$\\
$\ln \prot^j$ (day) & $\mathcal{U}(\ln \prot^{\mathrm{5th}},\ln \prot^{\mathrm{95th}})$\\
\multicolumn{2}{l}{{\it ($\amp$--$\ro$ analysis in Section~\ref{ssec:amp_ro})}}\\
$\ln\robreak$ 	     & $\mathcal{U}(\ln 0.5, \ln 2)$\\
$\sigma_{\ln\ro}$    & $\mathcal{N}_\mathrm{half}(0.1)$\\
$\ln \ro^j$  & $\mathcal{U}(\ln \ro^{\mathrm{5th}},\ln \ro^{\mathrm{95th}})$\\
\enddata
\tablecomments{$\mathcal{N}(\mu,\sigma)$ is the normal distribution with mean $\mu$ and standard deviation $\sigma$. $\mathcal{U}(a,b)$ is the uniform distribution between $a$ and $b$. $\mathcal{N}_\mathrm{half}(\sigma)$ is the half-normal distribution with scale $\sigma$. $x^{n\mathrm{th}}$ denotes $n$th percentile value of $x$ in each subsample.
The lower (upper) bound for $\pbreak$ was set to $\prot^{\mathrm{5th}}$ ($ \prot^{\mathrm{95th}}$) when the latter was shorter (longer) than 3 (30) days. The same is true for $\robreak$, with the corresponding thresholds being 0.5 and 2.}
\end{deluxetable}

The solid orange line and shaded region in Figure~\ref{fig:r_prot_example} shows the mean and 5th-95th percentile of the prediction by the broken power-law model in Equation~\ref{eq:plmodel}. The inferred $\pbreak$ (mean and 5th-95th percentile of the marginal posterior PDF) are also shown with vertical dotted line and shade. Our model fitting locates $\pbreak$ that is seen visually, which decreases with increasing $\teff$. A similar pattern was not found robustly for the lowest $\teff$ bin, presumably due to a small number of data points. The break is implied but the pattern is different for stars with $\teff>6,250\,\mathrm{K}$; this is not surprising either because they are the stars above the Kraft break \citep{1967ApJ...150..551K} that follow different $\prot$ evolution from cooler stars with convective envelopes. 
Below we omit these bins and consider stars with $\teff=4,250\,\mathrm{K}$--$6,250\,\mathrm{K}$, unless otherwise noted.

\begin{figure}
    \epsscale{1.15}
    \plotone{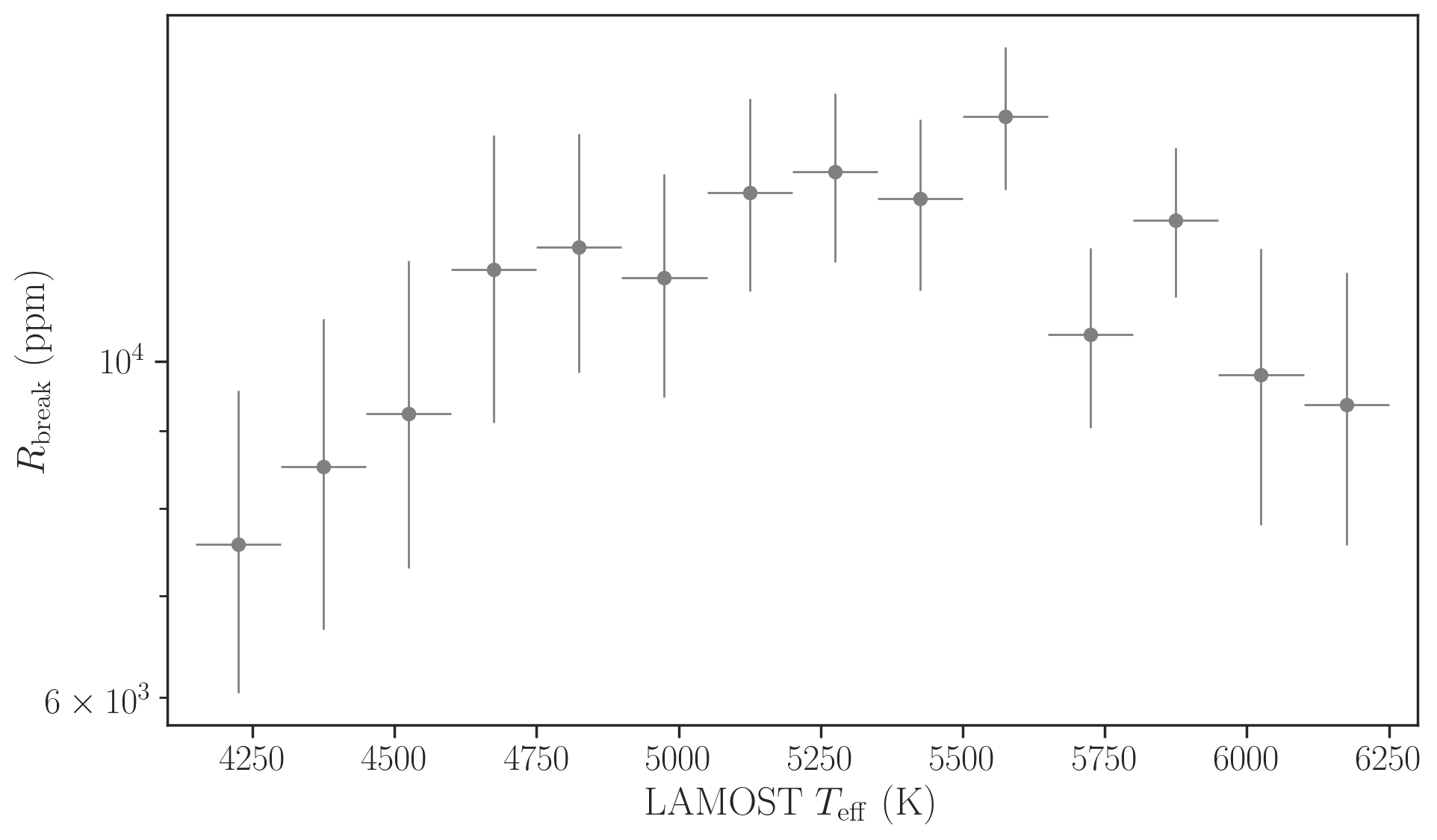}
    \plotone{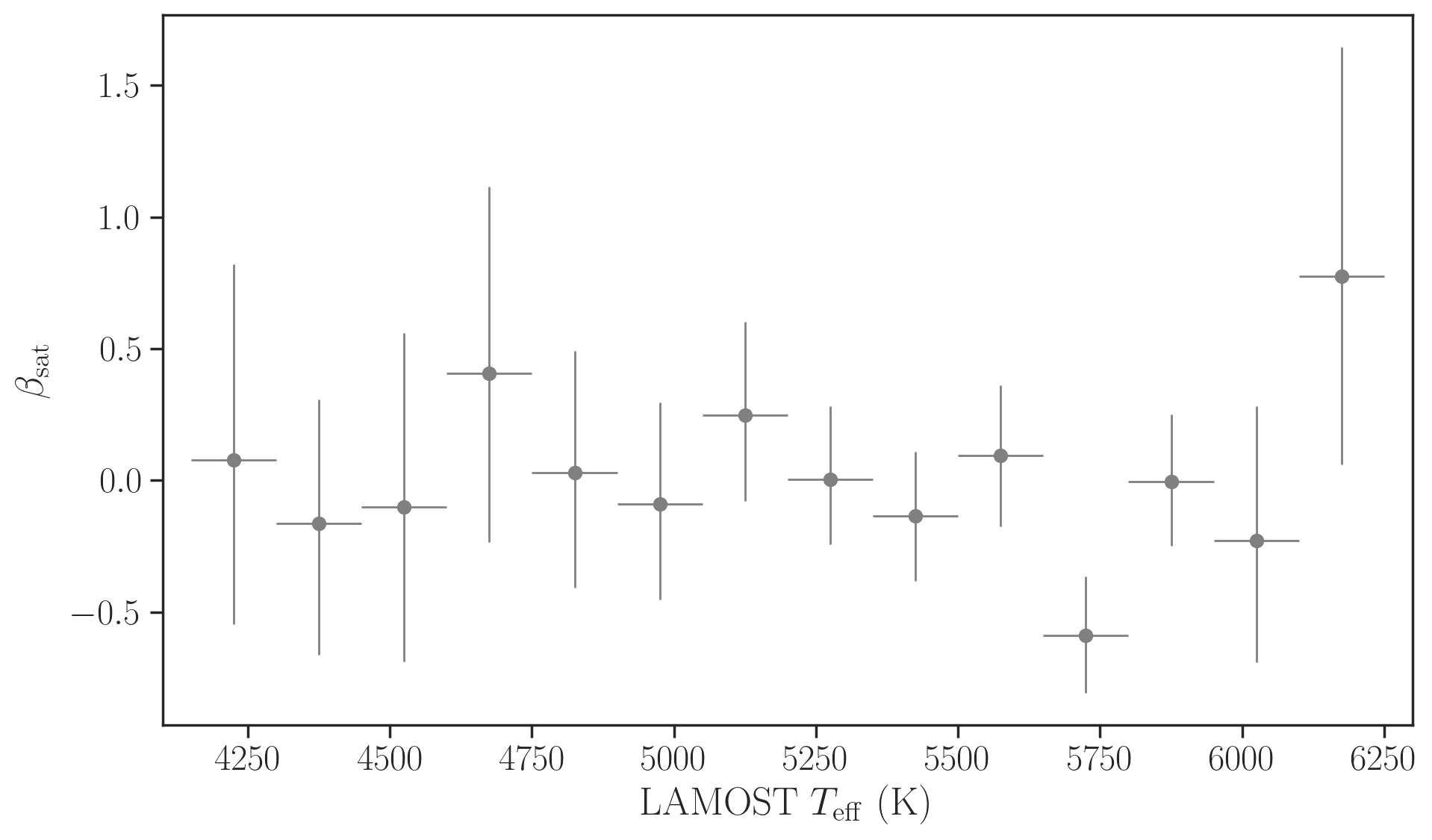}
    \plotone{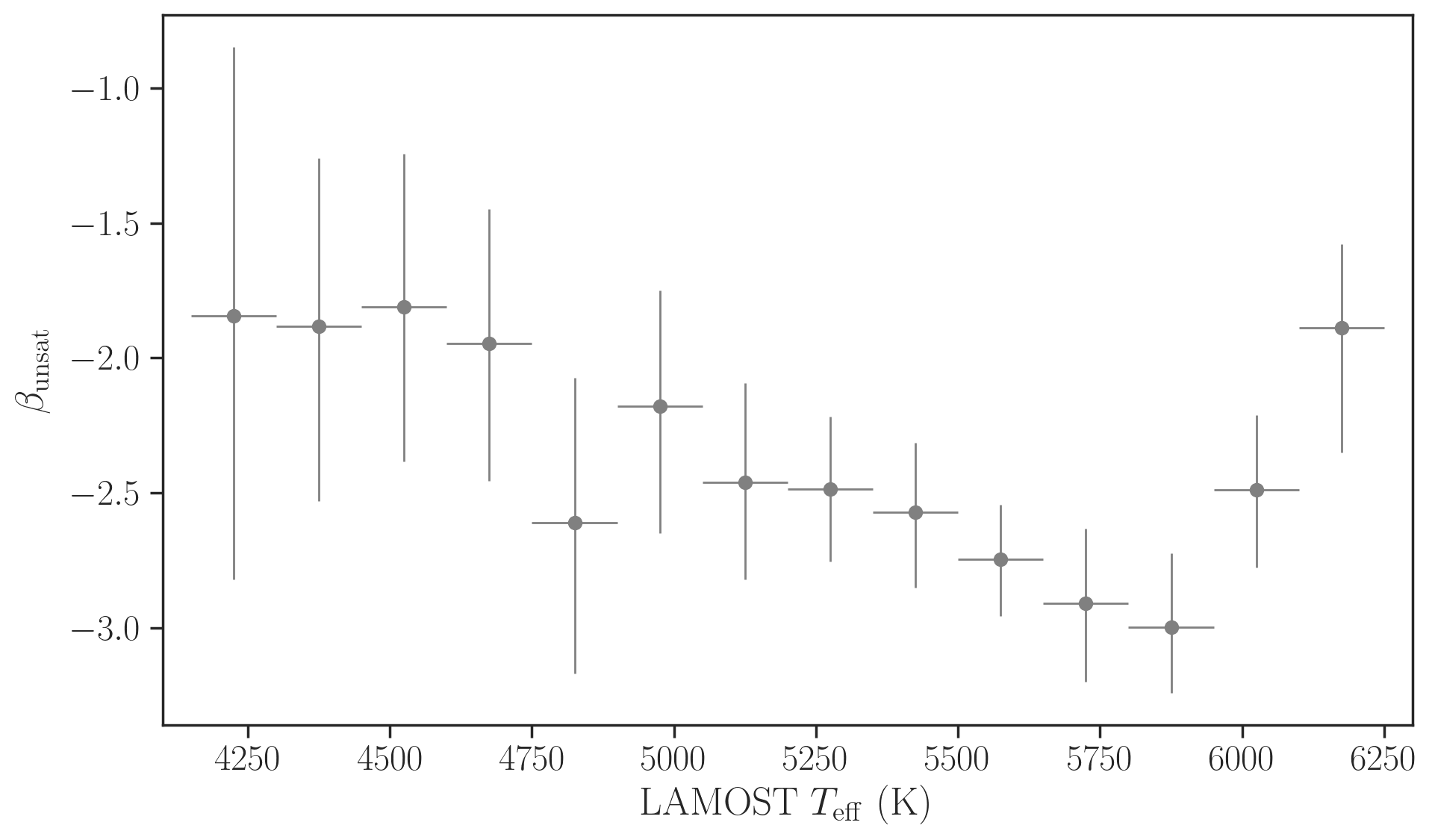}
    \plotone{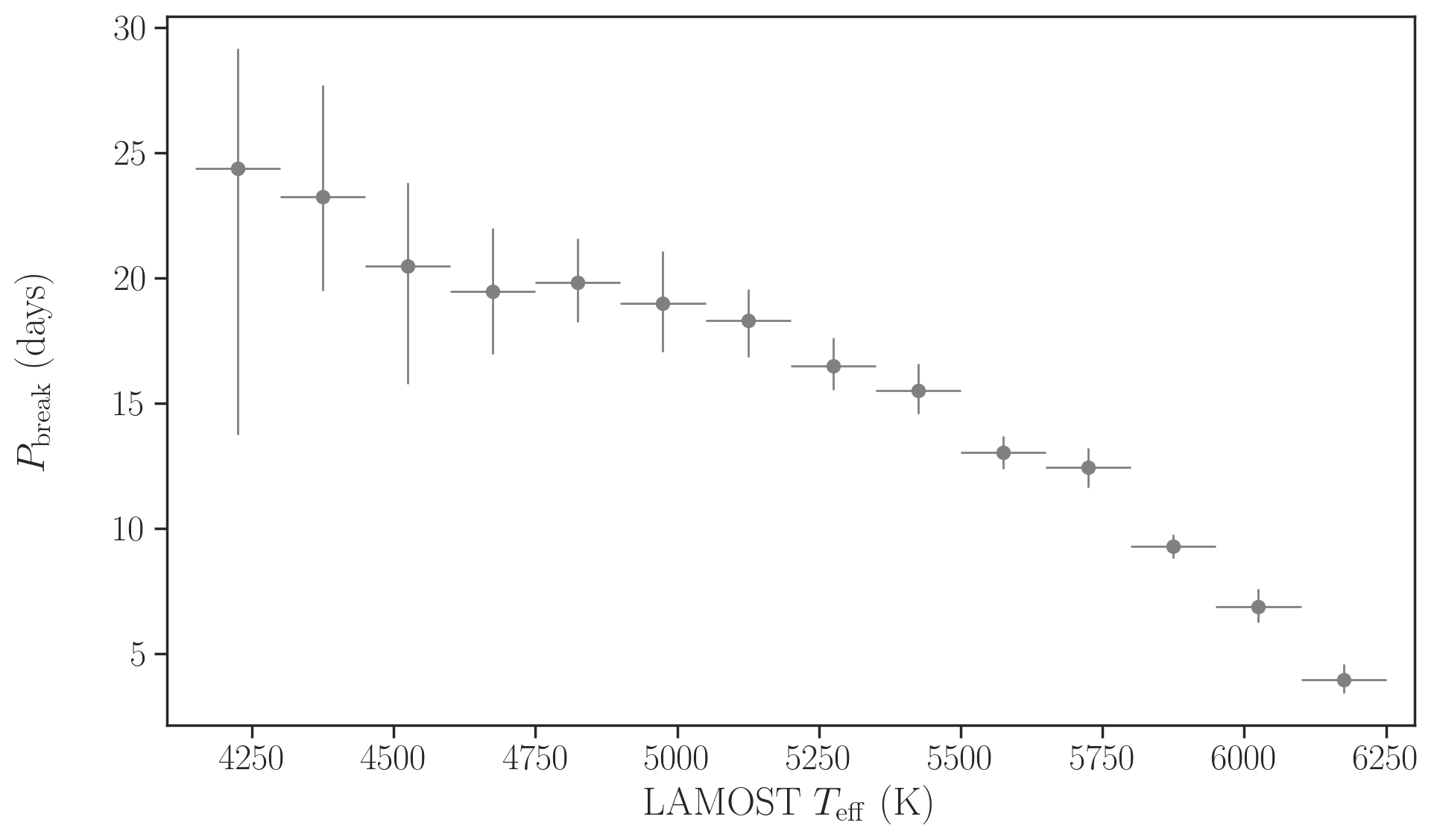}
    \caption{
    The means (circles) and 90\% intervals (vertical error bars) of the parameters in the $\amp(\prot)$ model (Section~\ref{ssec:amp_prot}) in different $\teff$ bins, whose widths are shown as horizontal error bars.
    }
    \label{fig:params_teff}
\end{figure}

Figure~\ref{fig:params_teff} shows the values of $\rbreak$, $\pbreak$, $\betaunsat$, $\betasat$ inferred in each $\teff$ bin. The values of $\rbreak$, $\betaunsat$, and $\betasat$ are largely insensitive to $\teff$, although the presence of possible trends is not excluded for $\rbreak$ and $\betaunsat$ (see also Section~\ref{ssec:r_ro_bias}).
A clear $\teff$ dependence is seen for $\pbreak$, which shortens quickly with increasing $\teff$ in agreement with the visual appearance of Figure~\ref{fig:r_prot_example} and Figure~\ref{fig:r_prot_all}.
This suggests that $\pbreak$ scales with $\tauc$, and motivates the analysis in the next subsection.
The strong curvature in the $\pbreak$--$\teff$ relation does not favor power-law scaling with parameters that depend more weakly on $\teff$, such as stellar mass and radius \citep[cf.][]{2014ApJ...794..144R}.

\subsection{Spot-Modulation Amplitude vs Rossby Number}\label{ssec:amp_ro}

The exact value of $\tauc$ depends on how it is estimated. 
The formula by \citet{1984ApJ...279..763N} has widely been used, which is based on theoretical evaluation of local turnover timescale near the bottom of the convective envelope as described in \citet{1980HiA.....5...91G}
and has been calibrated to minimize the scatter in the $\log R'_\mathrm{HK}$--$\ro$ relation.
Other scales have been proposed based on up-to-date stellar models and direct inference of the thickness of convective envelopes from asteroseismology  \citep{2010A&A...510A..46L,2021ApJ...912..127S, 2021ApJ...910..110L, 2021A&A...652L...2C}. These works generally yield $\tauc$ values that are larger by a factor of a few than \citet{1984ApJ...279..763N} for solar-mass main-sequence stars and have different dependence on $\teff$.
In this work, we use the formula in \citet{2011ApJ...741...54C} based on theoretical models of \citet{1998MNRAS.296..150G},
simply because it relates $\tauc$ directly to $\teff$ which we work on, and because it is close to the traditional scale by \citet{1984ApJ...279..763N} for dwarf stars \citep{1998MNRAS.296..150G} and allows for easier comparisons with other works.
We find that the typical $\amp$--$\ro$ relation in our sample remains unchanged for the $\tauc$ prescriptions in \citet{2021ApJ...910..110L} and \citet{2021A&A...652L...2C}.\footnote{The same appears to be the case for the analysis by \citet{2021ApJ...912..127S}, according to their Figure~3.} Thus the following discussion in Sections~\ref{sec:detection} and \ref{sec:discussion} is insensitive to which of these prescriptions is adopted, as long as $\ro$ is scaled adequately; see Appendix~\ref{ssec:tauc_amp} for details of these analyses.

Here we repeat almost the same analysis as in Section~\ref{ssec:amp_prot}, but replacing $\prot$ with the Rossby number $\ro = \prot/\tauc$ evaluated for each star; consequently, the model parameters $\pbreak$, $\sigma_{\ln\prot}$, and $\ln \prot^j$ in Equation~\ref{eq:plmodel}
are also replaced with $\robreak$, $\sigma_{\ln\ro}$, and $\ln \ro^j$, respectively (bottom part of Table~\ref{tab:priors}).
As noted above, the $\tauc$ value was calculated using the $\tauc$--$\teff$ relation ( Equation~36) given in \citet{2011ApJ...741...54C}.

\begin{figure}
    \epsscale{1.15}
    \plotone{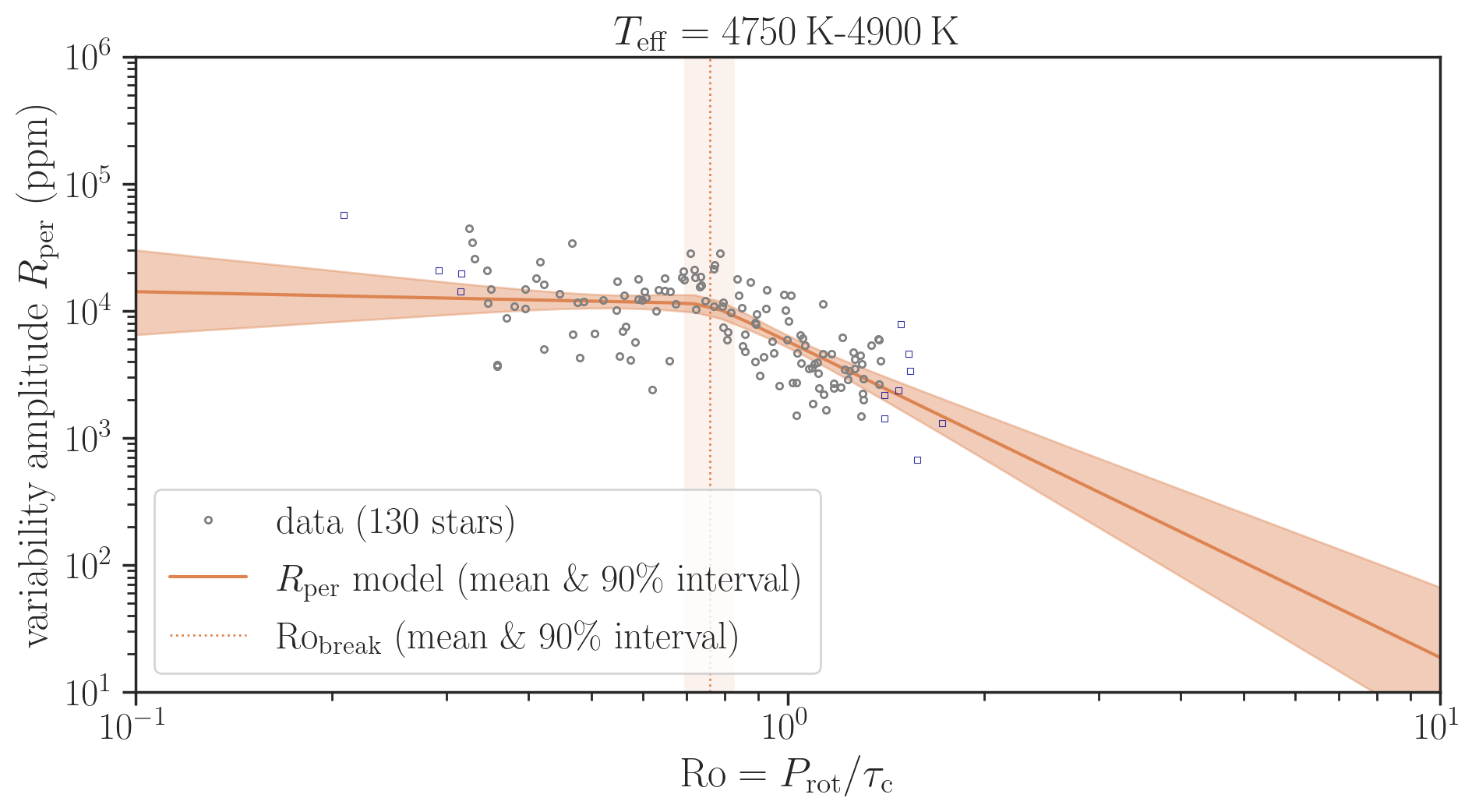}
    \plotone{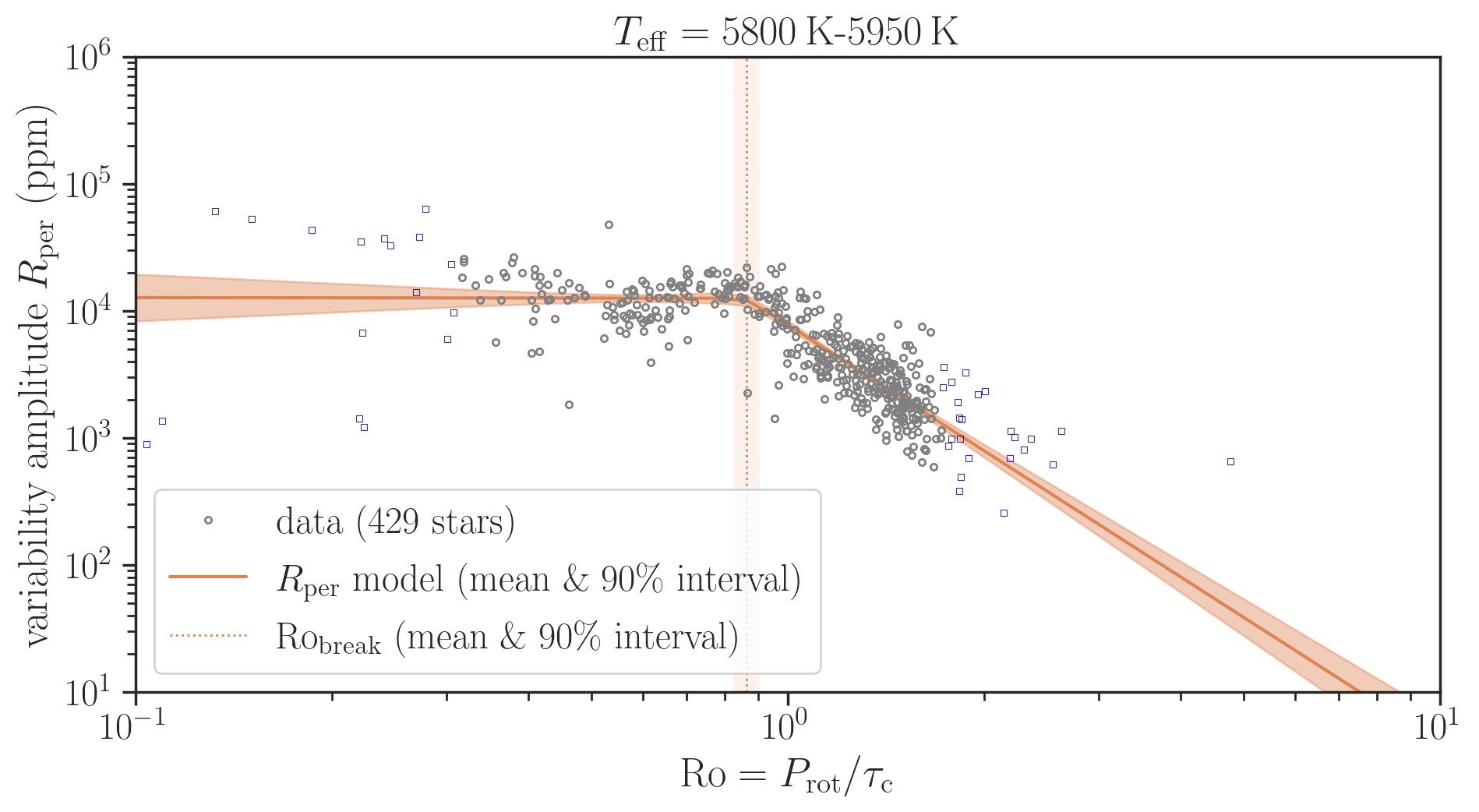}
    \caption{
    Spot-modulation amplitudes $\amp$ and Rossby numbers $\ro=\prot/\tauc$ for stars with $\teff=4,750$--$4,900\,\mathrm{K}$ (top) and with $\teff=5,800$--$5,950\,\mathrm{K}$ (bottom).
    Here $\tauc$ is based on the formula in \citet{2011ApJ...741...54C}.
    {\it Gray circles}:  Data points. 
    Blue open squares show the ones that were not used for modeling.
    {\it Orange solid line and shade}: Broken power-law model. 
    {\it Vertical orange dotted line and shade}: Inferred location of the break, $\robreak$. See Section \ref{ssec:amp_ro} for details.
    }
    \label{fig:r_ro_example}
\end{figure}

\begin{figure}
    \epsscale{1.15}
    \plotone{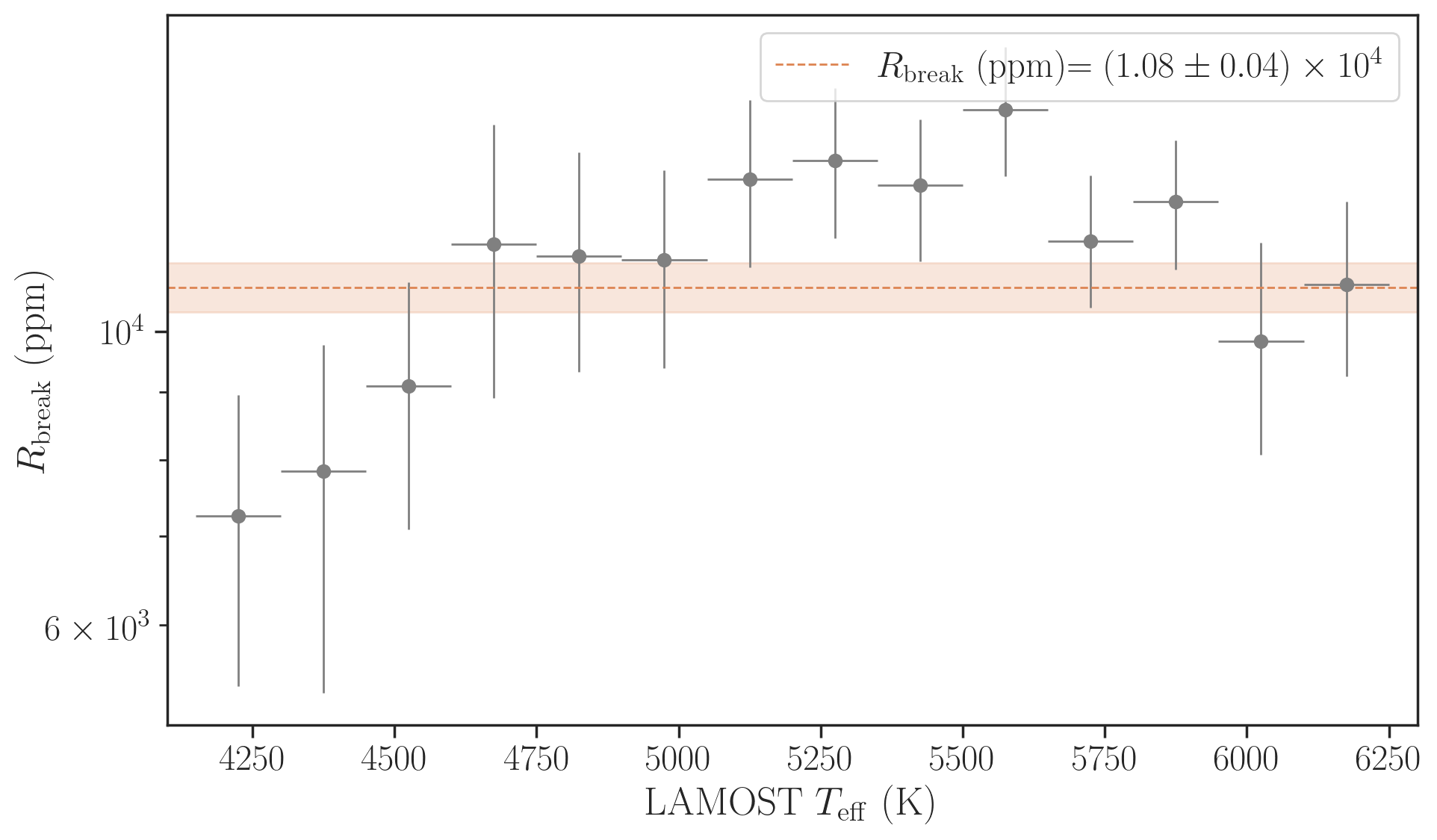}
    \plotone{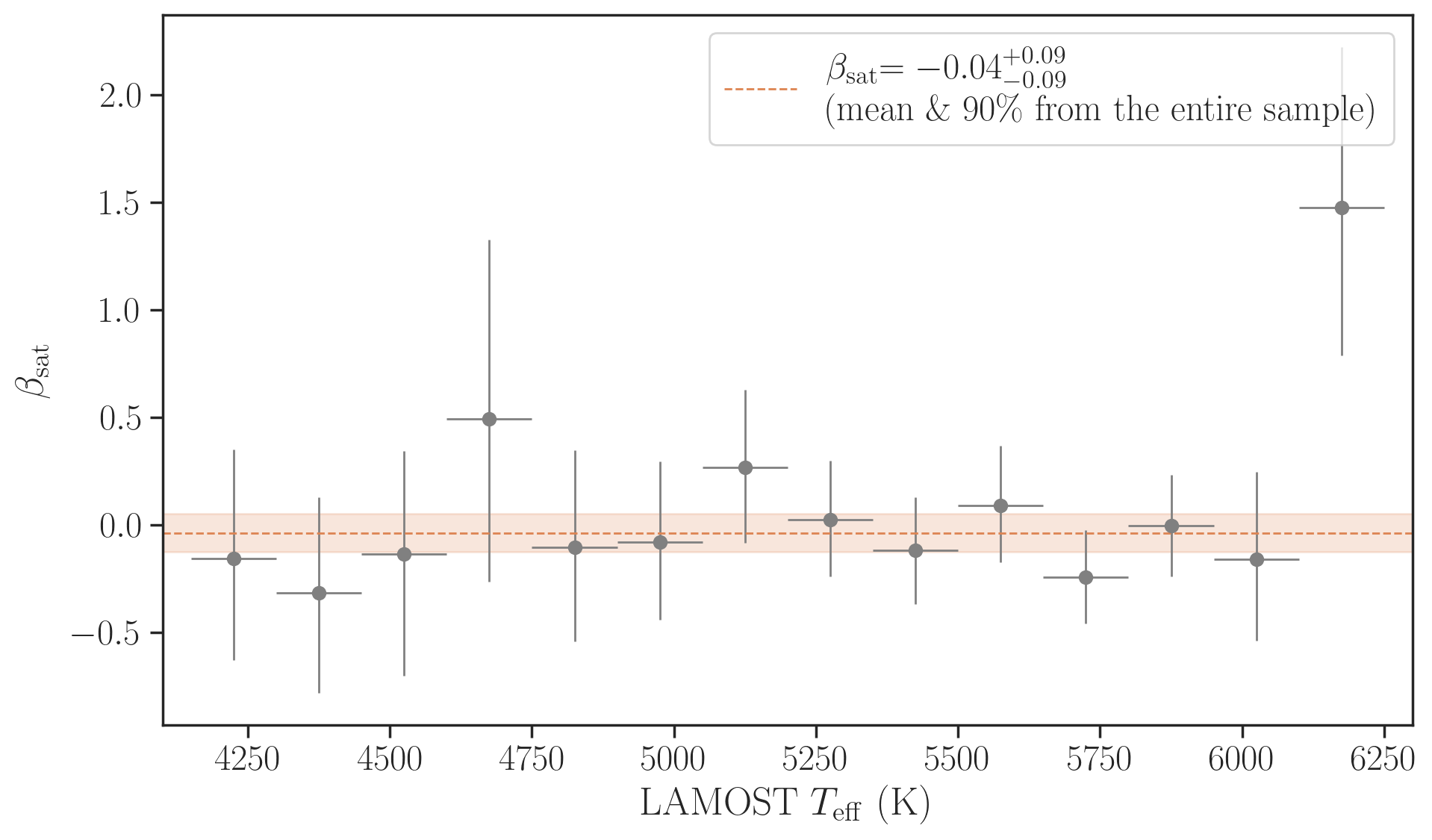}
    \plotone{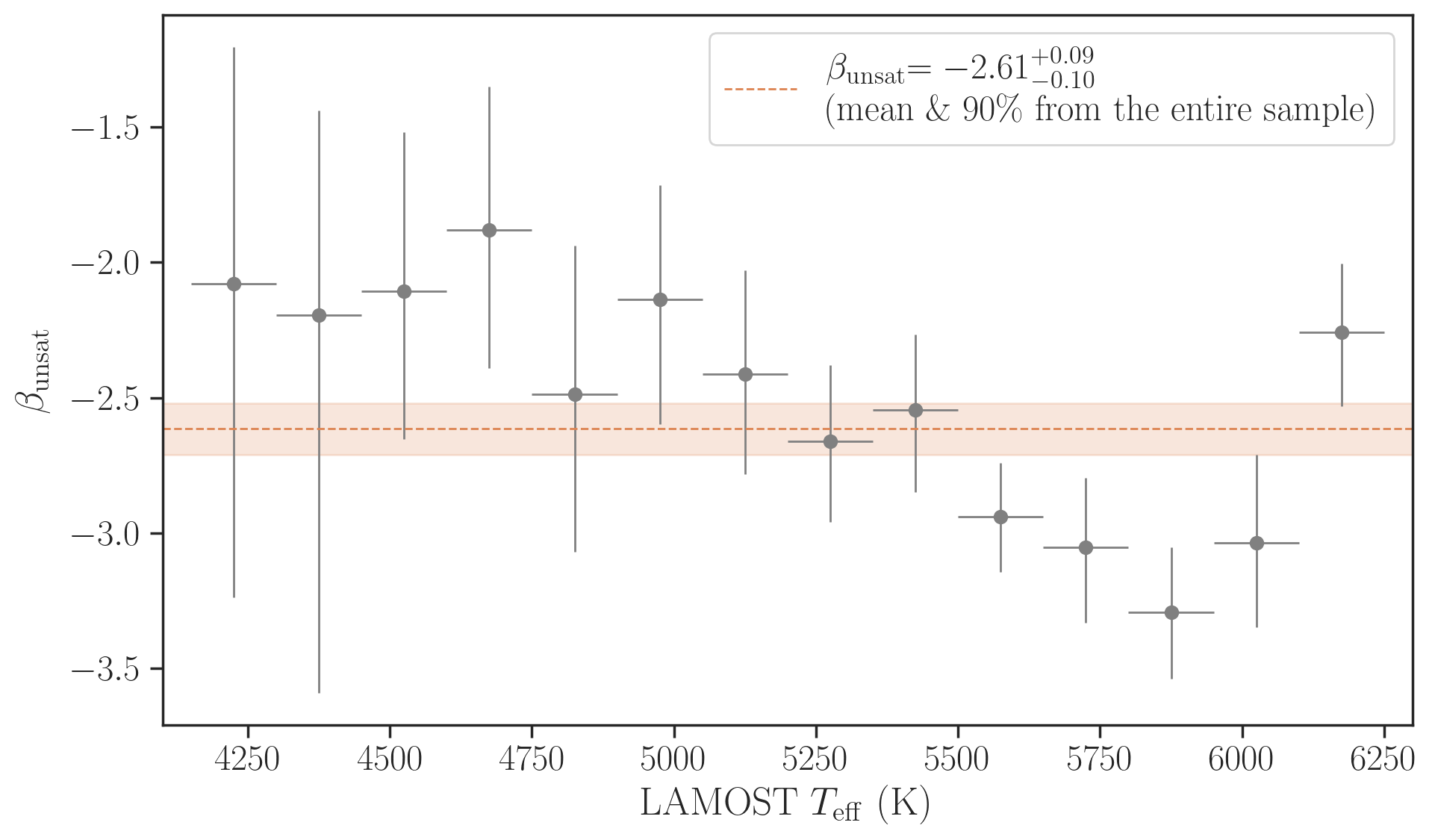}
    \plotone{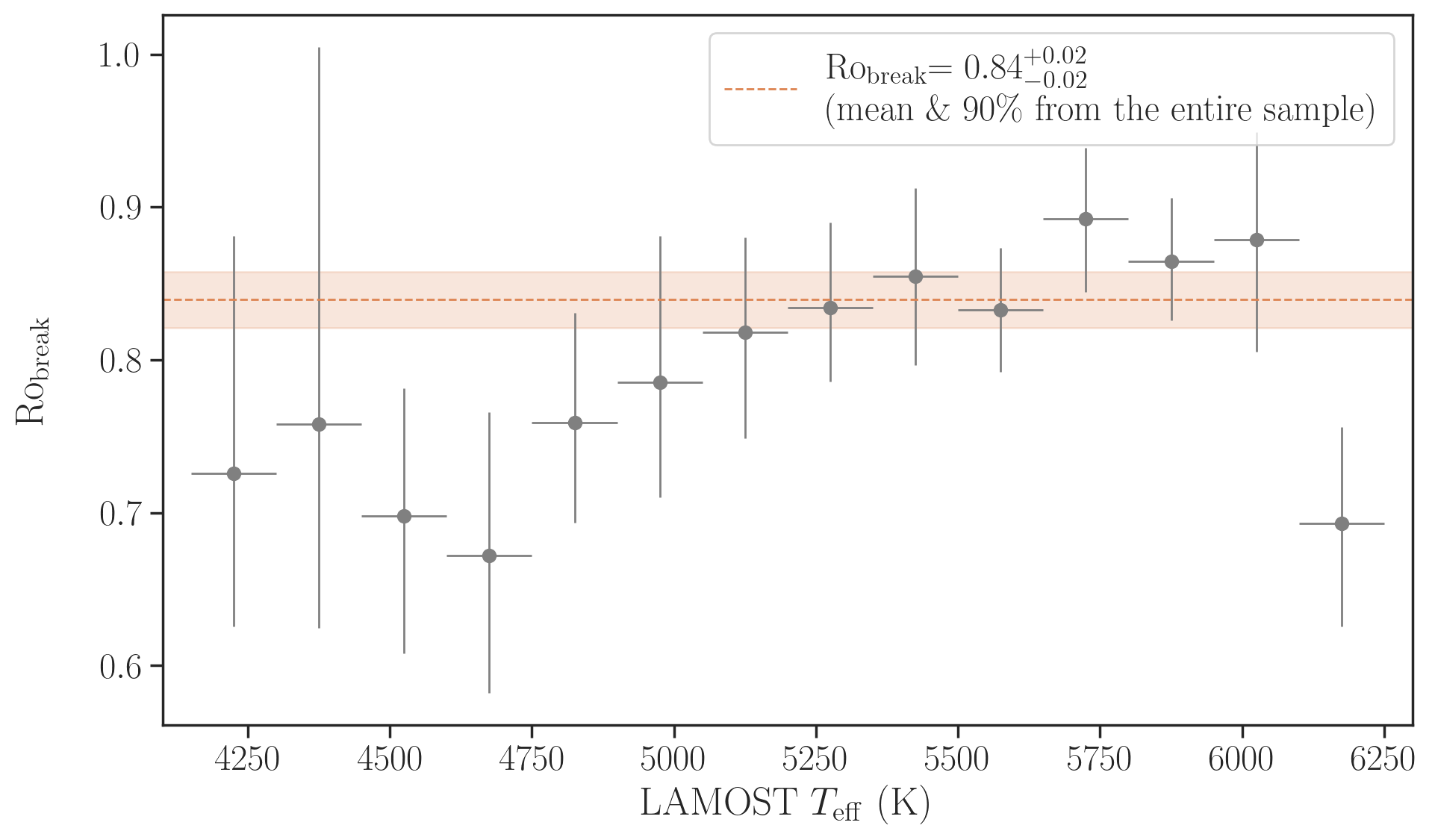}
    \caption{
    The means (circles) and 90\% intervals (vertical error bars) of the parameters in the $\amp(\ro)$ model (Section~\ref{ssec:amp_ro}) in different $\teff$ bins, whose widths are shown as horizontal error bars.
    }
    \label{fig:params_teff_ro}
\end{figure}

\begin{figure*}
    \epsscale{1.05}
    \plotone{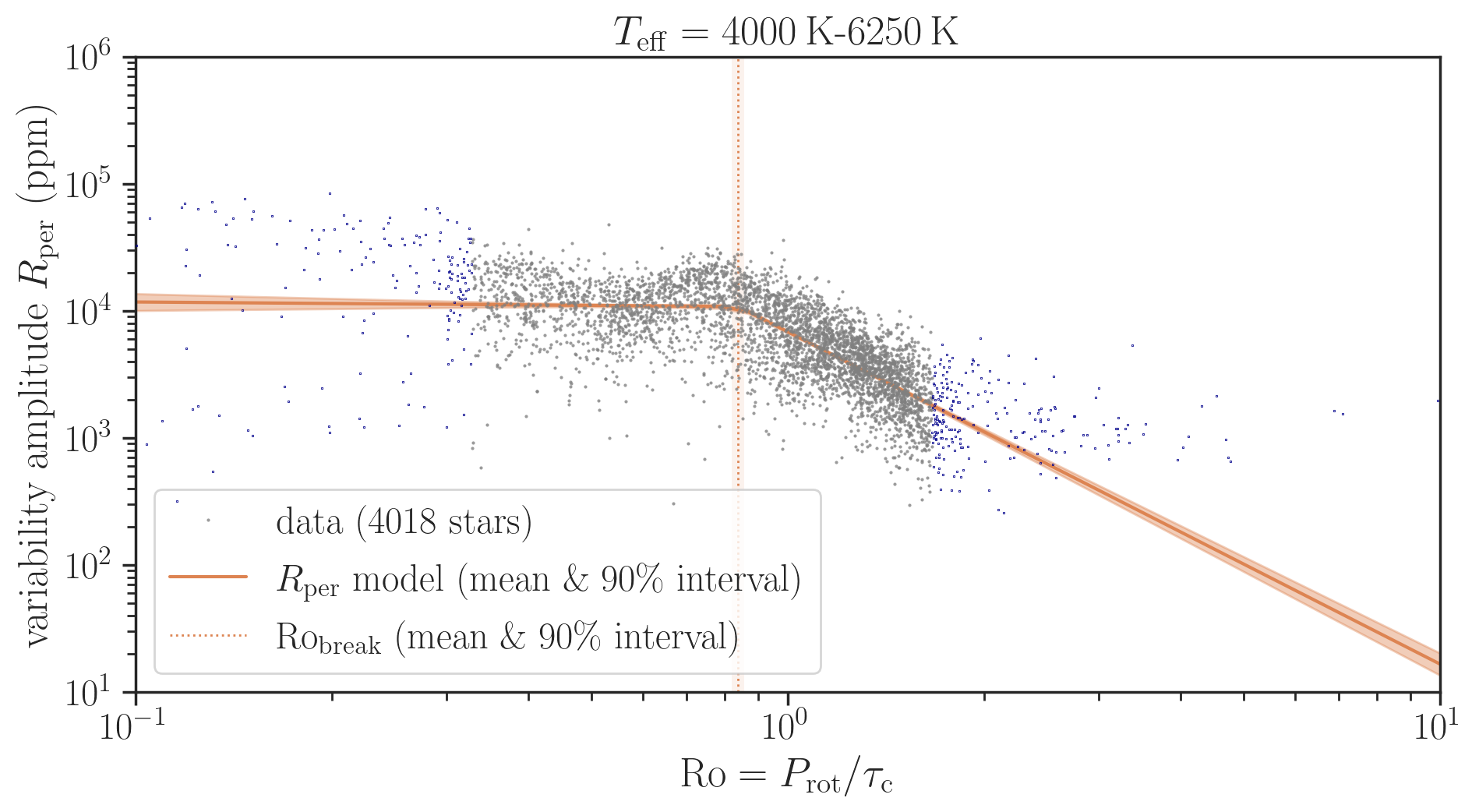}
    \caption{
    Spot-modulation amplitudes $\amp$ and Rossby numbers $\ro=\prot/\tauc$ for stars with $\teff=4,000$--$6,250\,\mathrm{K}$ in our sample, where $\tauc$ is based on the formula in \citet{2011ApJ...741...54C}.
    {\it Gray dots}:  Data points. 
    Blue dots show the ones that were not used for modeling.
    {\it Orange solid line and shade}: Broken power-law model. 
    {\it Vertical orange dotted line and shade}: Inferred location of the break, $\robreak$. See Section \ref{ssec:amp_ro} for details.
    }
    \label{fig:r_ro}
    
    \epsscale{1.05}
    \plotone{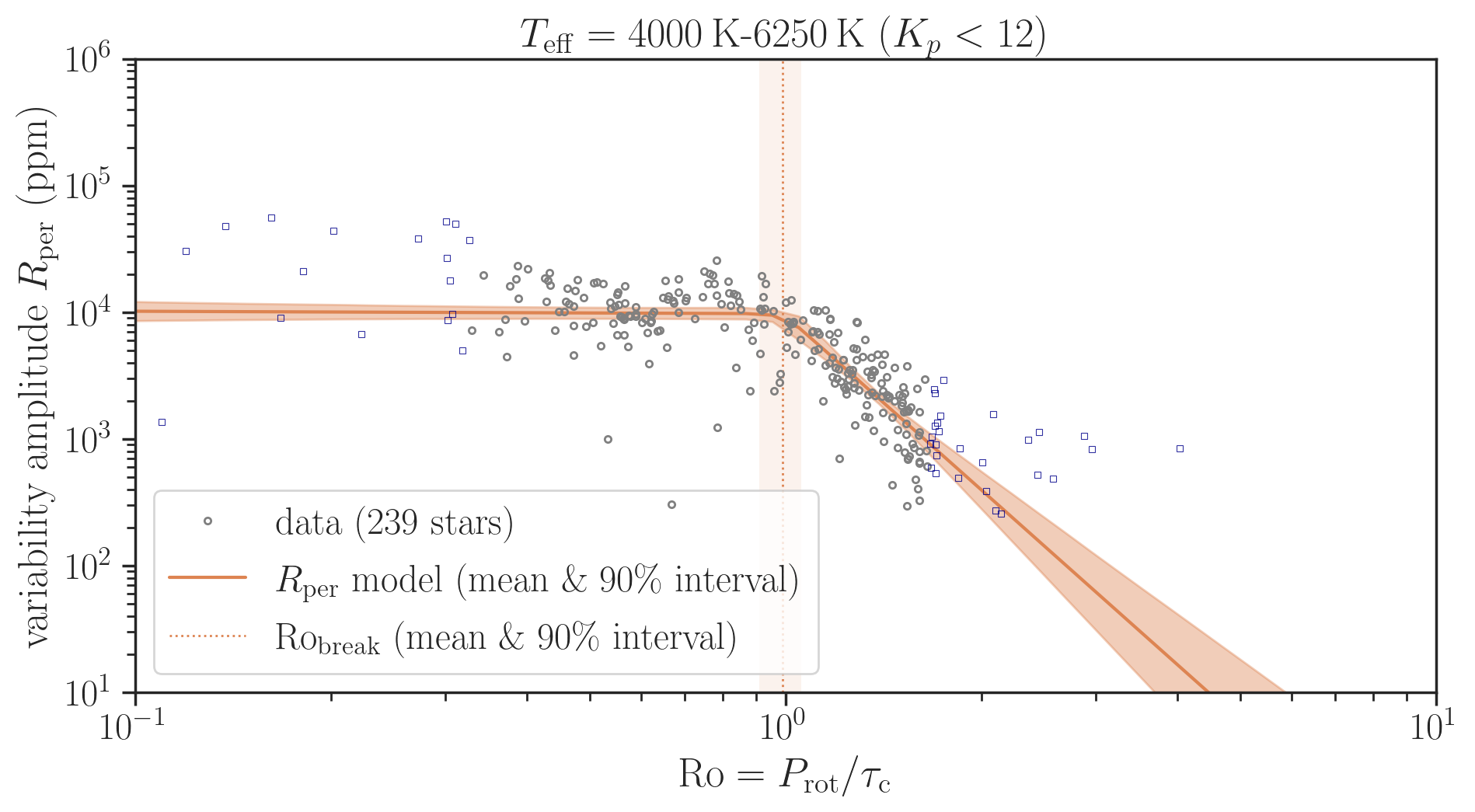}
    \caption{
    Spot-modulation amplitudes $\amp$ and Rossby numbers $\ro=\prot/\tauc$ for stars with $\teff=4,000$--$6,250\,\mathrm{K}$ and $K_p<12$ in our sample, where $\tauc$ is based on the formula in \citet{2011ApJ...741...54C}.
    {\it Gray circles}:  Data points. 
    Blue open squares show the ones that were not used for modeling.
    {\it Orange solid line and shade}: Broken power-law model. 
    {\it Vertical orange dotted line and shade}: Inferred location of the break, $\robreak$. See Section \ref{ssec:amp_ro} for details.
    }
    \label{fig:r_ro_kp12}
    
\end{figure*}

Figure~\ref{fig:r_ro_example} is analogous to Figure~\ref{fig:r_prot_example}; note that the $x$-axis is now $\ro$. We see a similar broken power-law pattern as seen in the $\amp$--$\prot$ plane, but now the break occurs at similar $\ro$ in different $\teff$ bins. The same is also true in other $\teff$ bins as shown in Figure~\ref{fig:r_ro_all}, which is analogous to Figure~\ref{fig:r_prot_all}.

Figure~\ref{fig:params_teff_ro} is analogous to Figure~\ref{fig:params_teff}, where now the break period $\pbreak$ is replaced with the break Rossby number $\robreak$.
The break location $\robreak$ now depends much less on $\teff$ than $\pbreak$ did, while the other parameters remain similar to those in Section~\ref{ssec:amp_prot}; this is reasonable because $\tauc$ in each narrow temperature bin is almost the same, and so the transition from $\prot$ to $\ro$ shift the whole data by almost the same amount in the $x$-direction.
On the other hand,
we also see some correlated pattern; for example, $|\betaunsat|$ and $\robreak$ may be systematically larger at higher $\teff$.\footnote{The hottest bin does not follow this trend, but in Figure~\ref{fig:r_ro_all} the data at $\ro\gtrsim 1$ are visually consistent with a steeper slope, which corresponds to larger $\robreak$. We suspect that the current result might be biased due to a larger fraction of outliers at low and high $\ro$.}
This could be of astrophysical origin, may be an artifact due to our imperfect knowledge of $\tauc(\teff)$ (see Appendix~\ref{ssec:tauc_amp}), may be due to detection bias (see Section~\ref{ssec:r_ro_bias}), or a combination of these effects. It is beyond the scope of this work to account for this possible dependence. 

Motivated by the (roughly) $\teff$-independent nature of the $\amp$--$\ro$ relation, in Figure~\ref{fig:r_ro} we show all stars with $\teff=4,000\,\mathrm{K}$--$6,250\,\mathrm{K}$ in the $\amp$--$\ro$ plane, and fit a single broken power-law relation $\amp(\ro)$ to the entire data (orange solid line and shade).
We find $\amp=(1.08\pm0.04)\times10^4\,\mathrm{ppm}$, $\robreak=0.84\pm0.02$, $\betasat=-0.04\pm0.09$, and $\betaunsat=-2.6\pm0.1$ (mean and 90\%). These values are also shown in Figure~\ref{fig:params_teff_ro} with horizontal orange dashed lines, which broadly agree with the values derived in separate $\teff$ bins. Here we see that the result from the entire sample tend to be closer to those of hotter stars, simply because they are more numerous in the sample. 
We discuss possible systematic errors in the inferred parameters due to detection bias against weak modulation further in Section~\ref{ssec:r_ro_bias}.

The kink in the $\amp$--$\ro$ relation has been noted in other works. The value of $\ro$ corresponding to the kink has been found to be $\sim 0.4$ in \citet{2021ApJ...912..127S} who estimated $\tauc$ using stellar models by \citet{2019A&A...631A..77A}; 0.23 in \citet{2021A&A...652L...2C} for their $\tauc$ calibration using seismic stars; and 0.82 when \citet{2021A&A...652L...2C} adopted the prescription by \citet{1984ApJ...279..763N}. 
When scaled by $\ro_\odot$, all these values roughly agree with what we found, $\ro/\ro_\odot \sim 0.4$ (since $\ro_\odot\approx 2$ in our scale).
We also note a wiggle for stars with $\ro\sim 0.4$--$0.8$ in our scale (or $0.2$--$0.4\,\ro_\odot$), which 
has also been noted by \citet{2021ApJ...912..127S} and is seen in our analyses using other $\tauc$ prescriptions (see Figure~\ref{fig:tauc} in Appendix~\ref{ssec:tauc_amp}).
While we do not understand its origin, we see a hint of a similar structure in the X-ray data, suggesting that this may not be an artifact related to the calibration of $\tauc$. See Section \ref{ssec:comparison} for further discussion.

\subsection{On the Impact of Detection Bias}\label{ssec:r_ro_bias}

Figure~\ref{fig:r_ro} shows that there exists some dispersion in $\amp$ at a fixed value of $\ro$: the dispersion is inferred to be $\approx 0.24\,\mathrm{dex}$ from our modeling in Section~\ref{ssec:amp_ro}. Some of the dispersion 
may be due to difference in spin-axis inclinations and/or activity cycles. Any systematics in the adopted $\tauc$--$\teff$ relation can also affect the scatter. In particular, we did not take into account its possible dependence on [Fe/H]. Indeed, we find that the residuals $\Delta\log_{10}\amp$ of the fit in Figure~\ref{fig:r_ro} is correlated with the LAMOST [Fe/H], with the Pearson R coefficient being $\approx 0.4$ or $\Delta\log_{10}\amp \approx 0.5\mathrm{[Fe/H]}$. This is qualitatively consistent with the finding of \citet{2021ApJ...912..127S} that metal-rich stars have enhanced activities, although the dependence may be weaker than was found to be typical by these authors.

As will be discussed in Section~\ref{sec:detection}, we find evidence that rotational modulation of the fainter stars has been missed due to their larger photometric noise.
In the presence of such a threshold, the dispersion in the $\amp$--$\ro$ relation --- regardless of its origin --- makes modulation with smaller $\amp$ more likely to be missed, and thus makes the observed slope to appear shallower. Correspondingly, $\robreak$ is inferred to be smaller.



\begin{figure*}
    \epsscale{1.05}
    \plotone{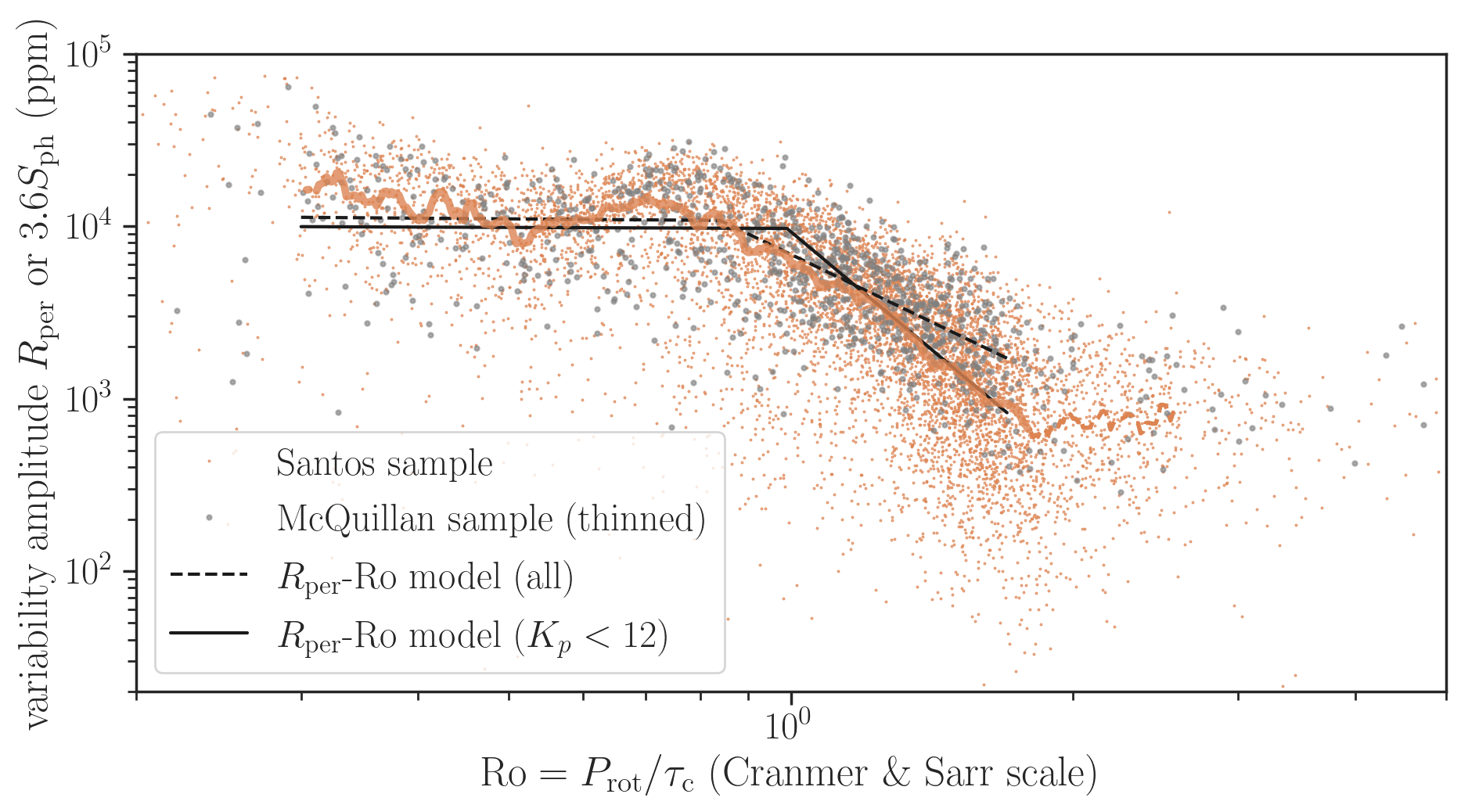}
    \caption{
    Spot-modulation amplitudes and Rossby numbers of stars from \citet[][{\it orange dots}]{2019ApJS..244...21S, 2021ApJS..255...17S} 
    compared with those of \citet[][{\it gray circles}: sample thinned by a factor of 10 to improve the visibility]{2014ApJS..211...24M}. The stars were selected in the same ways as described in Section~\ref{sec:sample}, and the modulation amplitudes in the Santos sample $S_\mathrm{ph}$ is multiplied by 3.6 so that the median values of $S_\mathrm{ph}$ and $R_\mathrm{per}$ match. The $\amp$--$\ro$ models derived from the \citet{2014ApJS..211...24M} stars using the entire sample in Section~\ref{ssec:amp_ro} (dashed) and using the brightest stars with $K_p<12$ in Section~\ref{ssec:r_ro_bias} (solid) are also shown with solid lines.
    The thick orange line is obtained by applying a median filter with the width of 0.02~dex to the Santos sample. 
    This curve matches well with the $\amp$--$\ro$ relation from the brightest stars (black solid line) up to $\ro \sim 2 \sim \ro_\odot$.
    Although the relation in the Santos sample appears to flatten at $\ro \gtrsim 2$, we note that this region --- shown with a dashed curve instead --- is also most sensitive to the detection bias against weak modulation.
    }
    \label{fig:santos}
\end{figure*}

To fully understand the impact of the detection bias, we need the complete knowledge of detection function {\it as well as $\ro$ values of all the observed stars with and without $\prot$ detections}, which is impractical. 
Instead, here we repeat the same analysis as in Section~\ref{ssec:amp_ro} for the brightest stars with $\teff=4,000$--$6,250\,\mathrm{K}$, using only the stars with the \kepler\ magnitudes $K_p<12$ for which detection bias appears to be minimal (see also Section~\ref{sec:detection}). Due to the correlation between $K_p$ and $\teff$, the resulting sample is mostly limited to stars with $\teff>5,500\,\mathrm{K}$.
The result is shown in Figure~\ref{fig:r_ro_kp12}; note that the data typically extend down to lower $\amp$ than in Figure~\ref{fig:r_ro}. From this analysis, we find $\robreak = 0.99^{+0.07}_{-0.08}$ and $\betaunsat=-4.6\pm0.9$ (mean and $90\%$ interval). These values differ by $\sim 0.1$ and $\sim 1$ from those inferred for stars with $\teff\gtrsim 5,500\,\mathrm{K}$ (Figure~\ref{fig:params_teff_ro}), in the directions consistent with what we expect from the detection bias as discussed above. 
This should be considered as systematics unaccounted for in our analysis in Section~\ref{ssec:amp_prot} and Section~\ref{ssec:amp_ro}.

As will be discussed in Section~\ref{sec:detection} further, cooler \kepler\ stars tend to have larger apparent magnitudes than the hotter ones due to their lower intrinsic luminosities. This causes cooler dwarfs to have higher $\amp$ thresholds, as is evident in Figure~\ref{fig:r_prot_all} (and Figure~\ref{fig:pr_teff}). Therefore the above bias is more severe for cooler stars, and the trend we see in Figure~\ref{fig:params_teff_ro} for $\betaunsat$ and $\robreak$ may in part be explained by this $\teff$-dependent bias. As expected, on the other hand, the $\amp$--$\ro$ relation inferred from the brightest stars remains unchanged at $\ro < \robreak$. Thus the $\teff$ dependence of $\rbreak$ seems real. This could be due to smaller spot coverage fractions in cooler stars, or smaller spot contrasts, or both.

\subsection{The Santos Sample}\label{ssec:santos}

While this paper mainly focuses on the \citet{2014ApJS..211...24M} sample, it is useful to consult other $\prot$ catalogs to better understand the applicability and limitations of the results based on this specific catalog.
We thus analyze the catalog by \citet{2019ApJS..244...21S, 2021ApJS..255...17S} that provided a larger number of $\prot$ measurements than \citet{2014ApJS..211...24M}.
We applied the same selection as described in Section~\ref{sec:sample} to the stars in \citet{2019ApJS..244...21S, 2021ApJS..255...17S} and found 8,713 (7,621) stars with $4,000\,\mathrm{K}<\teff<6,500\ (6,250)\,\mathrm{K}$. 

Figure~\ref{fig:santos} compares the photometric modulation amplitudes and Rossby numbers of the stars in this Santos sample (orange dots) against those in the McQuillan sample (gray circles), where the latter sample is thinned by a factor of 10 to improve the visibility while showing the main trend. Because the modulation amplitudes in the Santos catalog are given using the proxy $S_\mathrm{ph}$ \citep{2010Sci...329.1032G, 2014A&A...562A.124M}, here the values of $S_\mathrm{ph}$ are scaled uniformly by 3.6, which is the median of $\amp/S_\mathrm{ph}$ for stars in both samples.
Despite the simpleness of the conversion, the figure shows that the distributions of modulation amplitudes and $\ro$ in the two samples are very similar, except that the Santos sample reports $\prot$ for more stars with smaller amplitudes than in the McQuillan sample.
Quantitatively, the median-filtered $\amp$--$\ro$ relation in the Santos sample (thick orange line) follows more closely to the $\amp$--$\ro$ model derived from the $K_p<12$ stars in the McQuillan sample (black solid line), at least at larger $\ro$. This supports our argument on the detection bias in Section~\ref{ssec:r_ro_bias}: we argued that the $\amp$--$\ro$ relation based on all the stars (black dashed line) is shallower than that derived from the brightest stars (black solid line) because the former is biased against stars with weaker modulation, and here we do see that the Santos catalog that is apparently less biased against stars with weaker modulation follows the steeper relation.


Interestingly, the amplitude in the Santos sample appears to plateau again at $\ro\gtrsim \ro_\odot$. This hints that the modulation amplitude may not keep decreasing in the same way as in $\ro < \ro_\odot$. We note, however, that a more careful assessment of the detection function is required to confirm whether this is a typical behavior or not.
As will be discussed in detail below, the measured photometric amplitudes in this region are close to the photometric noise level of {\it Kepler}. It is thus conceivable that only the highest variability stars at given $\ro$ are seen here and/or that the measured amplitudes may be sensitive to how one corrects for the photon noise; although the latter is taken into account in the analysis of \citet{2021ApJS..255...17S}, the authors also comment on difficulties associated with small-amplitude modulation. We leave the detailed study of the amplitude--$\ro$ relation in this region for a future work. 
The following discussion is not affected by this ambiguity, because we will show that those stars mostly fall below the detection limit in the \citet{2014ApJS..211...24M} sample anyway --- unless the $\amp$--$\ro$ relation turns up at larger $\ro$.

\section{Detection Edge in the McQuillan Sample}\label{sec:detection}

Now we attempt to clarify how the $\amp$--$\prot$ ($\ro$) relation discussed in Section~\ref{sec:analysis}, when combined with the detection bias, sculpts the longer-period edge of the $\prot$--$\teff$ distribution in the {\it Kepler} sample.

\begin{figure}
    \epsscale{1.15}
    \plotone{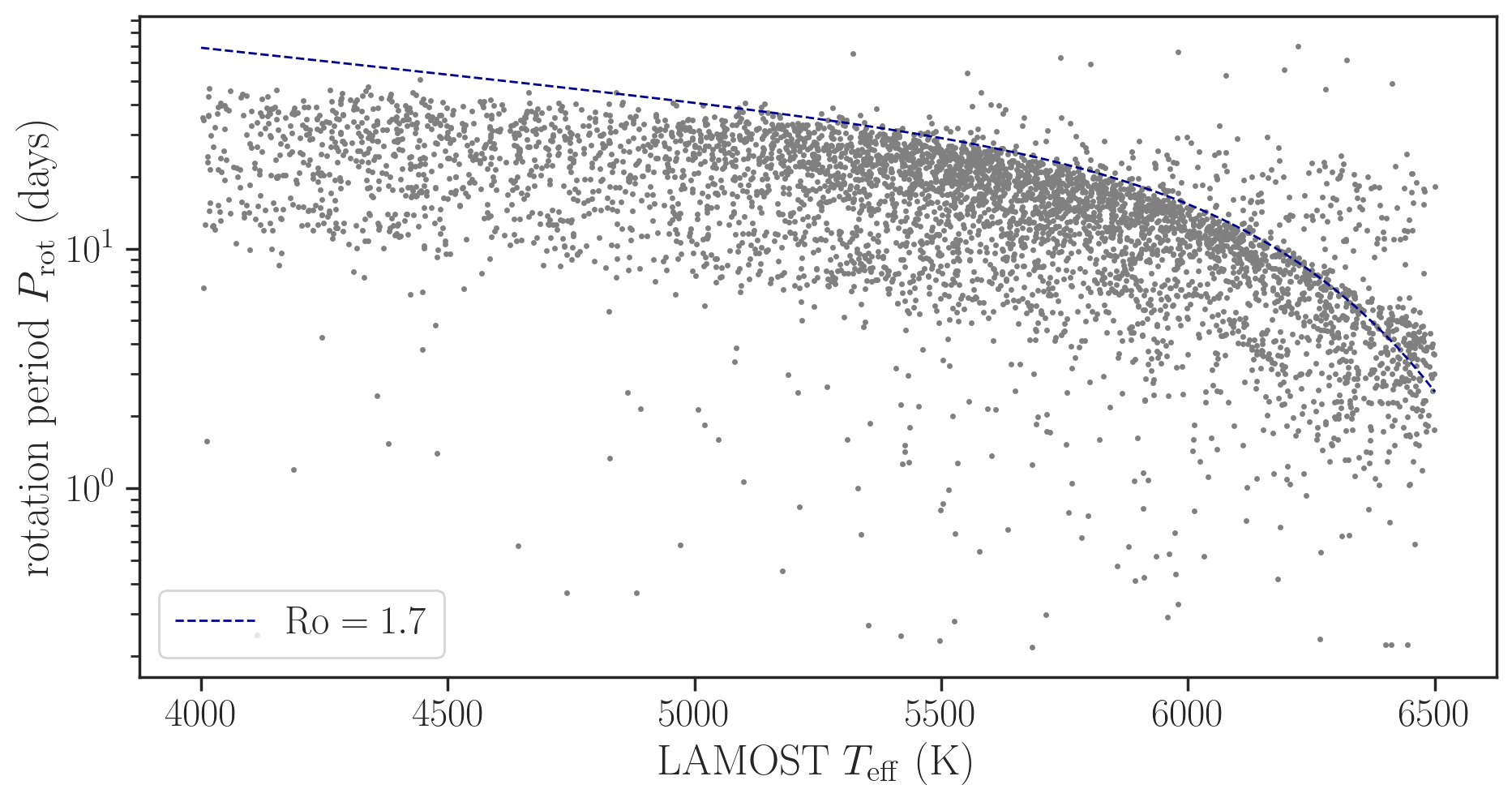}
    \plotone{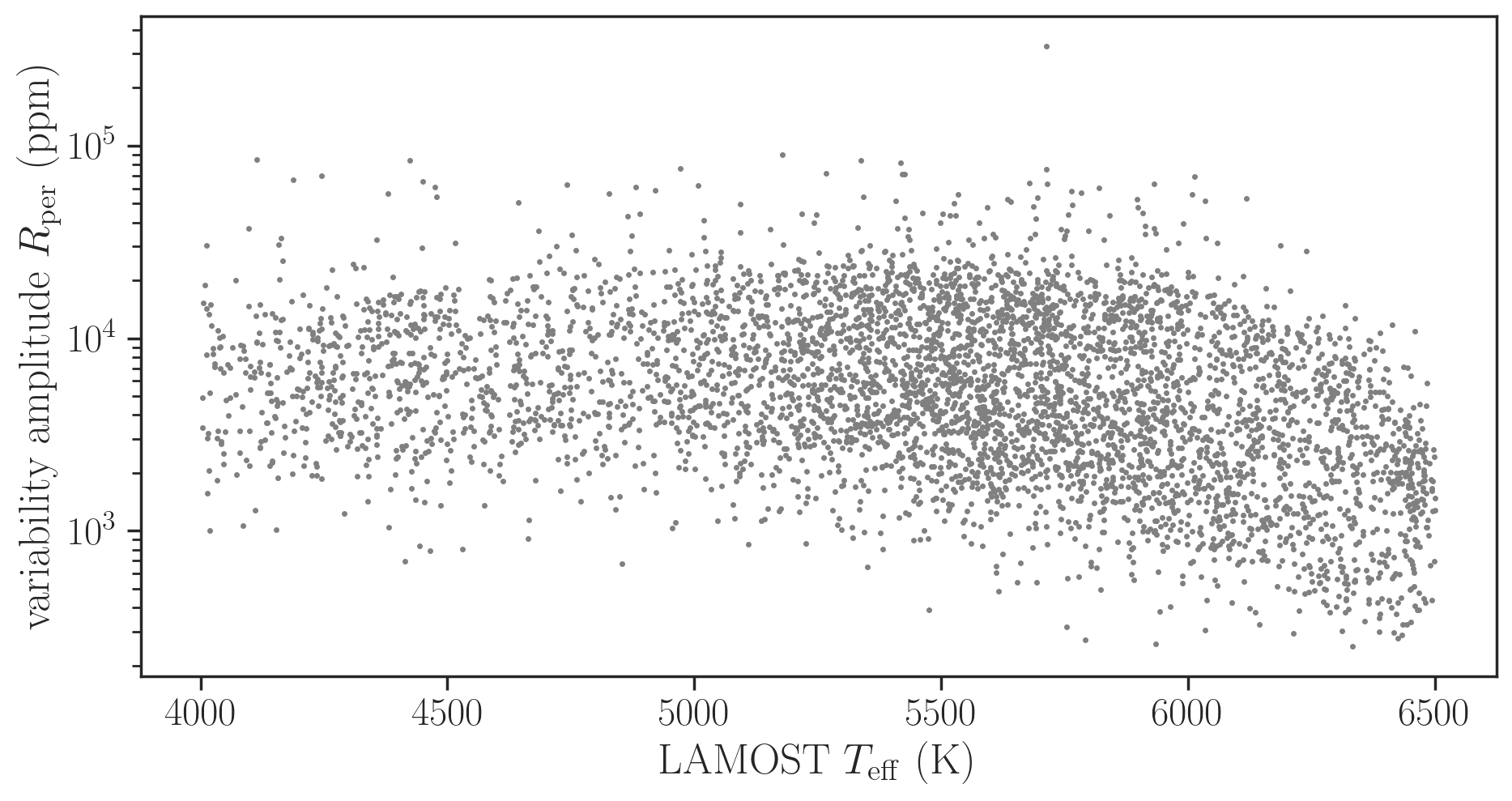}
    \caption{
    $\prot$ (top) and $\amp$ (bottom) as a function $\teff$ of the stars in our sample.
    The blue dashed line in the top panel corresponds to $\ro=\prot/\tauc=1.7$, where $\tauc$ is from the formula in \citet{2011ApJ...741...54C}. See Section~\ref{sec:detection} for details.
    }
    \label{fig:pr_teff}
\end{figure}

The $\prot$--$\teff$ distribution of \kepler\ stars has been known to exhibit a rather well-defined upper edge. 
One perplexing aspect of this upper edge is that it does not correspond to a constant variability amplitude, i.e., the {\it lower} edge of the $\amp$--$\teff$ distribution is not flat. These features are also apparent in our sample (Figure \ref{fig:pr_teff}).

\citet{2014ApJS..211...24M} pointed out that the upper $\prot$--$\teff$ edge lies roughly around a gyrochrone of the solar age.
\citet{2019ApJ...872..128V} pointed out that the upper edge is around $\ro\sim 2$ and discussed the possibility that this is related to detection bias.
Another explanation they proposed is that the edge is due to stalled spin down: the stars stop spinning down once they reach $\ro\sim2$ \citep{2016Natur.529..181V} and stay around the edge.
The presence of stalled spin down has now been supported by multiple studies as mentioned in Section \ref{sec:intro}; more recently,
the pile-up in the $\prot$-$\teff$ distribution around its upper edge (also apparent in the top panel of Figure \ref{fig:pr_teff}) has been reported and argued to provide a further support for the stalled spin down scenario \citep{2022arXiv220308920D}.
Nevertheless, these arguments for the presence of stalled spin down do not necessarily exclude the possibility that the edge in the \citet{2014ApJS..211...24M} sample is shaped by detection bias.

\begin{figure}
    \epsscale{1.15}
    \plotone{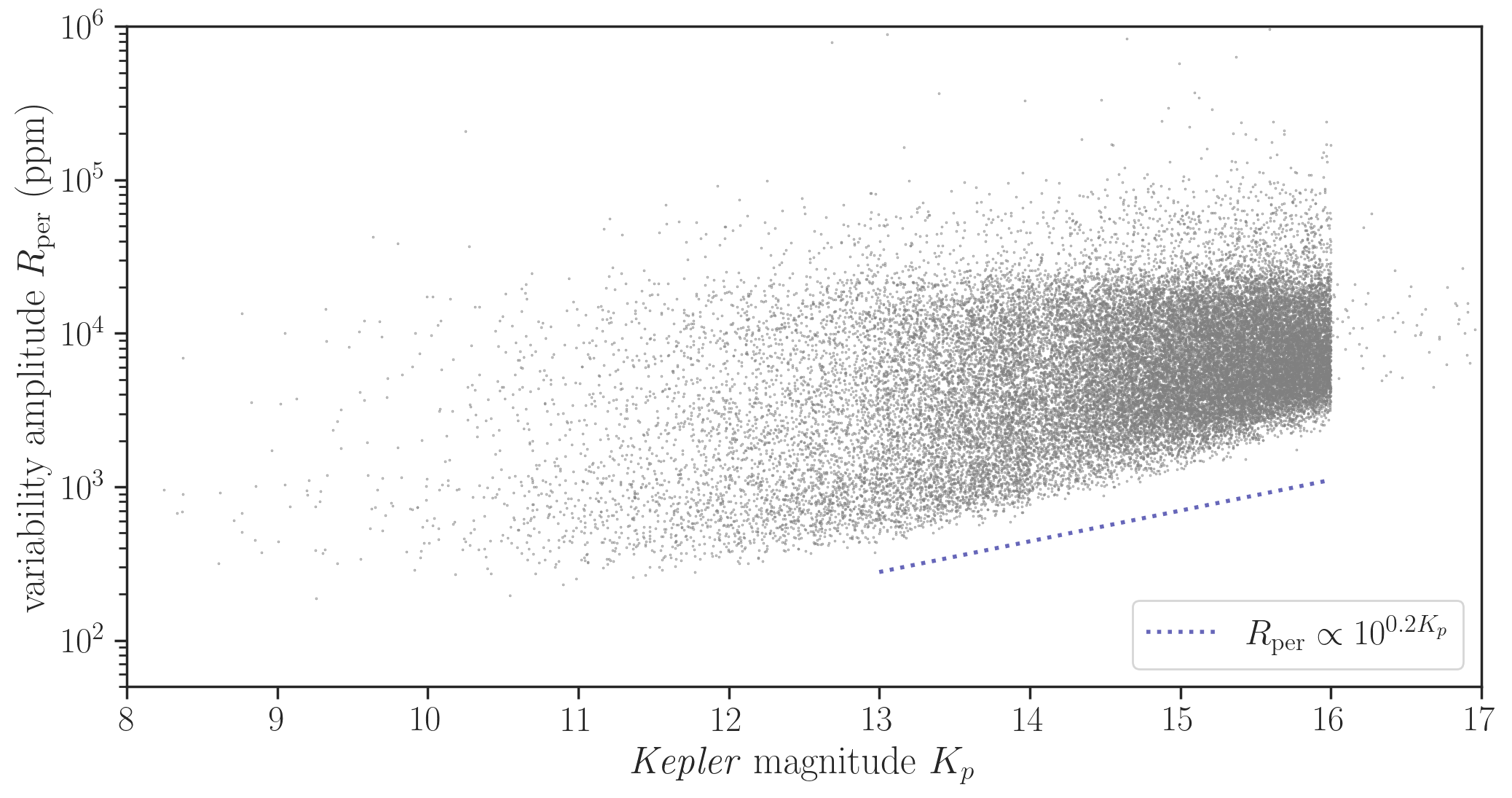}
    \caption{
    Modulation amplitudes $\amp$ vs {\it Kepler} magnitudes $K_p$ of all the stars with rotational modulation detected in \citet{2014ApJS..211...24M}. The blue dotted line shows the scaling $\amp \propto 10^{K_p/5}$ excepted for pure photon noise.
    }
    \label{fig:rper_kepmag}
\end{figure}

\begin{figure}
    \epsscale{1.15}
    \plotone{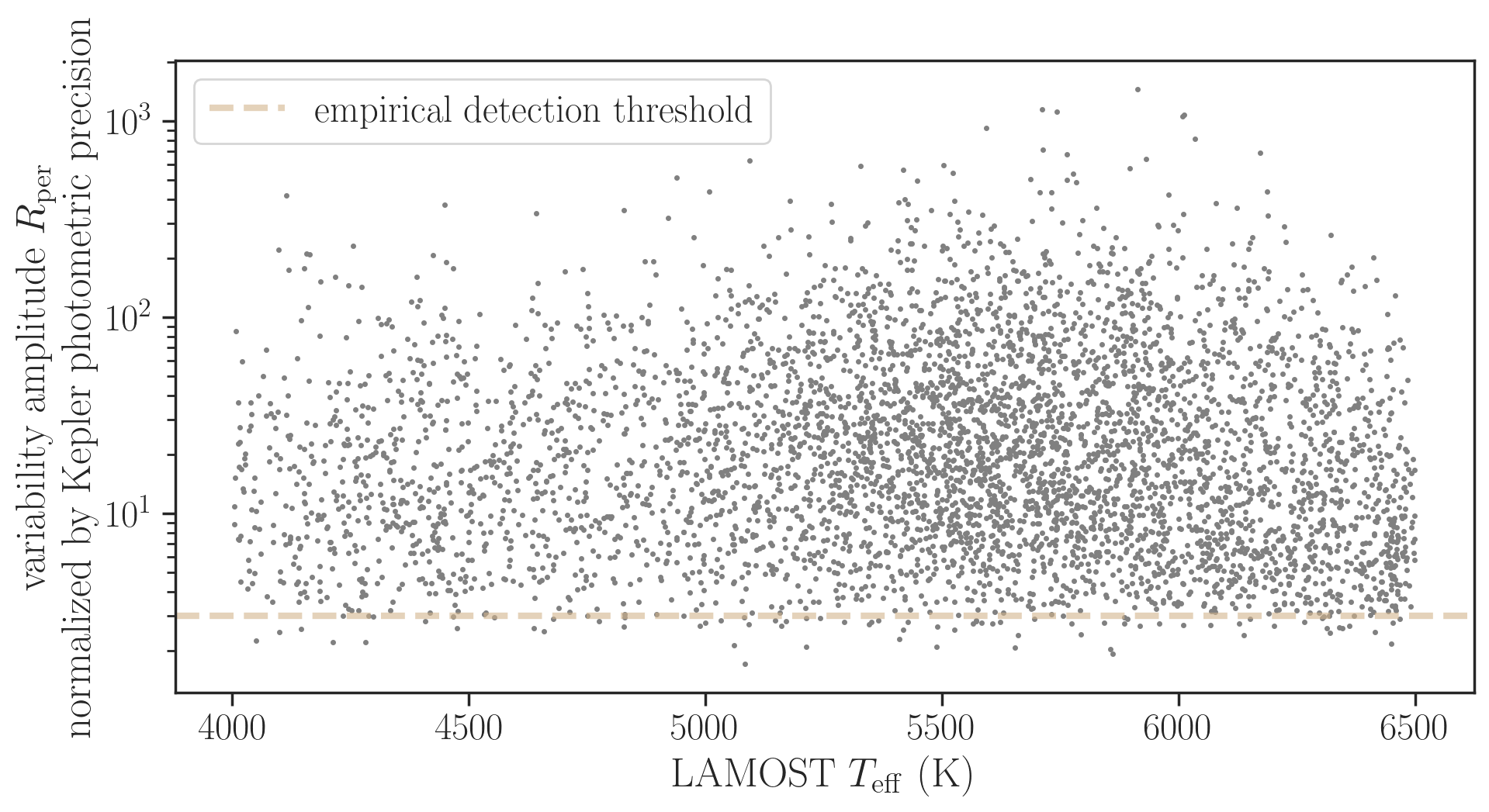}
    \plotone{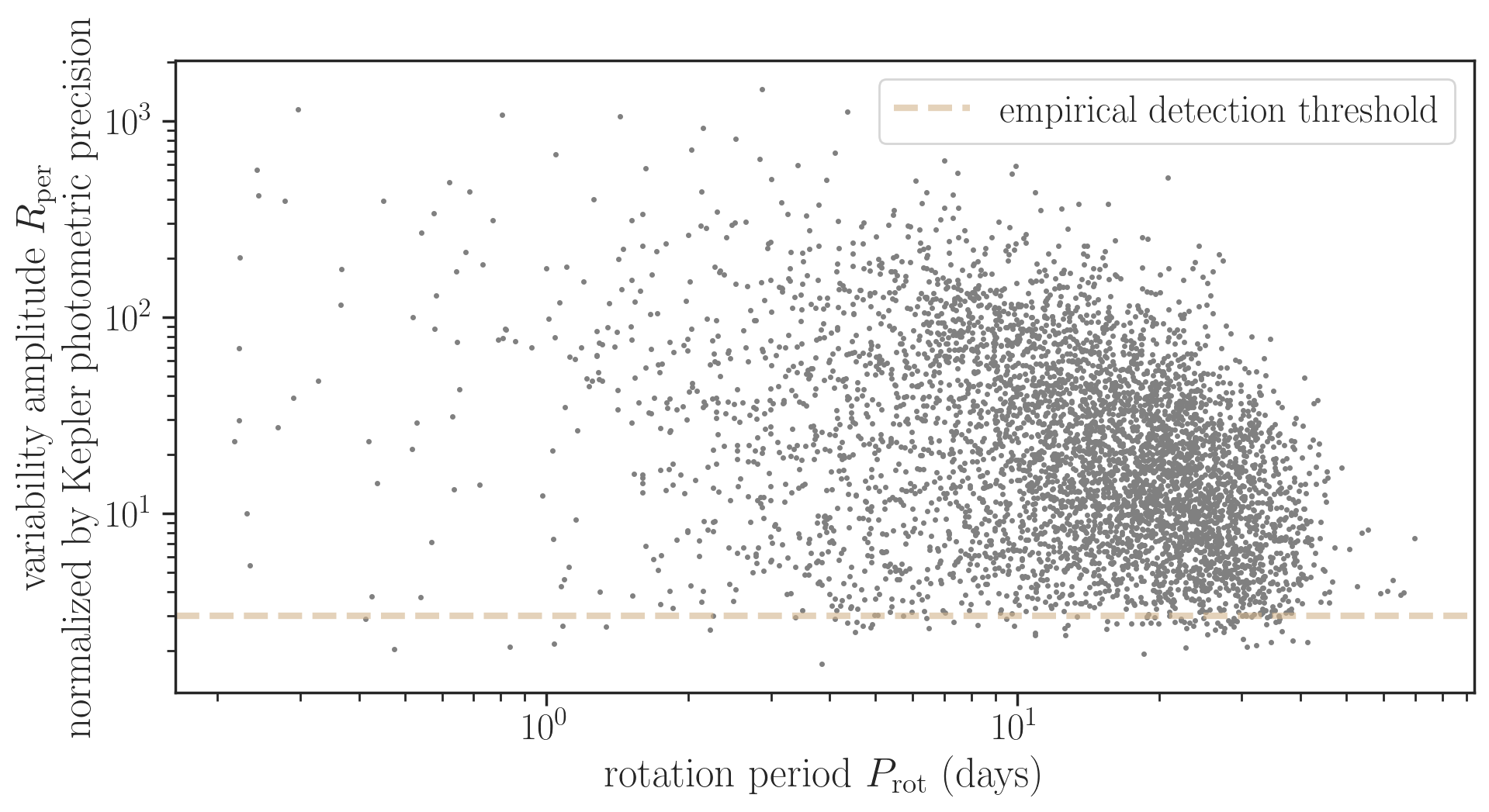}
    \plotone{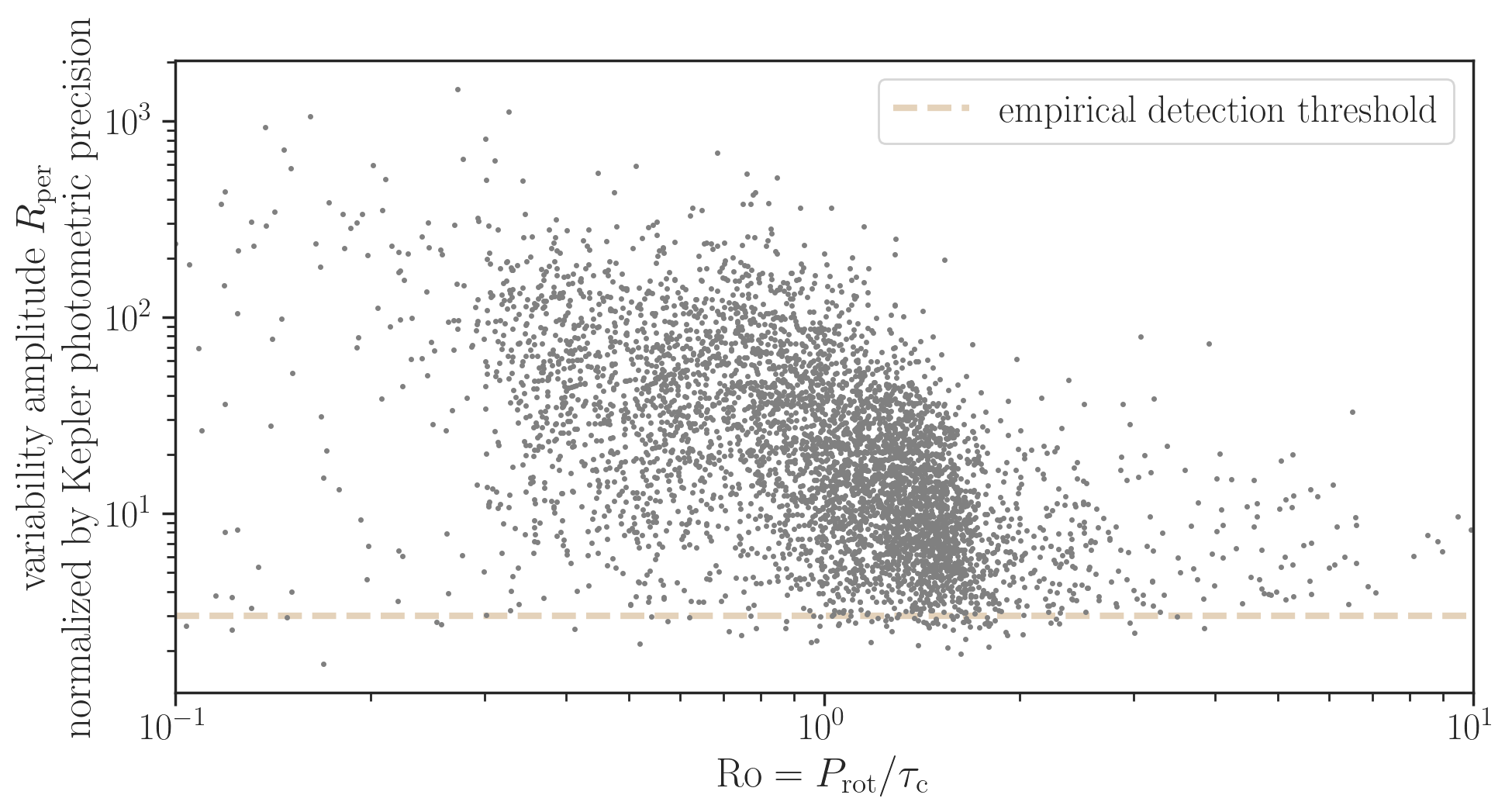}
    \caption{
    Spot-modulation amplitude $\amp$ normalized by photometric precision of {\it Kepler} (long-cadence exposure), as a function of $\teff$ {\it (Top)}, $\prot$ {\it (Middle)}, and $\ro$ {\it (Bottom)} of our sample stars. Note that the photometric precision is evaluated for each star based on the star's {\it Kepler} magnitude.
    The horizontal dashed line shows the empirical detection threshold adopted in this paper (see Section~\ref{sec:detection}).
    }
    \label{fig:rnorm_teff}
\end{figure}

\begin{figure}
    \epsscale{1.15}
    \plotone{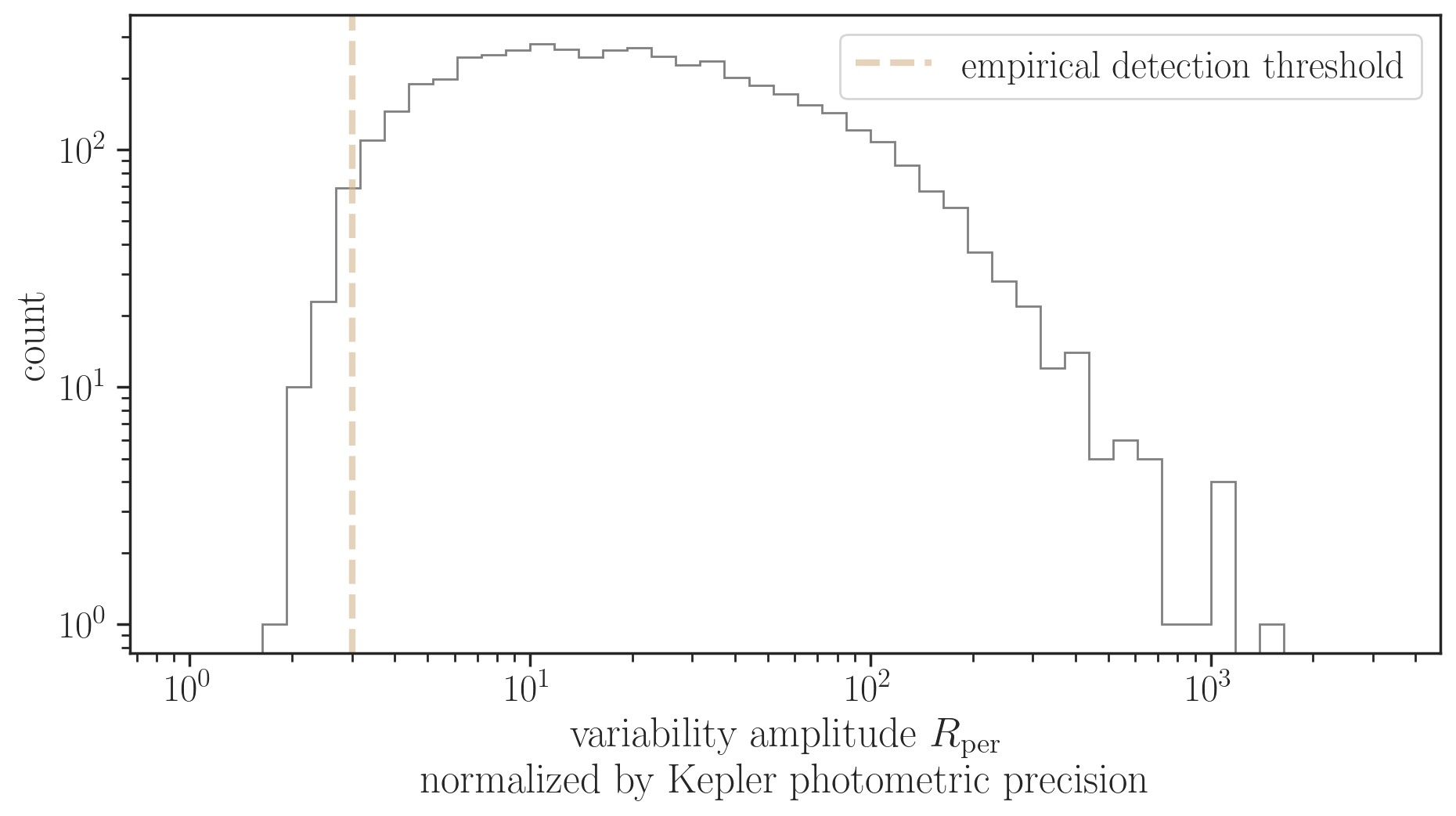}
    \caption{
    Histogram of the spot-modulation amplitude $\amp$ normalized by the magnitude-dependent photometric precision of {\it Kepler} in our sample.
    The vertical dashed line shows the empirical detection threshold
    (see Section~\ref{sec:detection}).
    }
    \label{fig:rnorm_hist}
\end{figure}

\begin{figure}
    \epsscale{1.15}
    \plotone{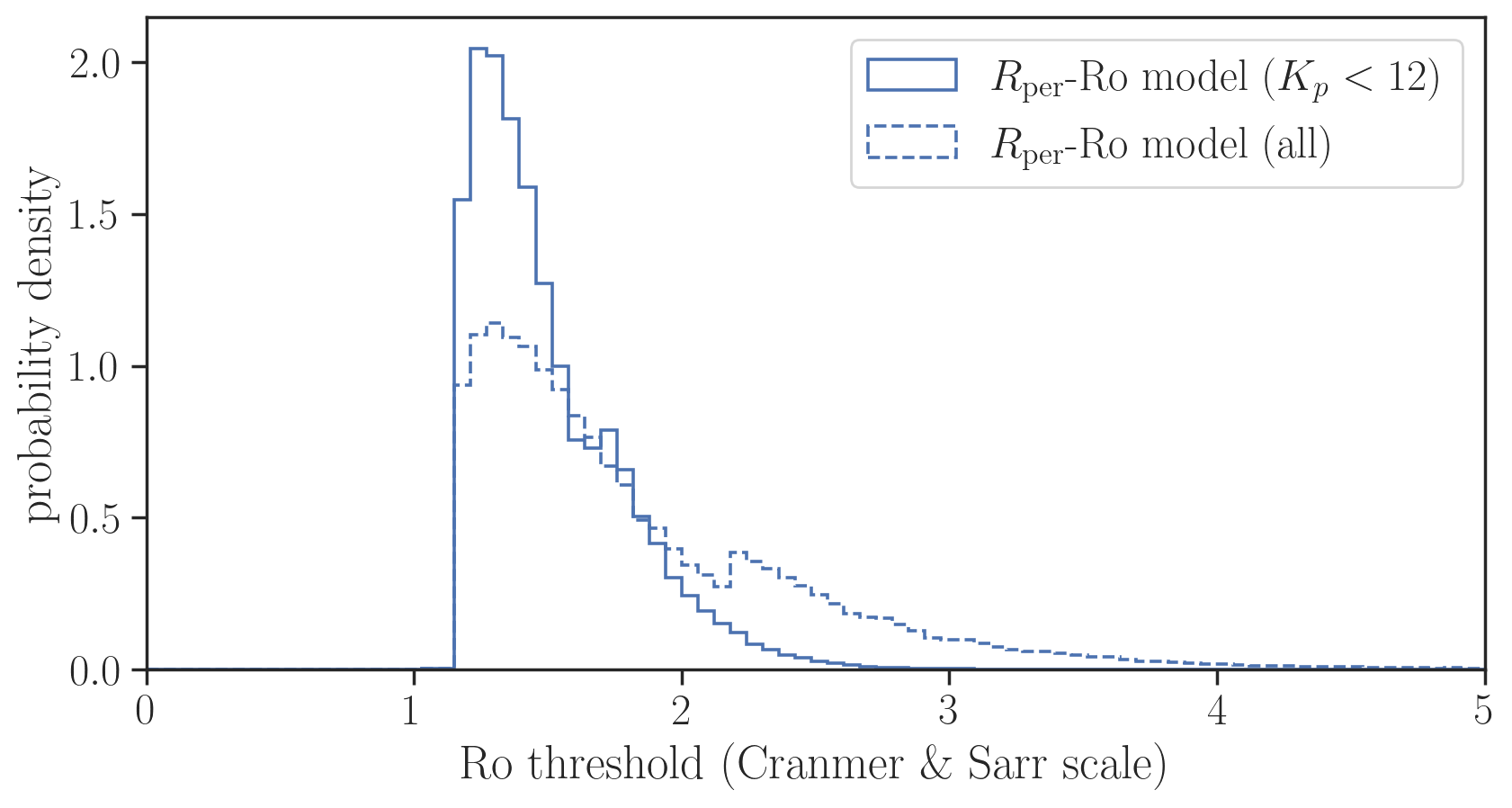}
    \plotone{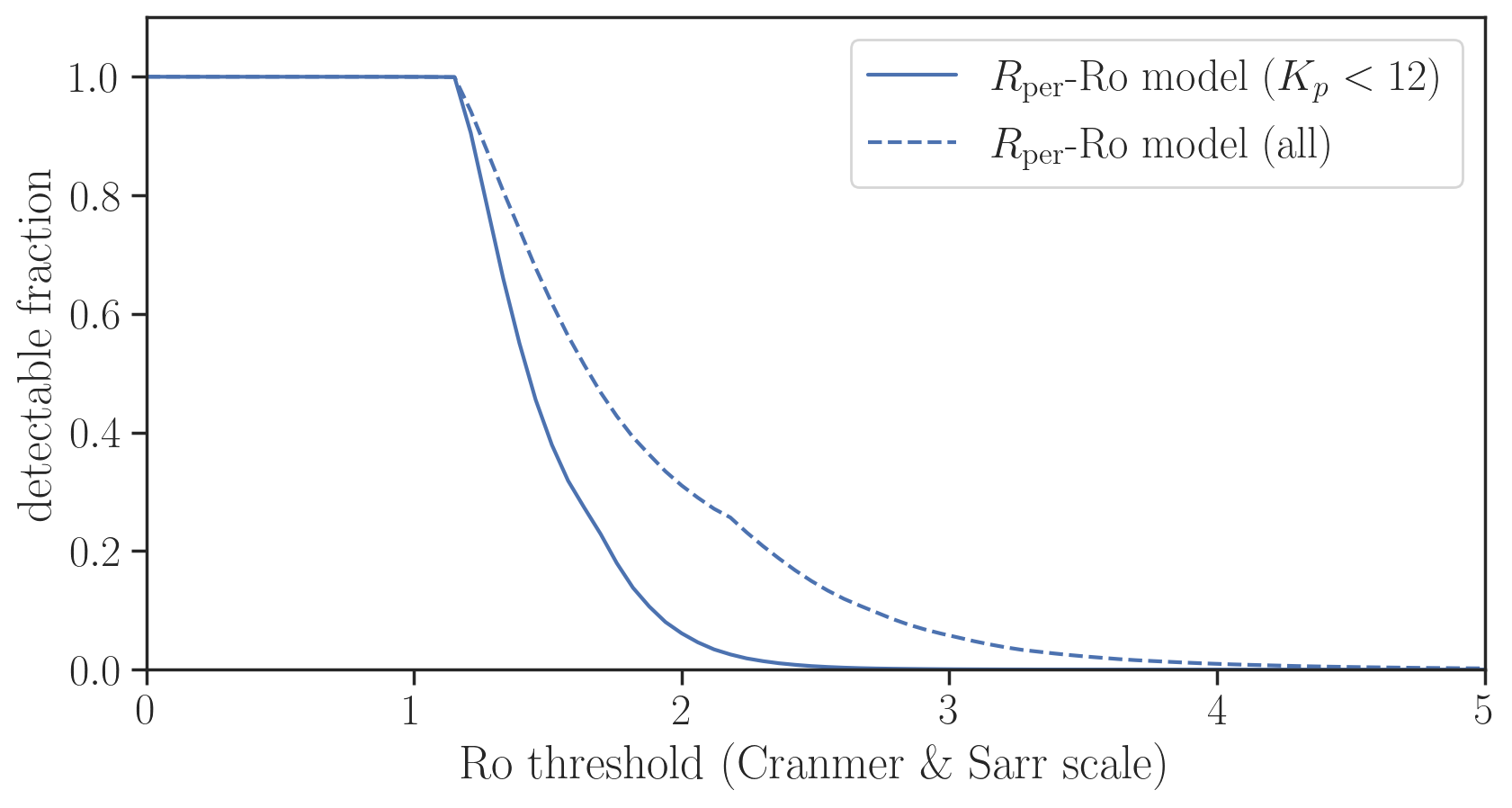}
    \caption{
    {\it (Top)} The distribution of $\ro$ corresponding to the magnitude-dependent detection threshold for all the stars for which rotatioanl modulation has been {\it searched} by \citet{2014ApJS..211...24M}.  
    The solid histogram is based on the $\amp$--$\ro$ relation derived using the brightest sample stars with $\teff=4,000$--$6,250\,\mathrm{K}$ and $K_p<12$ (Section~\ref{ssec:r_ro_bias}). The dashed one 
    is based on the $\amp$--$\ro$ relation derived using all the sample stars with $\teff=4,000$--$6,250\,\mathrm{K}$ (Section~\ref{ssec:amp_ro}).
    {\it (Bottom)}
    The normalized inverse cumulative functions of the distributions in the top panel, which provides the fraction of stars in the searched sample for which rotational modulation of a star with a given value of $\ro$ would have been detectable. 
    }
    \label{fig:roth_hist}
\end{figure}

\subsection{Evidence for the Noise-Dependent Cutoff}

Here, we argue that this edge results from the detection threshold set by photometric precision of {\it Kepler} that depends on apparent magnitudes of stars in the {\it Kepler} band, $K_p$. 
We first note, in Figure~\ref{fig:rper_kepmag}, that the distribution of $\amp$ and $K_p$ for all the stars with $\prot$ detection in \citet{2014ApJS..211...24M} has a sharp lower edge with a positive slope, whose value at $K_p\gtrsim 13$ is not far from the scaling for pure photon noise: $\amp \propto 10^{K_p/5}$. This indicates that the detectability is limited by photometric precision for those fainter stars, which comprise the majority of the sample.
In the top panel of Figure~\ref{fig:rnorm_teff}, we show $\amp$ normalized by the photometric precision $\sigmakep$ for long-cadence (29.4~min) exposure of {\it Kepler} for each star against $\teff$. Here $\sigmakep$ was evaluated using the photometric precision estimated by the {\it Kepler} team as a function of $K_p$,\footnote{\url{https://nexsci.caltech.edu/workshop/2012/keplergo/CalibrationSN.shtml}} which takes into account noise sources other than the photon noise and is applicable to $K_p\gtrsim 12$.\footnote{The $K_p$ dependence also agrees with the estimates by \citet{2010ApJ...713L.120J} using the Quarter 1 data, at least in the fainter end that is relevant to our discussion.}
We do not use the Combined Differential Photometric Precision \citep[CDPP,][]{2010ApJ...713L..79K} commonly used for evaluating noise levels relevant to planet search, because here we need to evaluate the noise that does not include intrinsic stellar variabilities. 
In this plane, the lower edge of the $\amp/\sigmakep$ distribution is flat across $\teff$, again indicating that the sample is limited by photometric precision: the lower edge in the $\amp$--$\teff$ distribution (Figure~\ref{fig:pr_teff}, bottom panel) is higher for cooler stars because they tend to be apparently (and intrinsically) fainter than the hotter ones.
The lower edge of $\amp/\sigmakep$ is also flat as a function of $\prot$ and $\ro$, as shown in the middle and bottom panels of Figure~\ref{fig:rnorm_teff}.

The histogram of $\amp/\sigmakep$ (Figure~\ref{fig:rnorm_hist}) shows a sharp cutoff around 3 (bottom $\approx2\%$), which we adopt as an empirical detection threshold of the sample to guide the present discussion. This value, shown as the tan horizontal dashed line, agrees visually with the lower edge of the distributions in Figure~\ref{fig:rnorm_teff}, and also well explains the difference between the $\amp$ distribution of stars with $\teff=4,000$--$6,500\,\mathrm{K}$ in the \citet{2014ApJS..211...24M} sample and of stars with $K_p<12$.
We note that this threshold value is specific to the \citet{2014ApJS..211...24M} sample, as well as to the timescale for which photometric precision is defined.
We suspect that this rather steep cutoff is associated with the threshold on the weight parameter $w$ that was used by \citet{2014ApJS..211...24M} to distinguish between periodic and false detections. This parameter is related to the local peak height of the autocorrelation function that would explicitly depend on the noise level.
We also note again that Figure~\ref{fig:rper_kepmag} shows {\it all} the stars with significant $\prot$ detections in \citet{2014ApJS..211...24M}; thus the lower edge is not due to our sample selection.

\subsection{Rossby Number Cutoff}

Given the presence of the detection edge, the next question is what value of $\ro$ (or $\prot$) this edge corresponds to --- and we find the detection edge should correspond to $\ro\sim 1$--$2$ in the \citet{2011ApJ...741...54C} scale, or $\ro \sim 0.5$--$1\,\ro_\odot$.
This value is derived by equating the roughly $\teff$-independent $\amp(\ro)$ derived in Sections~\ref{ssec:amp_ro} and \ref{ssec:r_ro_bias} with $3\sigmakep(K_p)$ of each star and by solving for $\ro_\mathrm{threshold}(K_p)$: rotational modulation is detectable for a given star with the magnitude $K_p$ if its $\ro$ is lower than $\ro_\mathrm{threshold}(K_p)$. 
The distribution of $\ro_\mathrm{threshold}(K_p)$ computed in this way for all the stars for which rotational modulation has been {\it searched} by \citet{2014ApJS..211...24M} (i.e., stars with {\it and without} $\prot$ detection) is shown in the top panel of Figure~\ref{fig:roth_hist}. The solid histogram shows the result based on the $\amp$--$\ro$ relation derived 
in Section~\ref{ssec:r_ro_bias} using the brightest ($K_p<12$) stars with $\teff=4,000$--$6,250\,\mathrm{K}$, which is likely less affected by detection bias and more representative (see also Section~\ref{ssec:santos});
the dashed histogram shows the result based on the relation derived using all the sample stars (Section~\ref{ssec:amp_ro}), which may be more appropriate for cooler stars. 
Both distributions are sharply peaked around $\ro\approx 1.2 \sim 0.6\,\ro_\odot$, which corresponds to the faintest (and hence most abundant) stars with $K_p\sim16$.
The peak is also narrow because of the strong $\prot$ dependence of $\amp$: for $\amp \sim \ro^{-3}$--$\ro^{-4}$, $\ro_\mathrm{threshold}(K_p)$ increases only by a factor of two for the $K_p$ difference of five.
By definition of $\ro_\mathrm{threshold}$,
its normalized inverse cumulative distribution function, shown in the bottom panel of Figure~\ref{fig:roth_hist}, provides the detectability function $p_\mathrm{det}(\mathrm{Ro})$, the fraction of stars in the searched sample for which rotational modulation of a star with a given value of $\ro$ would have been reported as a robust detection by \citet{2014ApJS..211...24M}.
We see that faint stars start to be missed at $\ro\gtrsim1.2$, and that the detection becomes impossible for almost all stars at $\ro\sim 2 \approx \ro_\odot$ for the $\amp$--$\ro$ relation derived from the brightest stars (solid line), which is likely more representative than that from all stars (dashed line).
This is how the combination of the rapid drop of $\amp$ with increasing $\ro$ and roughly magnitude-limited nature of the \kepler\ sample imprints the detection edge around $\ro\sim 2 \approx \ro_\odot$;
the sample becomes roughly $\ro$-limited, and the longest detected $\prot$ increases with decreasing $\teff$ roughly as $\tauc$.
This agrees with what is observed, that the upper edge is close to a curve of $\ro=1.7$ (blue dashed line in the top panel of Figure~\ref{fig:pr_teff}) in our $\tauc$ scale.
It is beyond the scope of this work to understand this value more quantitatively:
the threshold depends on the unknown distribution of $\ro$ in all the searched stars (i.e., stars with {\it and without} detected $\prot$) as well as on exact dependence of the detectability on the signal to noise, both of which need to be modeled. It is also very sensitive to the steepness of the $\amp$--$\ro$ relation, as shown in Figure~\ref{fig:roth_hist}, as well as on its possible dependence on $\teff$ which is difficult to assess in the current sample (Section~\ref{ssec:r_ro_bias} and Appendix~\ref{ssec:tauc_amp}).

\begin{figure}
    \epsscale{1.15}
    \plotone{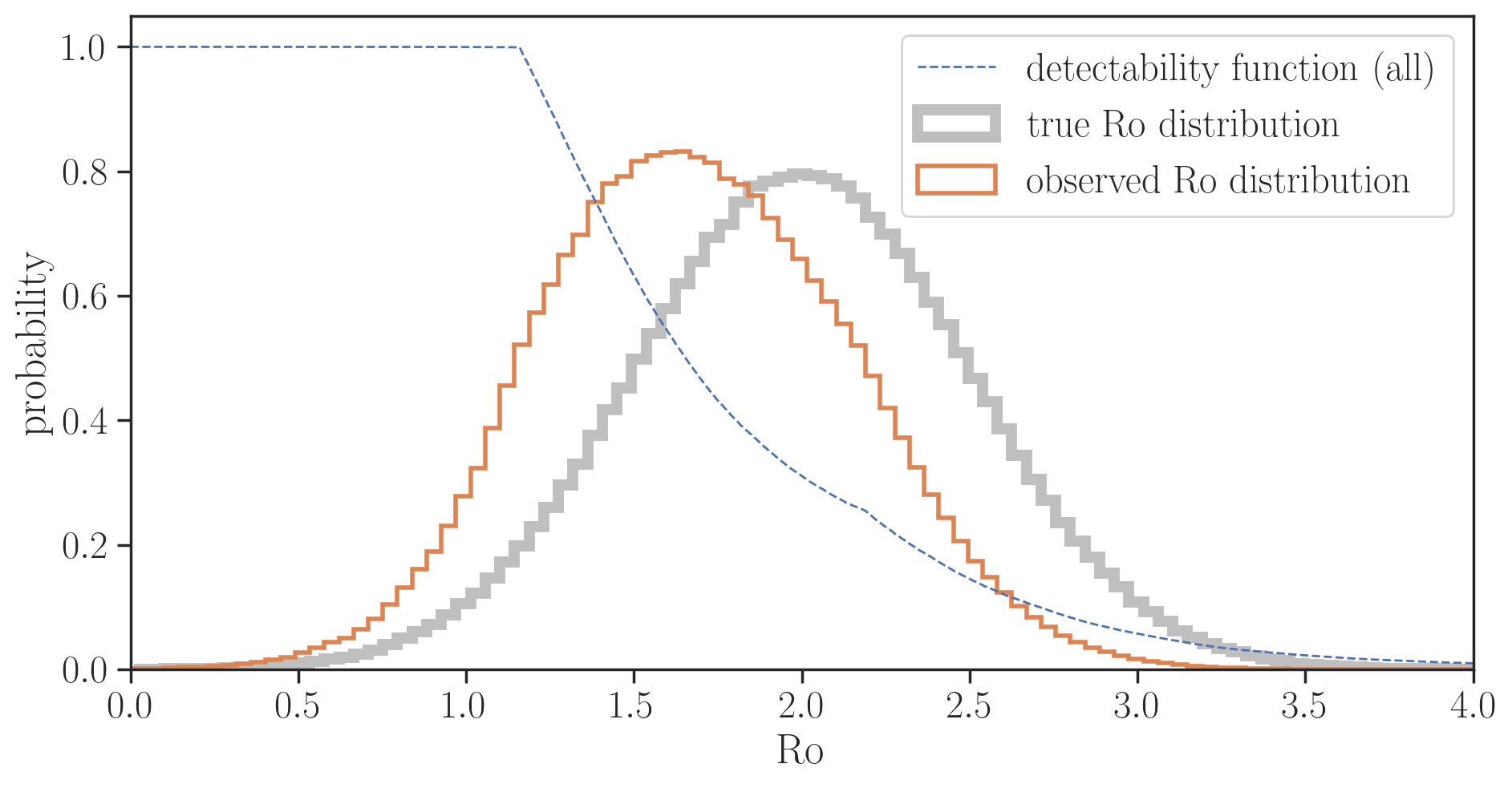}
    \plotone{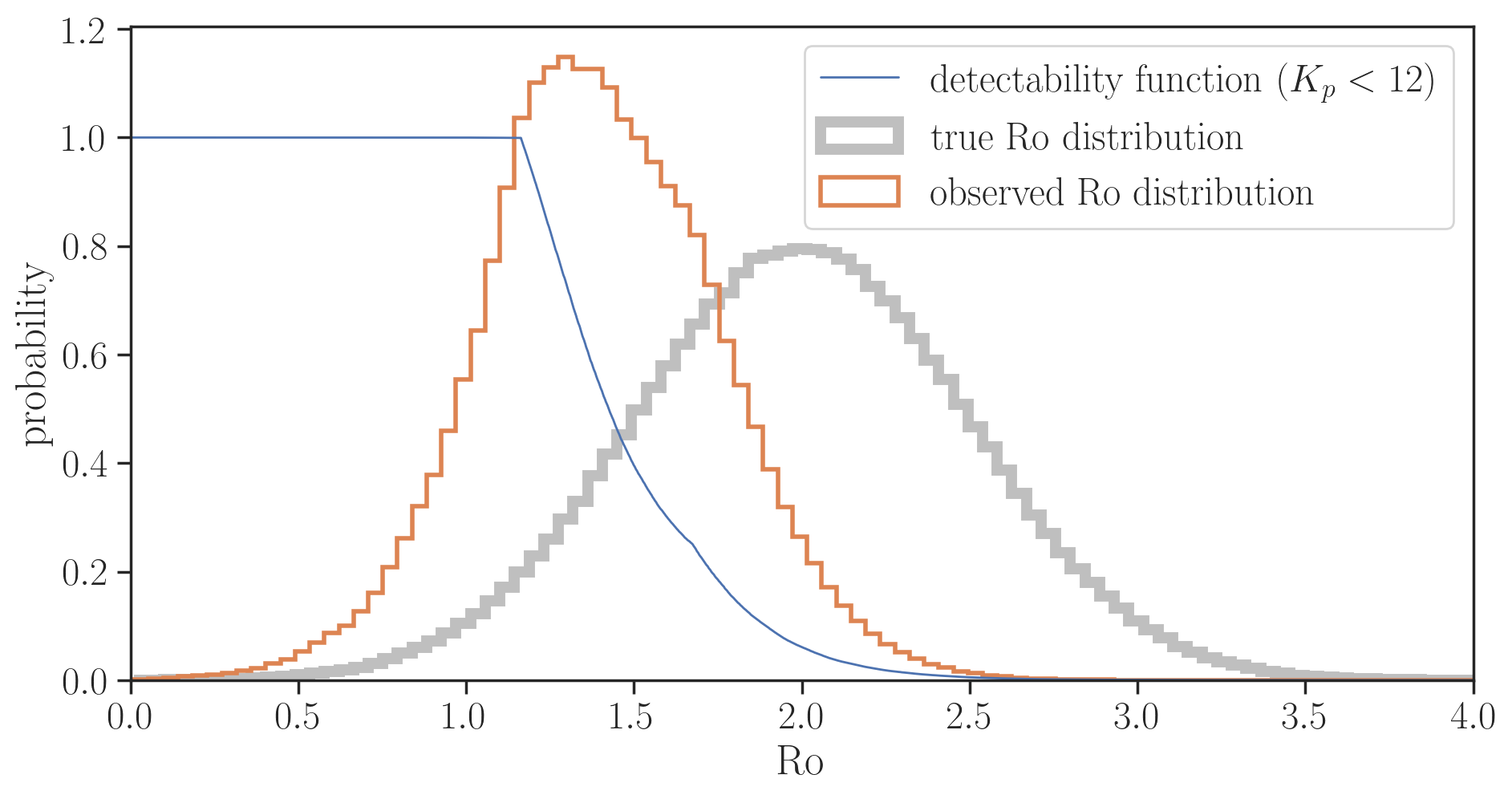}
    \plotone{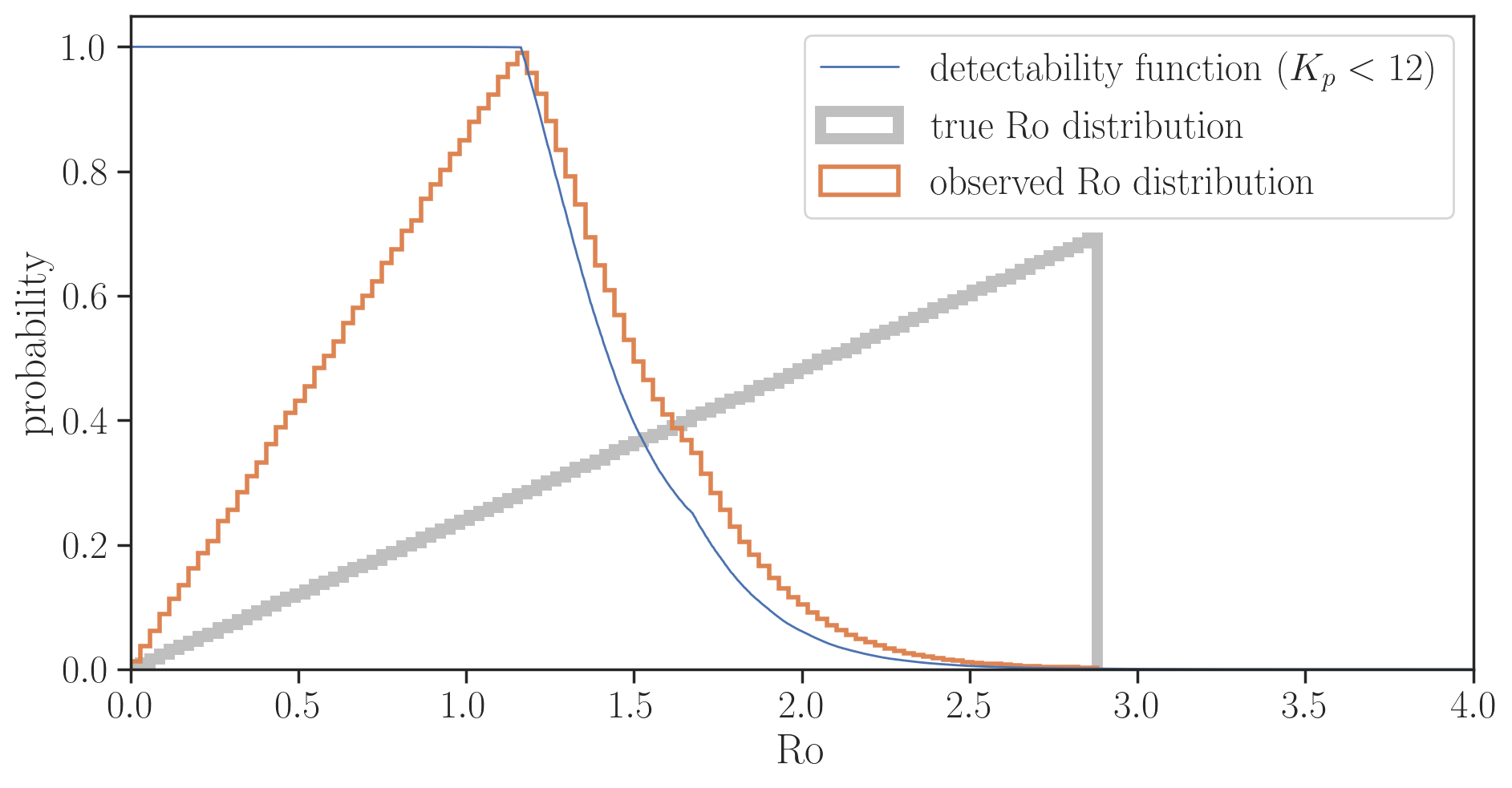}
    \caption{
    An illustration of how the true $\ro$ distribution (gray thick histogram) is deformed by the $\ro$-dependent detectability (blue curve) into the observed one (orange histogram).
    {\it (Top)} The detectability function is based on all stars with $\teff=4,000$--$6,250\,\mathrm{K}$, and the true distribution is a Gaussian.
    {\it (Middle)} The detectability function is based on brightest stars with $\teff=4,000$--$6,250\,\mathrm{K}$ and with $K_p<12$, and the true distribution is a Gaussian (same as above).
    {\it (Bottom)} The detectability function is based on brightest stars with $\teff=4,000$--$6,250\,\mathrm{K}$ and with $K_p<12$ (same as above), and the true distribution corresponds to Sun-like stars with the uniform age distribution and spin evolution following Skumanich's law.
    See Section~\ref{ssec:romodels} for details.
    }
    \label{fig:detmodel}
\end{figure}

The above argument, along with external evidence for stalled spin down, suggests that both effects discussed by \citet{2019ApJ...872..128V} are important in understanding the $\prot$--$\teff$ distribution around its upper edge.
Around the time when typical Sun-like stars observed by {\it Kepler} cease to spin down at $\ro\sim \ro_\odot$, their rotational modulation signals have already started to be buried under photometric noise and to be missed from the sample with $\prot$ detection. 
Therefore, the pile-up we see in the observed $\prot$--$\teff$ distribution may be just a tip of the iceberg:
the true pile-up may be located at longer periods, but have been capped due to the detection edge.
This interpretation is qualitatively consistent with the finding of \citet{2021NatAs...5..707H} and \citet{2022MNRAS.510.5623M}, who worked on the $\prot$ sample much less biased against slower rotators, and found that most stars in the sample are around {\it or above} the upper edge defined by stars with $\prot$ from rotational modulation \citep[see Figure~9 of][]{2022MNRAS.510.5623M}.

\subsection{Impact on the Observed $\ro$ Distribution}\label{ssec:romodels}

To demonstrate the impact of the detection edge further, we use simple models to show how the $p_\mathrm{det}(\ro)$ in the bottom panel of Figure~\ref{fig:roth_hist} works to bias the observed $\ro$ distribution. In the top panel of Figure~\ref{fig:detmodel}, 
we adopt $p_\mathrm{det}(\ro)$ derived from all the stars (dashed line in Figure~\ref{fig:roth_hist}) and simulate how this modifies the true $\ro$ distribution, here chosen to be a Gaussian with the mean of 2 and scale of $0.5$ (thick gray line): the result is the orange histogram. The location of the observed peak ($\ro\sim 1.5$) has shifted from the truth due to a rapid decrease of $p_\mathrm{det}$ at $\ro \gtrsim 1$.
The same is true but the bias is more severe when we adopt $p_\mathrm{det}(\ro)$ derived from the stars with $K_p<12$ (solid line in Figure~\ref{fig:roth_hist}), as shown in the middle panel. The difference from the top panel also illustrates how the resulting distribution is sensitive to the steepness of the $\amp$--$\ro$ relation above $\robreak$. 
In the bottom panel, $p_\mathrm{det}(\ro)$ is again from the stars with $K_p<12$, and the true $\ro$ distribution is constructed by sampling age $t$ from the uniform distribution between $0$ and $10\,\mathrm{Gyr}$, translating $t$ into $\prot = 25\,\mathrm{days}\,(t/4.6\,\mathrm{Gyr})^{1/2}$ and to $\ro=\prot/\tauc(5,777\,\mathrm{K})$ using the formula in \citet{2011ApJ...741...54C}: this simulates a collection of Sun-like stars that (i) have a uniform age distribution, (ii) have the same $K_p$ distribution as the {\it Kepler} stars, and (iii) keep spinning down following Skumanich's law during their entire main-sequence life.
This is merely another toy model but provides two useful insights. First, the sharp decrease of $p_\mathrm{det}$ at $\ro\gtrsim 1$ produces a peak in the observed distribution (orange thin histogram) as long as the true $\ro$ distribution keeps increasing across the threshold $\ro$,
even without stalled spin down. Second, the fraction of stars with detectable rotational modulation (i.e., mean value of $p_\mathrm{det}(\ro)$ in the sample) is computed to be 0.28, which is close to the observed value \citep{2014ApJS..211...24M}: a combination of the sharp detection edge and the top-heavy $\ro$ distribution provides a reasonable explanation for why $\prot$ has not been detected for the majority of solar-mass stars. This experiment suggests that stalled spin down, if real, should start operating at $\ro \gtrsim 2$ so that a significant fraction of solar-mass stars evade detection of rotaitonal modulation. 
It may also explain why the pile-up found by \citet{2022arXiv220308920D} corresponds to a lower $\ro\sim 1.5$ than that inferred from asteroseismology or $v\sin i$ ($\ro\sim 2$). 
The last model discussed here might even suggest that the pile-up of $\ro$ alone does not serve as a conclusive evidence for the stalled spin down, as it shows that the observed distribution is not very sensitive to the $\ro$ distribution above the detection edge. 
We note that this argument does not deny the importance of possible systematic offsets in $\teff$ as discussed by \citet{2022arXiv220308920D}. 
Nevertheless, these toy models demonstrate the importance of considering detection bias in interpreting the observed $\prot$ distribution.

\subsection{Comparison with the Santos Sample}\label{ssec:santos_noise}

\begin{figure}
    \epsscale{1.15}
    \plotone{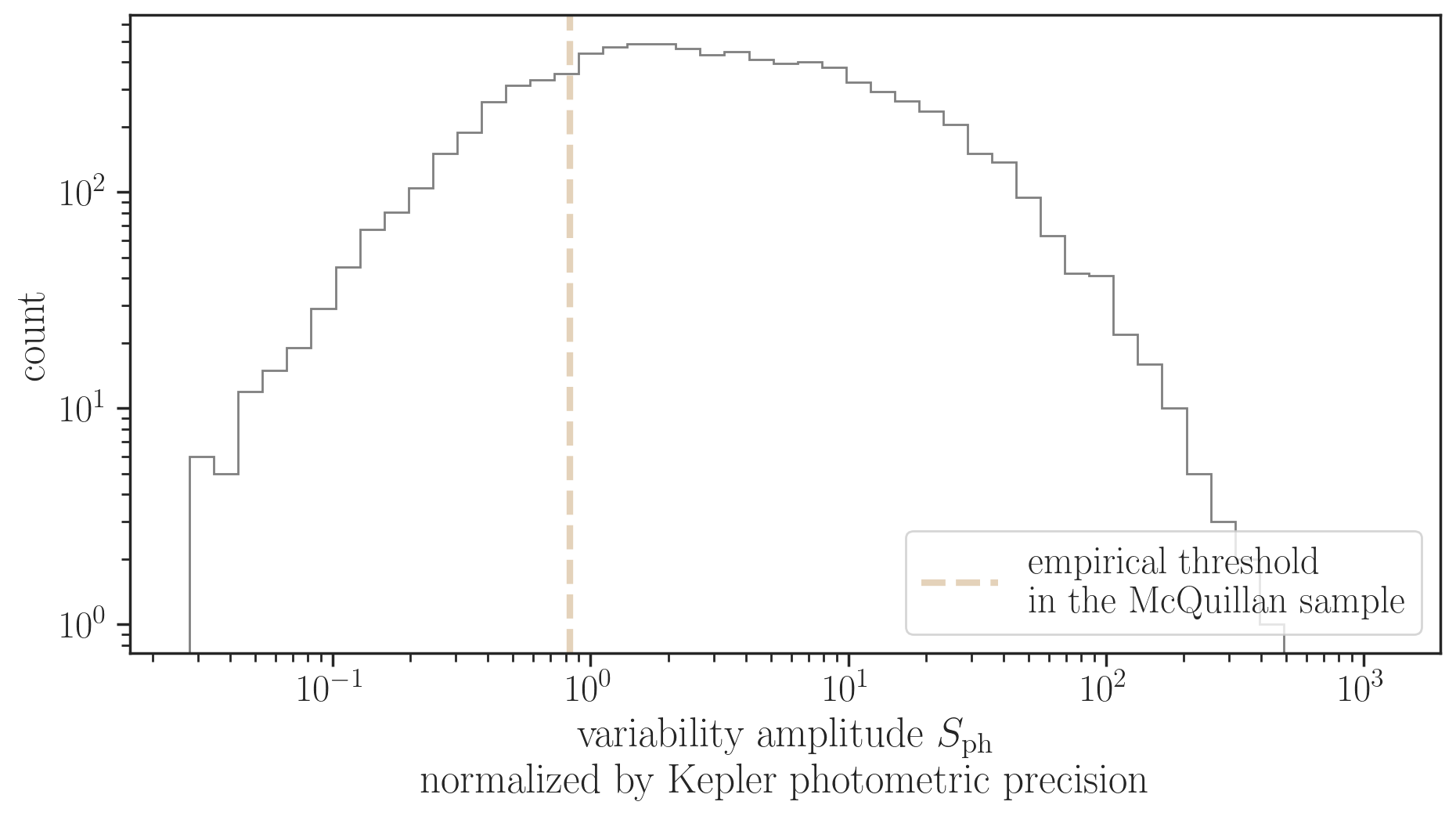}
    \plotone{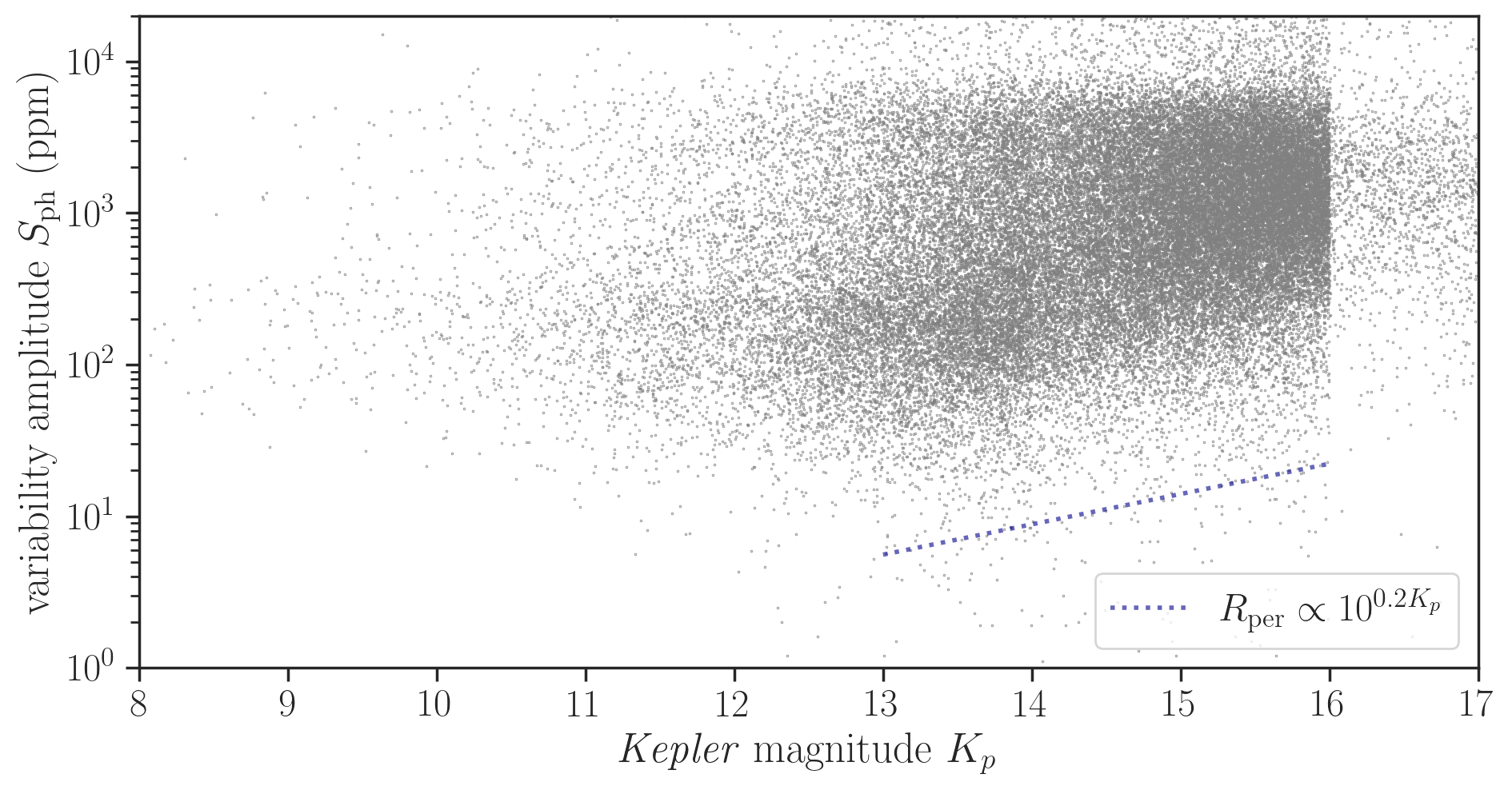}
    \caption{
    Same as Figure~\ref{fig:rnorm_hist}
    {\it (Top)} and Figure~\ref{fig:rper_kepmag}  {\it (Bottom)} but for the $\prot$ sample from 
    \citet{2019ApJS..244...21S, 2021ApJS..255...17S}.
    Here $S_\mathrm{ph}$ is used instead of $\amp$, and the empirical detection threshold in the {\it McQuillan} sample is shown in the scale of $S_\mathrm{ph}$ with the tan dashed line in the top panel.
    }
    \label{fig:santos_noise}
\end{figure}

We saw in Section~\ref{ssec:santos} that the $\prot$ catalog by \citet{2019ApJS..244...21S, 2021ApJS..255...17S} includes more stars with smaller modulation amplitudes than in \citet{2014ApJS..211...24M}. This suggests that the Santos sample is subject to a very different detection function from McQuillan's. This situation is shown in Figure~\ref{fig:santos_noise}; here we reproduce Figure~\ref{fig:rnorm_hist} and Figure~\ref{fig:rper_kepmag} for the Santos sample, where $S_\mathrm{ph}$ is used instead of $\amp$, and the empirical threshold in the {\it McQuillan} sample is shown in the scale of $S_\mathrm{ph}$ (see Section~\ref{ssec:santos}). Although the top panel does show the decrease in the detection rate at a  signal-to-noise corresponding to the McQuillan edge, we do not find such a sharply defined threshold as seen in the McQuillan sample. Correspondingly, we do not see a well-defined lower edge in the amplitude--magnitude plane in the bottom panel.
We also find that the number of stars ``leaking" below the McQuillan threshold increase toward higher $\teff$.
This comparison illustrates the importance of considering detection functions in a sample-specific manner.
While the Santos sample does include more period detections than in McQuillan's, the detection function might also be more difficult to quantify.

Although it is beyond the scope of this work to fully assess the impact of detection bias in the Santos sample,
the following arguments suggest that it is likely significant in the Santos sample too, at least for nearly solar-mass stars.
First, the detection fraction of $\prot$ is still $\approx 30\%$ for G stars even in \citet{2021ApJS..255...17S}. 
Second, the steep $\amp$--$\ro$ relation implies that the longest detectable $\prot$ (or largest detectable $\ro$) is not drastically changed by improving the detection threshold: for $\amp \sim \ro^{-4.6}$ derived from the brightest stars, a factor of 10 improvement in the detection threshold results in an only $\approx 60\%$ increase in the threshold $\ro$. This might explain why the detection fraction did not drastically increase even in the updated analysis of \citet{2021ApJS..255...17S}.
Therefore, similar features seen in both samples with different detection functions do not necessarily indicate their astrophysical origins.

\section{Discussion}\label{sec:discussion}

\begin{figure*}
    \epsscale{1.1}
    \plotone{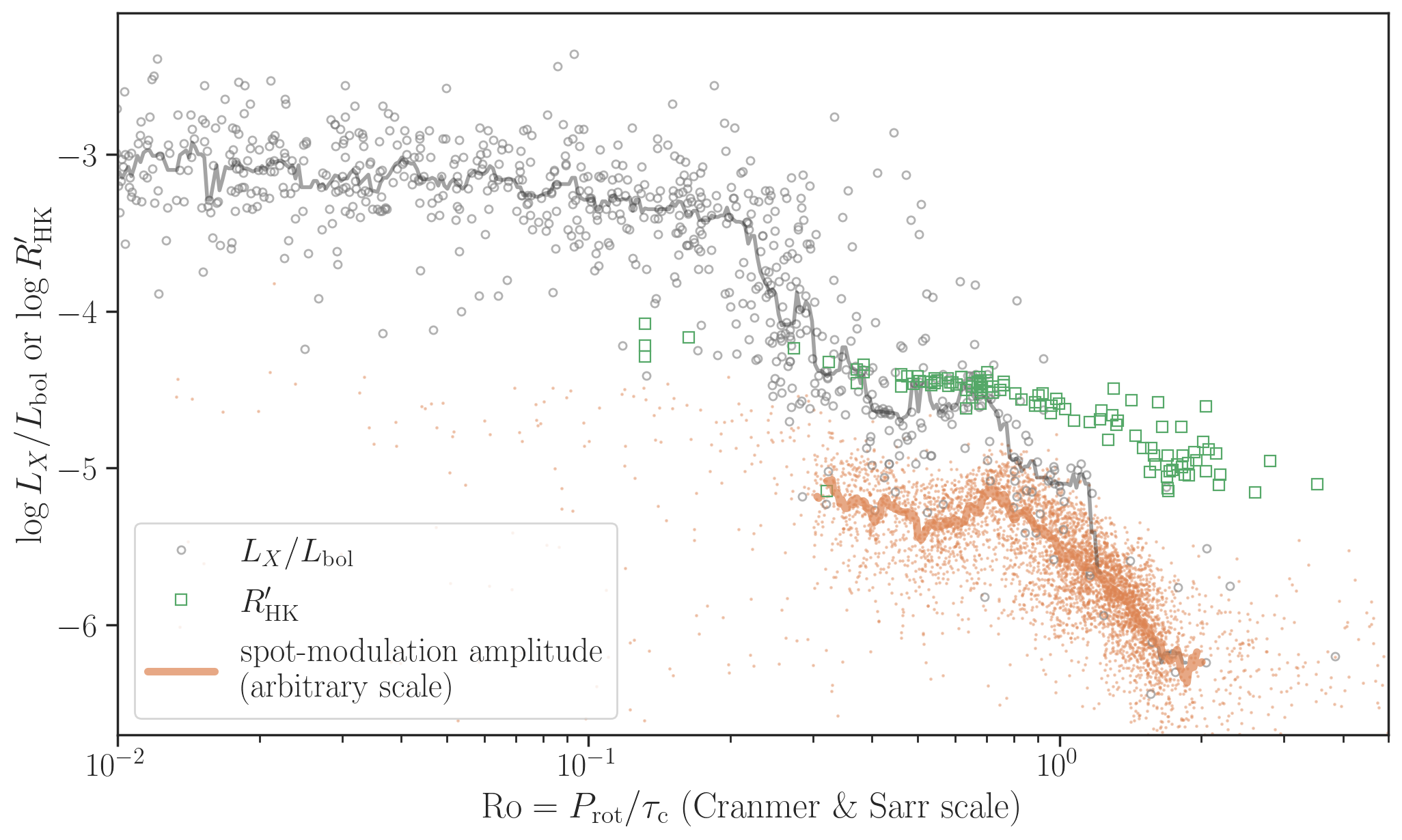}
    \caption{
    Rossby number dependence of spot-modulation amplitude (orange dots), X-ray to bolometric luminosity $L_X/L_\mathrm{bol}$ (open gray circles), and $\log R'_\mathrm{HK}$ (open green squares). For spot-modulation amplitude and $\lx$, the solid lines show the median filtered data with the width of 0.02 and 0.06~dex, respectively.
    }
    \label{fig:comparison}
\end{figure*}

\subsection{Comparison with Other Activity Indicators}\label{ssec:comparison}

The $\amp$--$\ro$ relation presented in Section~\ref{sec:analysis} is reminiscent of the relation known for X-ray luminosities normalized by the bolometric values $L_X/L_\mathrm{bol}$ \citep[e.g.][]{2003A&A...397..147P}: $L_X/L_{\rm bol}$ plateaus at $\ro\lesssim 0.1$, and decays as $L_X/L_{\rm bol} \sim \ro^{-2.7\pm 0.13}$ at least up to $\ro \sim 2$ \citep{2011ApJ...743...48W}.
The analysis of \citet{2011ApJ...743...48W} is based on the $\tauc$ scale from \citet{1984ApJ...279..763N} that is close to what we have adopted, and so the saturation of spot-modulation amplitude at $\ro \sim 0.8$ occurs within the so-called unsaturated regime of $\lx$, where it exhibits a power-law decay. Does this mean the X-ray activity and spot-modulation amplitude evolve differently as a function of $\ro$ despite their presumably common origin?

The analysis of \citet{2011ApJ...743...48W} assumes a two-piece power law and has captured a transition at $\ro \sim 0.1$, which therefore is insensitive to finer structures at larger $\ro$.
Thus here we seek for evidence of another transition in the ``unsaturated" X-ray regime. 
Figure~\ref{fig:comparison} shows the $\log \lx$--$\ro$ data (gray open circles) from \citet{2011ApJ...743...48W}
along with $\log\amp$--$\ro$ data in our sample (orange dots), where the scale of $\amp$ is shifted arbitrarily but that of $\ro$ is not.
Here we recomputed $\ro$ in the \citet{2011ApJ...743...48W} sample using their $\teff$ and the \citet{2011ApJ...741...54C} relation so that the comparison can be made using the same $\tauc$ scale. 
We see that the two data in fact follow the same pattern at $\ro\gtrsim0.3$ including a possible wiggle mentioned in Section~\ref{sec:analysis}: there is a hint of a ``shoulder" in the $\lx$ data beginning around $\robreak$ inferred from photometric modulation, which is also apparent in Figure~7 of \citet{2018A&A...618A..48M}.
The structure is more clearly seen in the median-filtered data with the width of 0.06~dex (solid gray line), i.e., a representation that does not assume a single power-law relation in this $\ro$ range. 
This reinforces the physical connection between surface spots and coronal X-ray emission, and suggests that the structure in the unsaturated X-ray regime is not an artifact; 
remember that $\amp$--$\prot$ relation shows a kink regardless of the prescription to compute $\tauc$.

A break at a similar value of $\ro$ in the chromospheric activities
has been noted \citep[e.g.,][]{1984ApJ...279..763N, 1987A&A...177..131R, 2021ApJ...910..110L},
which \citet{2021ApJ...910..110L} 
attributed to a transition of dominant dynamo regimes. Although \citet{2021ApJ...910..110L} reported a break at a lower value of $\ro$ than $\robreak$, the location agrees with what has been inferred from $\amp$ (and X-rays) if the common $\tauc$ scale is adopted, as is also confirmed by their conversion that the threshold is $\ro=0.91$ in the Noyes scale.
The data for main-sequence stars from \citet{2020NatAs...4..658L}, \kepler\ asteroseismic stars from \citet{2016ApJ...826L...2M}, along with the overlapping sample from \citet{2008ApJ...687.1264M} and \citet{2018A&A...618A..48M} for which both $\prot$ and $R'_\mathrm{HK}$ are readily available, are plotted with green open squares in Figure~\ref{fig:comparison}, which shows a kink at $\log R'_\mathrm{HK}\sim -4.5$ reported by \citet{2020NatAs...4..658L}.
Here again we use the \citet{2011ApJ...741...54C} formula to recompute $\tauc$ for those stars, where $\teff$ is estimated from $B-V$ using the table from \citet{2013ApJS..208....9P}.
We also note that the similarity between the chromospheric and X-ray fluxes has been noted by \citet{2018A&A...618A..48M}.

A transition at a similar $\ro$ might also been seen, though less clearly, in the photospheric filling factor $f_*$ of the magnetic flux. 
Note again that here we are focusing on the region around $\ro \sim 1$, rather than the saturation similar to that in X-ray around $\ro \sim 0.1$ \citep{2009ApJ...692..538R,2014MNRAS.441.2361V}.
The measurements for GKM stars presented in \citet{2011ApJ...741...54C} --- along with the Sun --- shows that $f_*$ decreases by roughly two orders of magnitudes between $\ro\sim0.2$ and $2$. Although the data do not densely cover $\ro$ around $\robreak$ seen in the spot-modulation amplitude, the empirical scaling they found, 
$f_*\propto \ro^{-2.5}$ to $f_*\propto \ro^{-3.4}$, is roughly in agreement with what we found for $\amp$.
\citet{2018A&A...618A..48M} also found a hint of a similar trend in the H$\alpha$ luminosity of M dwarfs studied by \citet{2017ApJ...834...85N}. These data are not shown in Figure~\ref{fig:comparison}, because the transition features are visually less clear.

In summary,
coronal and chromospheric fluxes (and perhaps magnetic and H$\alpha$ fluxes as well) show transitions at $\ro$ similar to $\robreak\sim 0.4\,\ro_\odot$ found for the photometric modulation amplitude,
thus 
suggesting that they share the same physical origin. 
We also confirmed that the same remains to be the case when the $\tauc$ prescription from \citet{2021ApJ...910..110L} is adopted instead; see Figure~\ref{fig:comparison_l21} in Appendix~\ref{ssec:tauc_comp}.
Although our sample does not constrain the $\amp$--$\ro$ relation at $\ro\lesssim 0.2\,\ro_\odot$, other ground- and space-based photometry works generally show even larger spot-modulation amplitudes up to $\sim 10\%$ for those younger stars with shorter rotation periods \citep[e.g.,][]{2009ApJ...691..342H, 2016AJ....152..113R, 2020ApJ...893...67M}.
Thus the evolution at lower $\ro$ may also be similar to the X-ray and chromospheric fluxes \citep[see also, e.g., Figure 7 of][for the latter]{2008ApJ...687.1264M}.

\subsection{Implications for Weakened Magnetic Braking}\label{ssec:wmb}

\citet{2016Natur.529..181V} proposed that magnetic braking ceases at a critical Rossby number of 
$\ro_\mathrm{WMB} \sim \ro_\odot$ based on comparison between their spin evolution models and the age/rotation measurements for $\sim 20$ stars. As we saw in Section~\ref{ssec:comparison}, the information on how various activity indicators evolve around $\ro\sim\ro_\odot$ is in general limited
(Figure~\ref{fig:comparison}), but some indicators may be showing hints of corresponding changes.
A transition at $\ro\sim\ro_\odot$ has been suggested in the chrmospheric fluxes \citep{2016ApJ...826L...2M}. 
A small number of measurements in the Santos sample hints that $\amp$ might also follow a similar pattern at 
$\ro \gtrsim \ro_\odot$ (Section~\ref{ssec:santos}), although a more careful analysis would be required to confirm weather this is a typical behavior or not, because the detection bias is significant here (Section~\ref{sec:detection}).

On the other hand, all the indicators show that the activity pattern changes in a continuous but non-monotonic manner up to $\ro_\odot$. 
Thus it also seems conceivable
that the departure from the standard spin evolution starts earlier than $\ro_\odot$ and proceeds gradually.
In particular, the decrease in $\amp$ at $\ro\gtrsim 0.5\,\ro_\odot$ may indicate that large spots suddenly start dissolving into smaller pieces. If so, this 
seems qualitatively consistent with a scenario that the concentration of the magnetic fields into smaller spatial scales and the associated reduction of angular momentum loss is responsible for weakened magnetic braking \citep{2016Natur.529..181V, 2015ApJ...798..116R}.
We note, though, that the relation between the photometric light curves and spot distribution is generally very complicated \citep[e.g.,][]{2021AJ....162..123L} and that the modulation amplitude may not be readily translated into the largest spot size.
More in-depth analyses of the light curve morphology as a function of $\ro$ may bring this hypothesis into sharper focus \citep[e.g.,][]{2017ApJ...851..116M, 2019A&A...621A..21R}.

Regardless of whether or not the pattern in the $\amp$--$\ro$ relation we derived is physically related to weakened magnetic braking, our finding has important implications for studies of weakened magnetic braking using rotation periods from photometric modulation. We presented evidence that detection bias becomes particularly important in the relevant $\ro$ range 
(Section~\ref{sec:detection}).
The subtlety arises from the fact that the $\ro$ corresponding to detection edge, $\ro_\mathrm{edge}$, happens to be close to $\ro_\mathrm{WMB}$.
This is in part a coincidence, because $\ro_\mathrm{edge}$ is determined by photometric precision of {\it Kepler}.
On the other hand, it is also true that the strong dependence makes $\ro_\mathrm{edge}$ insensitive to photometric precision --- so if $\robreak \sim \ro_\mathrm{WMB}$ due to physics, it is inevitable that $\ro_\mathrm{edge}\sim \ro_\mathrm{WMB}$. Given $\amp \sim \ro^{-4}$ found for solar-mass stars, the photometric precision needs to be improved by an order of magnitude to push $\ro_\mathrm{edge}$ up by a factor of two.
This argument explains why it has not been easy to find the signature of the weakened magnetic braking in the photometric sample, and indicates that it is crucial to consider detection bias when interpreting the sample quantitatively in terms of weakened magnetic braking scenario.

\subsection{Further Test of the View}

We argued in Section~\ref{sec:detection} that the detectability of photometric rotational modulation in \kepler\ stars is determined by the combination of the (roughly) $\teff$-independent steep $\amp$--$\ro$ relation and magnitude-dependent detection threshold.
If this is correct,
photometric rotational modulation should have been detected for Sun-like main-sequence stars 
{\it if and only if} a star is younger than a certain $\teff$-dependent threshold age.
In the other extreme case where the upper edge in the $\prot$--$\teff$ distribution is solely due to stalled spin down, on the other hand,
stars with photometrically detected $\prot$ around the edge should contain many stars older than the age corresponding to the onset of the stalled spin down.
We investigate this hypothesis
in a companion paper (Masuda 2022, in prep.) using the isochronal age estimates for a sample of {\it Kepler} stars with {\it and without} detected rotational modulation.

\section*{}
The data and the code underlying this article are available through GitHub.\footnote{\url{https://github.com/kemasuda/acheron/tree/main/kepler_prot_teff}}

\acknowledgements

The author thanks Shinsuke Takasao for helpful conversations on the subject of this paper.
This work was supported by JSPS KAKENHI Grant Number 21K13980.
This work has made use of data from the LAMOST project.
Guoshoujing Telescope (the Large Sky Area Multi-Object Fiber Spectroscopic Telescope LAMOST) is a National Major Scientific Project built by the Chinese Academy of Sciences. Funding for the project has been provided by the National Development and Reform Commission. LAMOST is operated and managed by the National Astronomical Observatories, Chinese Academy of Sciences.
This work has made use of data from the European Space Agency (ESA) mission {\it Gaia} (\url{https://www.cosmos.esa.int/gaia}), processed by the {\it Gaia} Data Processing and Analysis Consortium (DPAC, \url{https://www.cosmos.esa.int/web/gaia/dpac/consortium}). Funding for the DPAC has been provided by national institutions, in particular the institutions participating in the {\it Gaia} Multilateral Agreement.
Funding for the Kepler mission is provided by the NASA Science Mission Directorate. STScI is operated by the Association of Universities for Research in Astronomy, Inc., under NASA contract NAS 5–26555.

\software{
JAX \citep{jax2018github}, NumPyro \citep{bingham2018pyro, phan2019composable}
}

\appendix

\section{The Effective Temperature Dependence of the $\amp$--$\prot$ and $\amp$--$\ro$ Relations}

Figure~\ref{fig:r_prot_all} and \ref{fig:r_ro_all} 
show the $\amp$--$\prot$ relations and the $\amp$--$\ro$ relations
in different $\teff$ bins, respectively.

\begin{figure}
    \epsscale{1.15}
    \plottwo{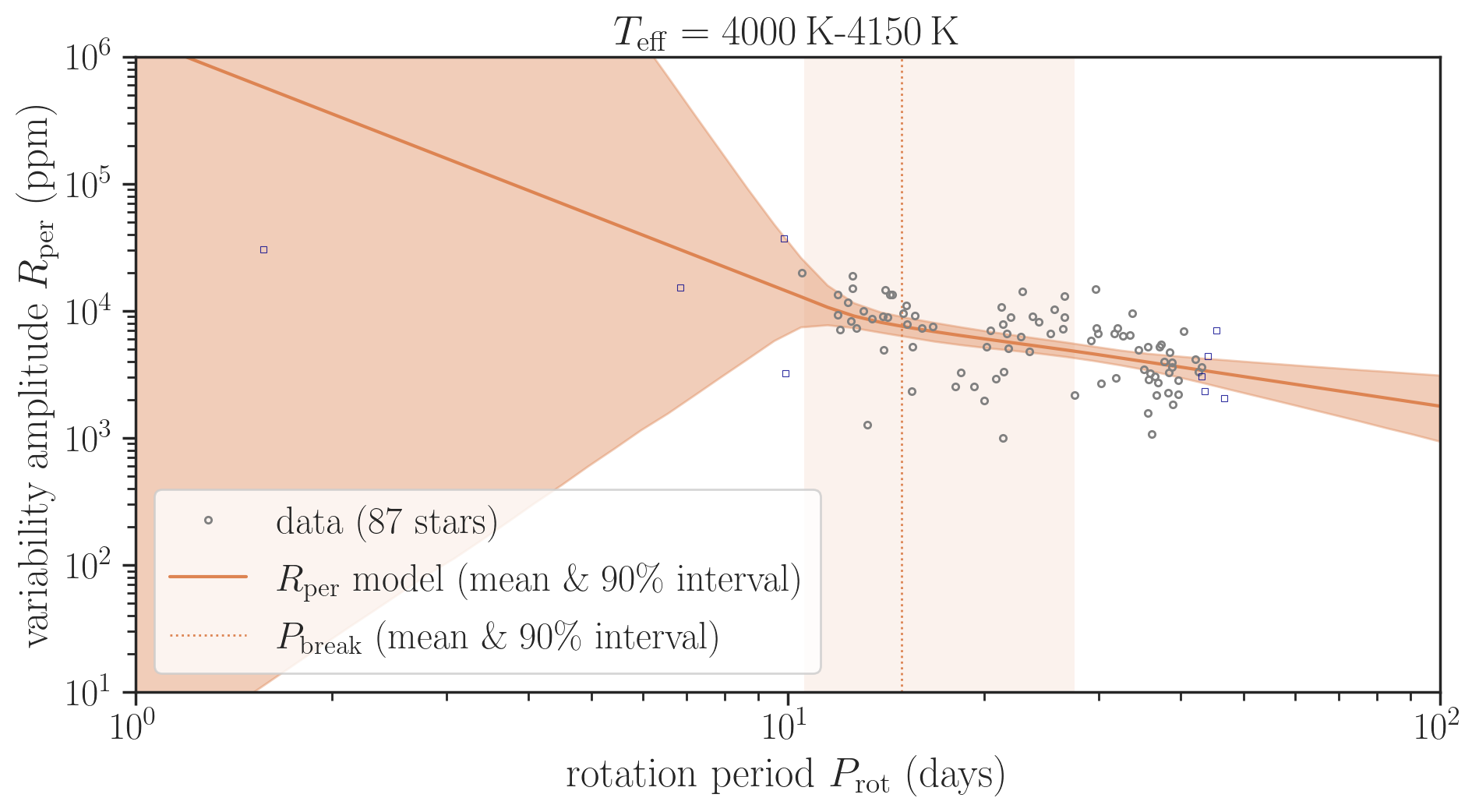}{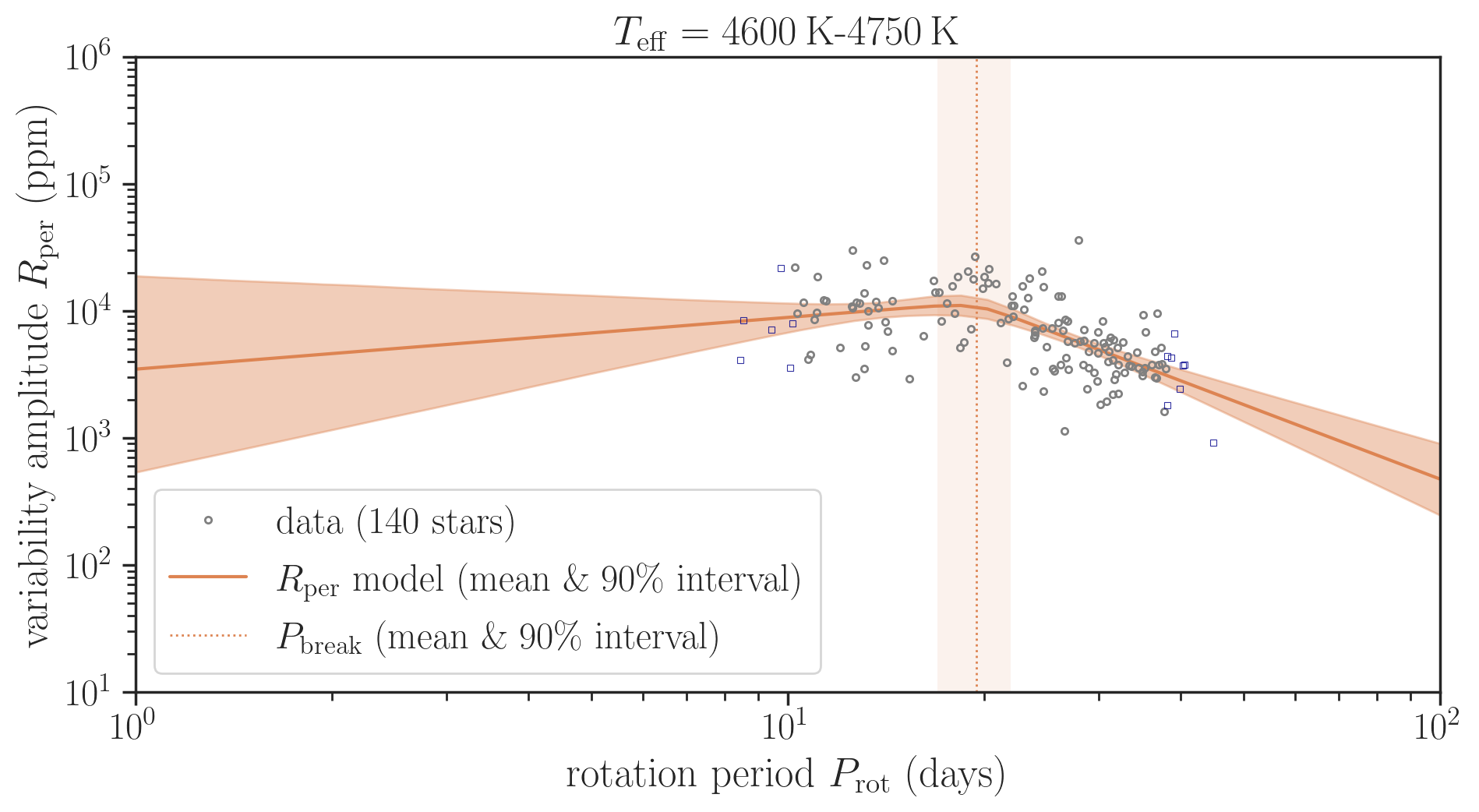}
    \plottwo{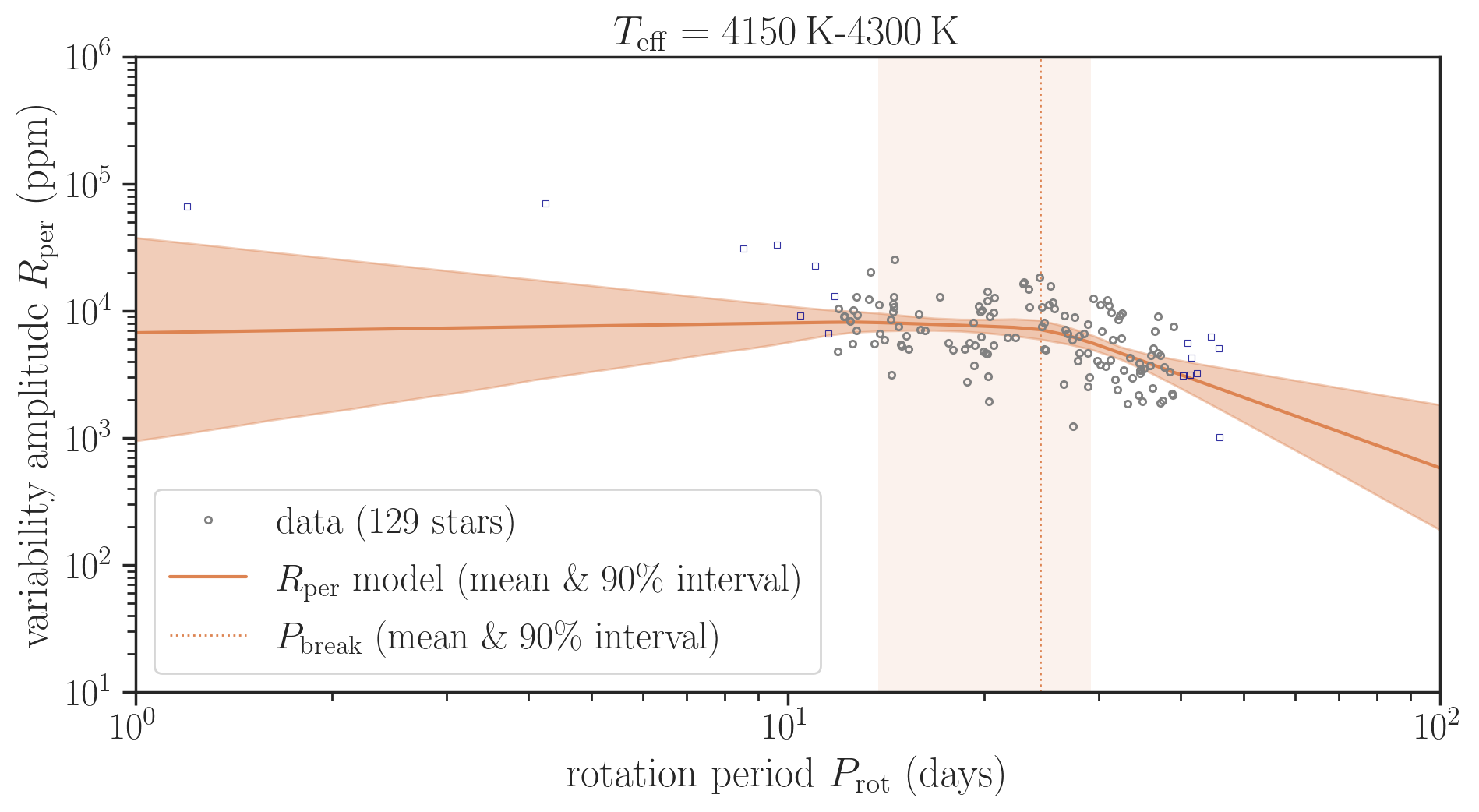}{teff4750-4900_model}
    \plottwo{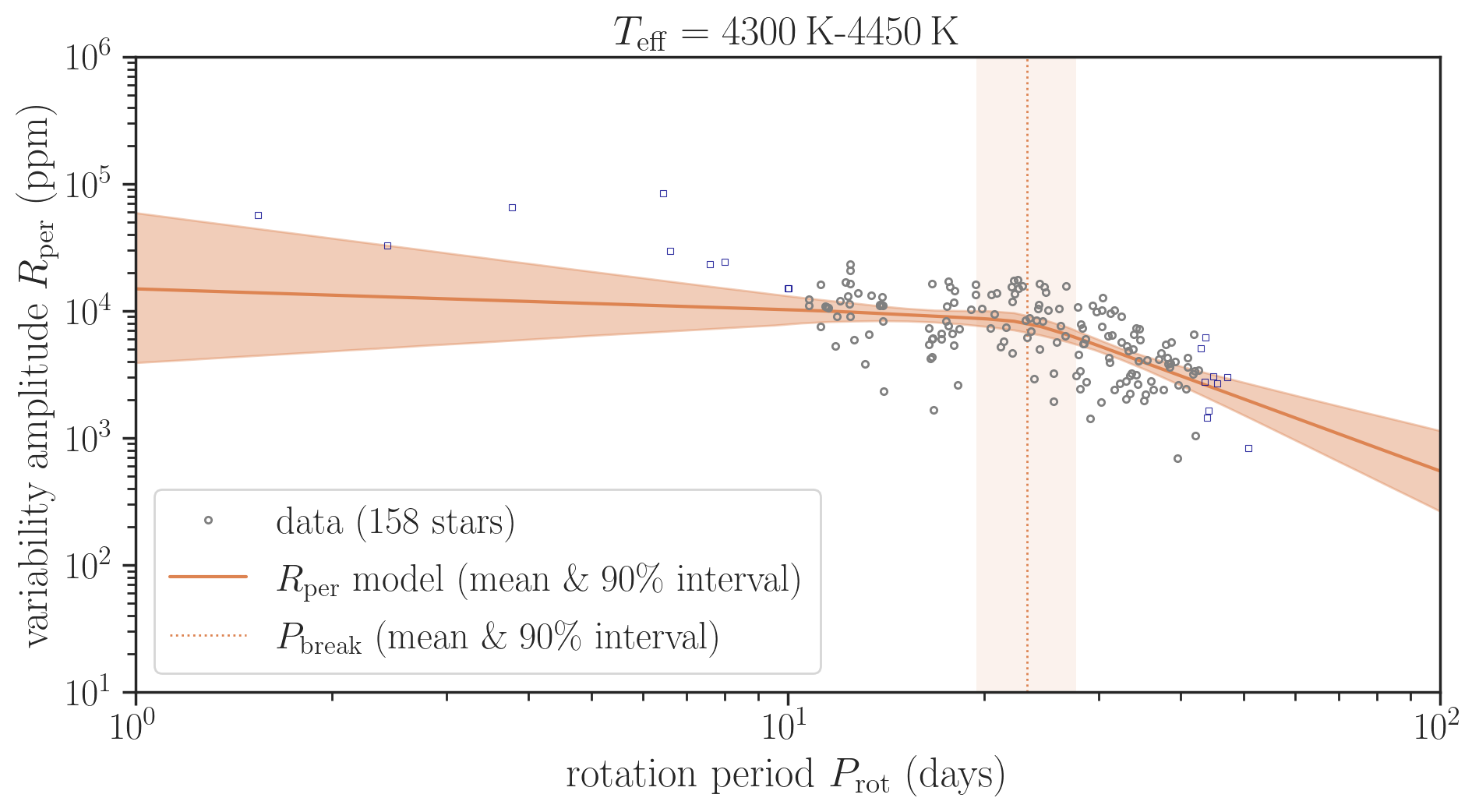}{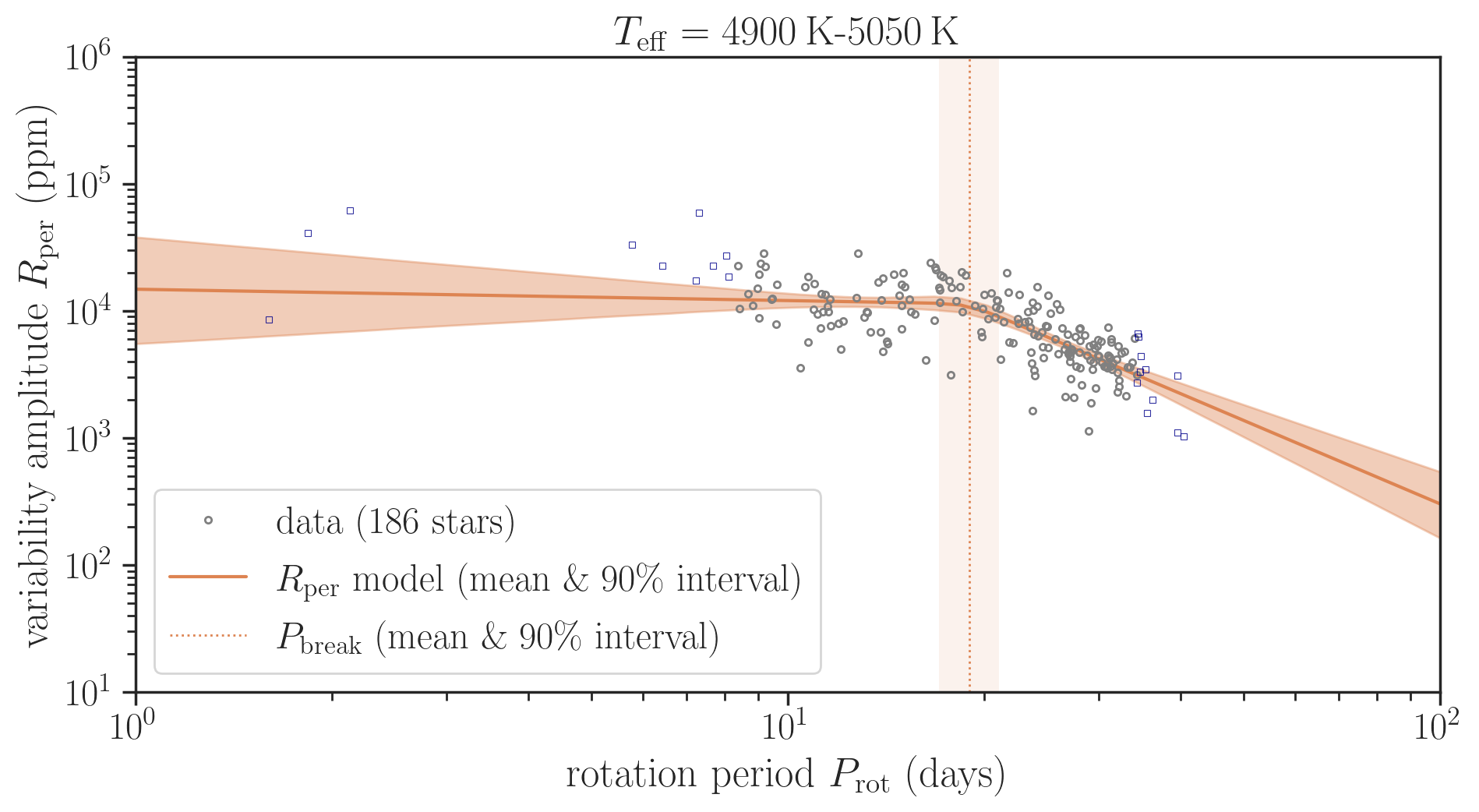}
    \plottwo{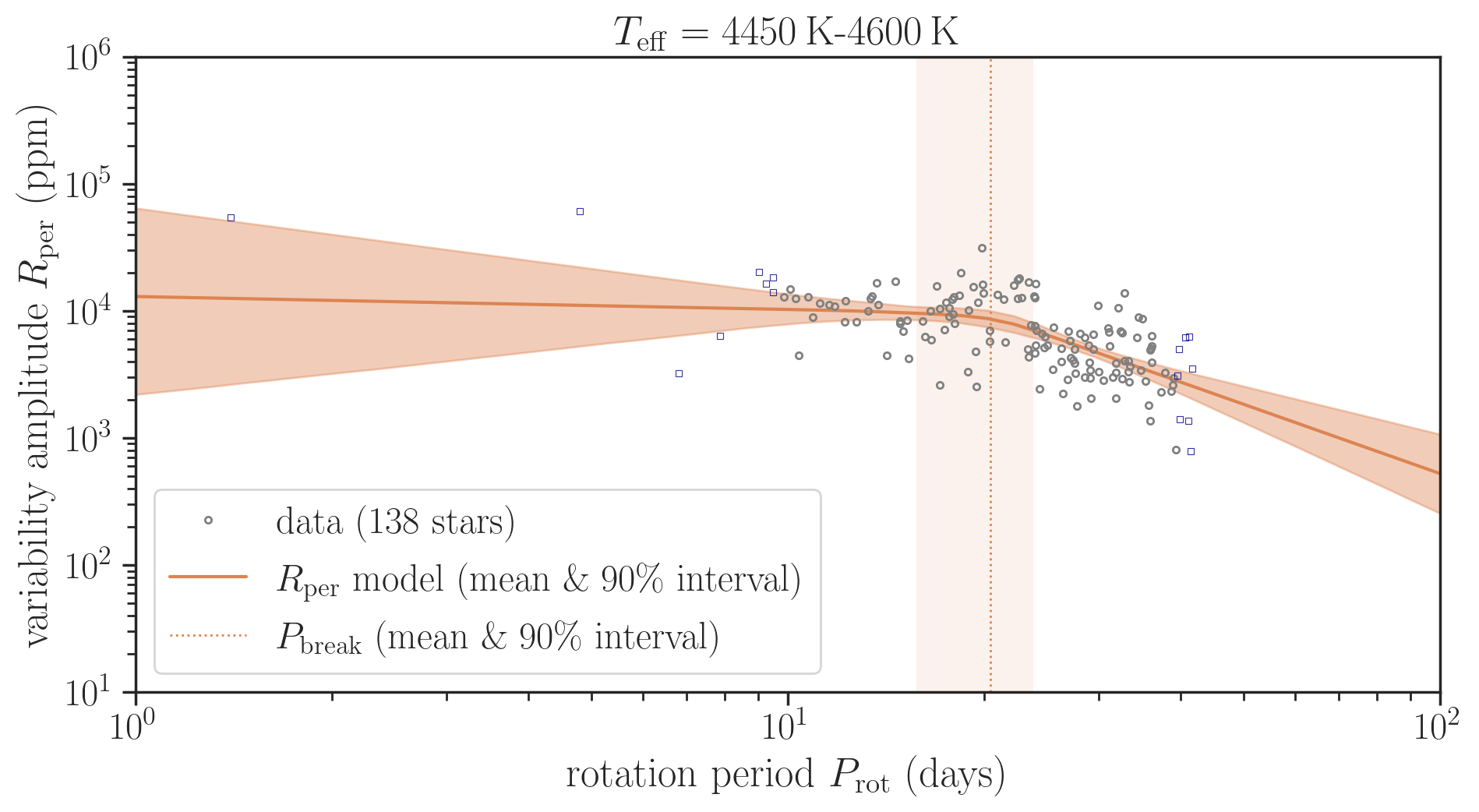}{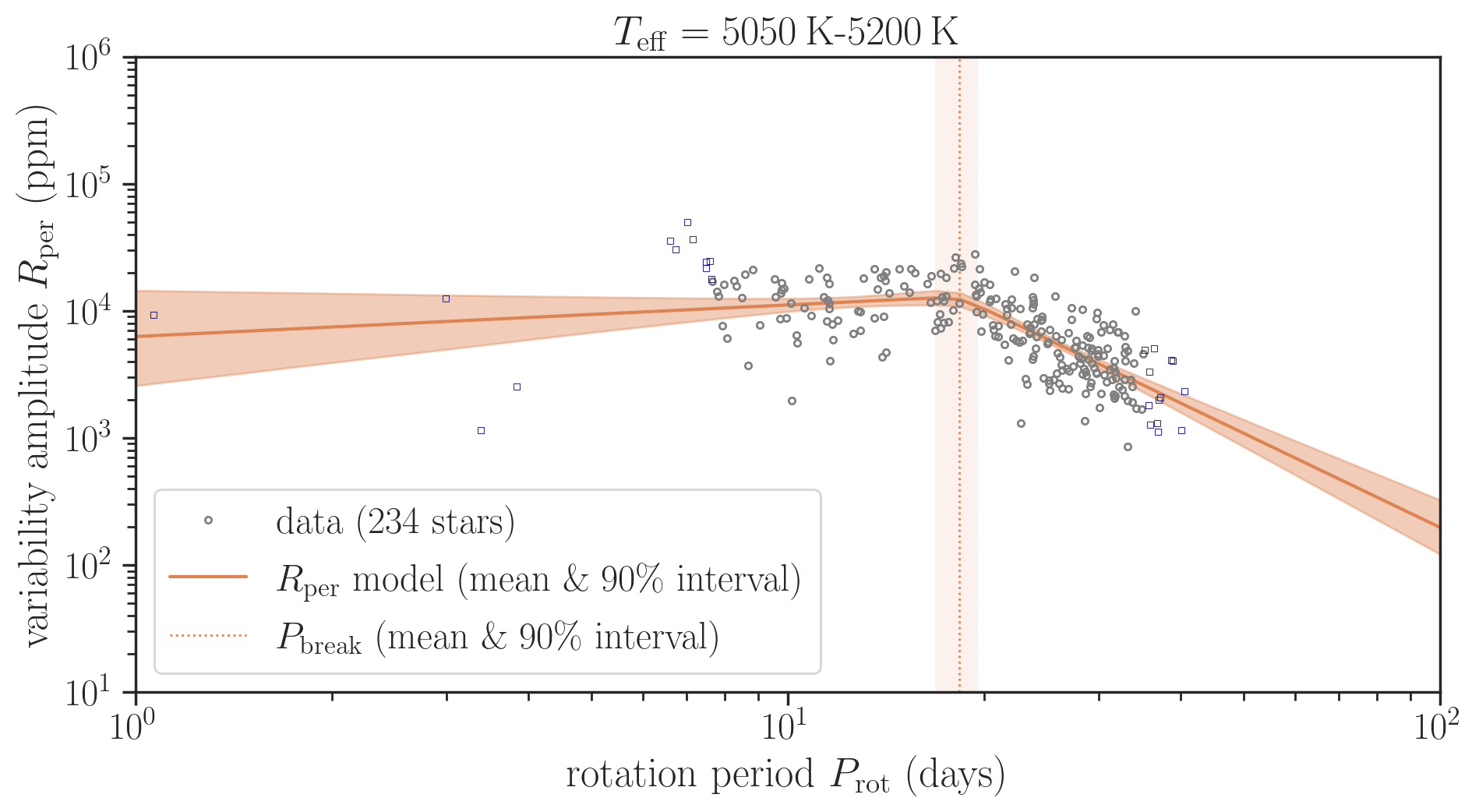}
    \caption{
    Spot-modulation amplitudes $\amp$ and rotation periods $\prot$ for stars in each $\teff$ bin.
    {\it Gray circles}:  Data points. Blue open squares show the ones that were not used for modeling.
    {\it Orange solid line and shade}: Broken power-law model. 
    {\it Vertical orange dotted line and shade}: Inferred location of the break, $\pbreak$. See Section \ref{ssec:amp_prot} for details.
    }
    \label{fig:r_prot_all}
\end{figure}
\addtocounter{figure}{-1}
\begin{figure}
    \epsscale{1.15}
    \plottwo{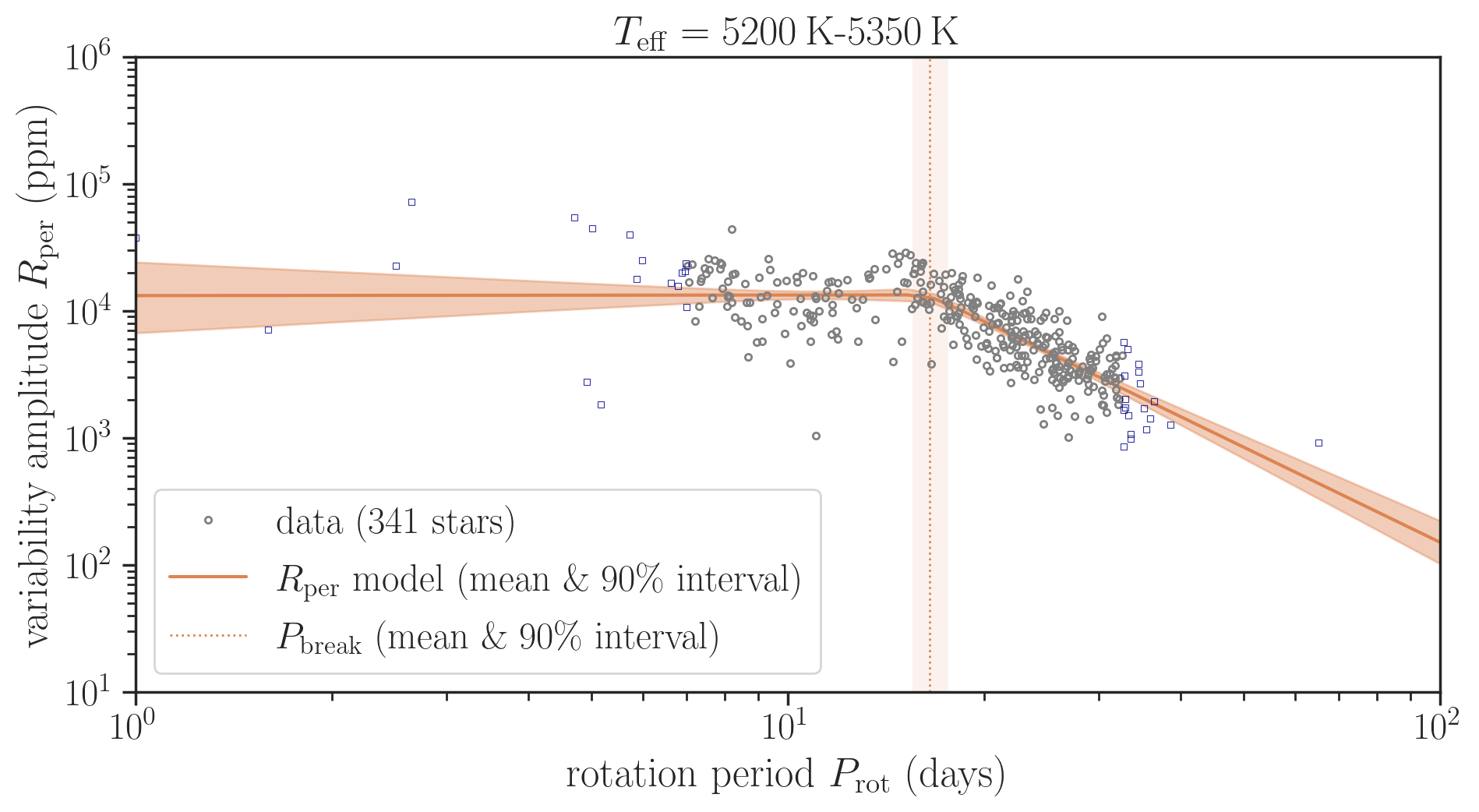}{teff5800-5950_model}
    \plottwo{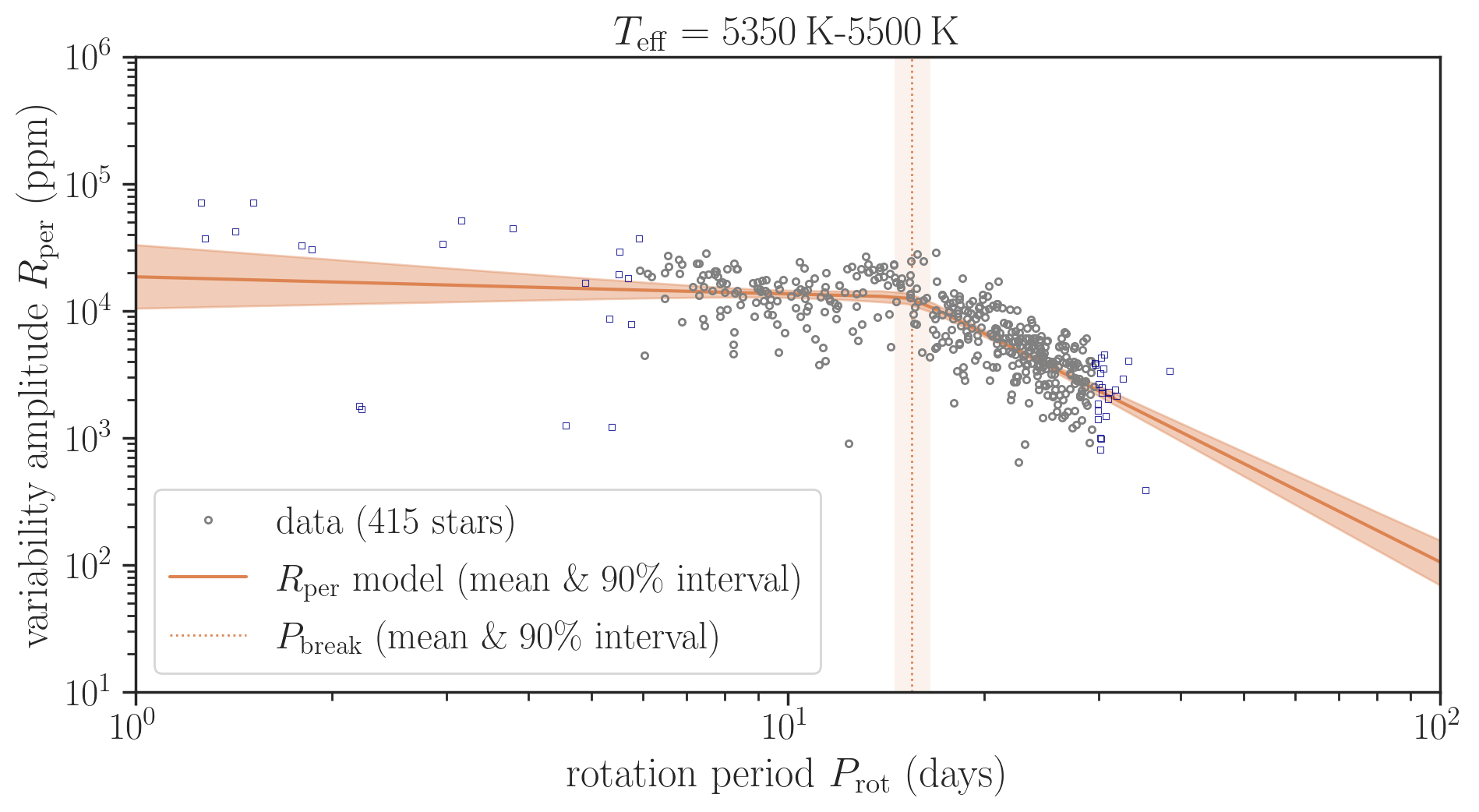}{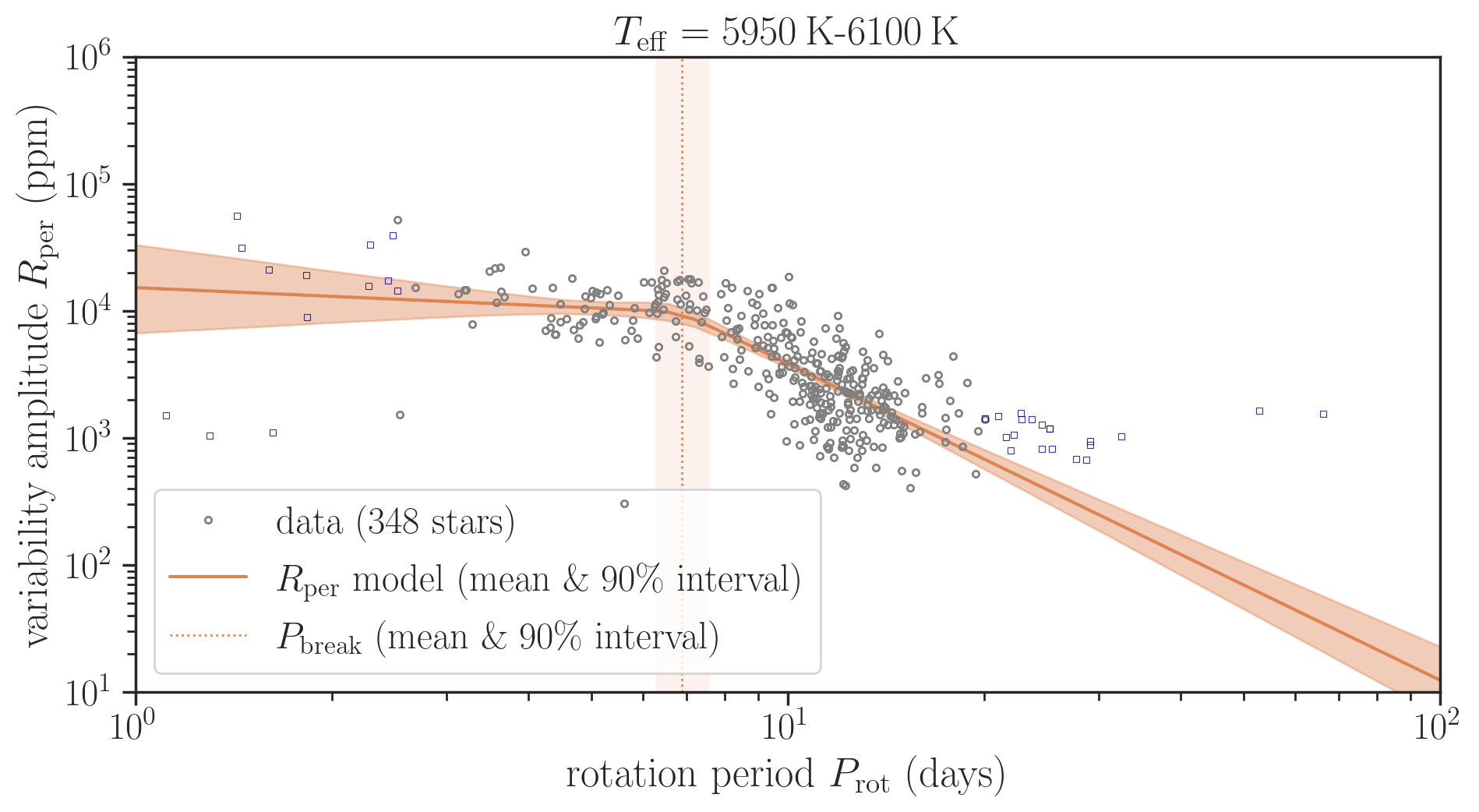}
    \plottwo{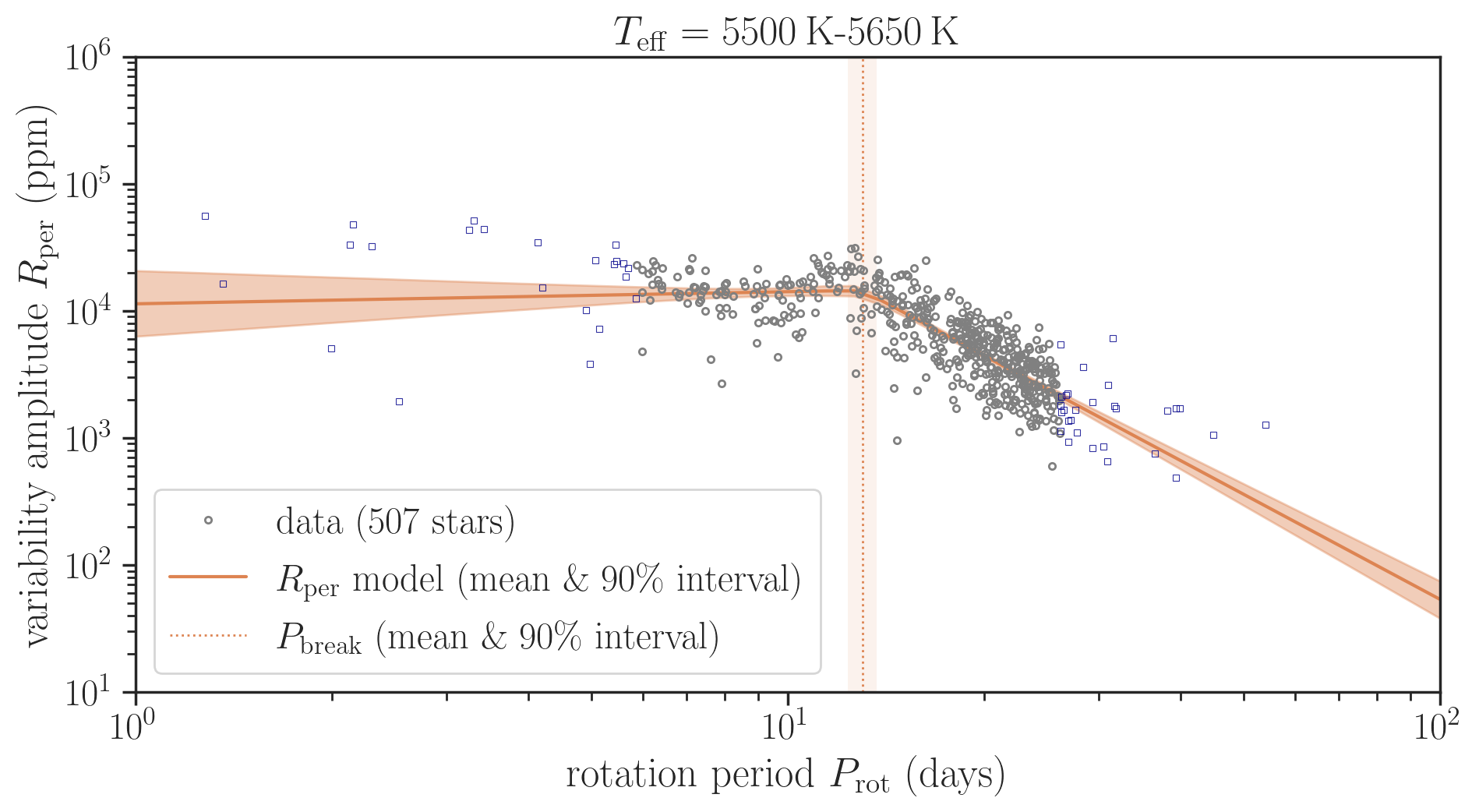}{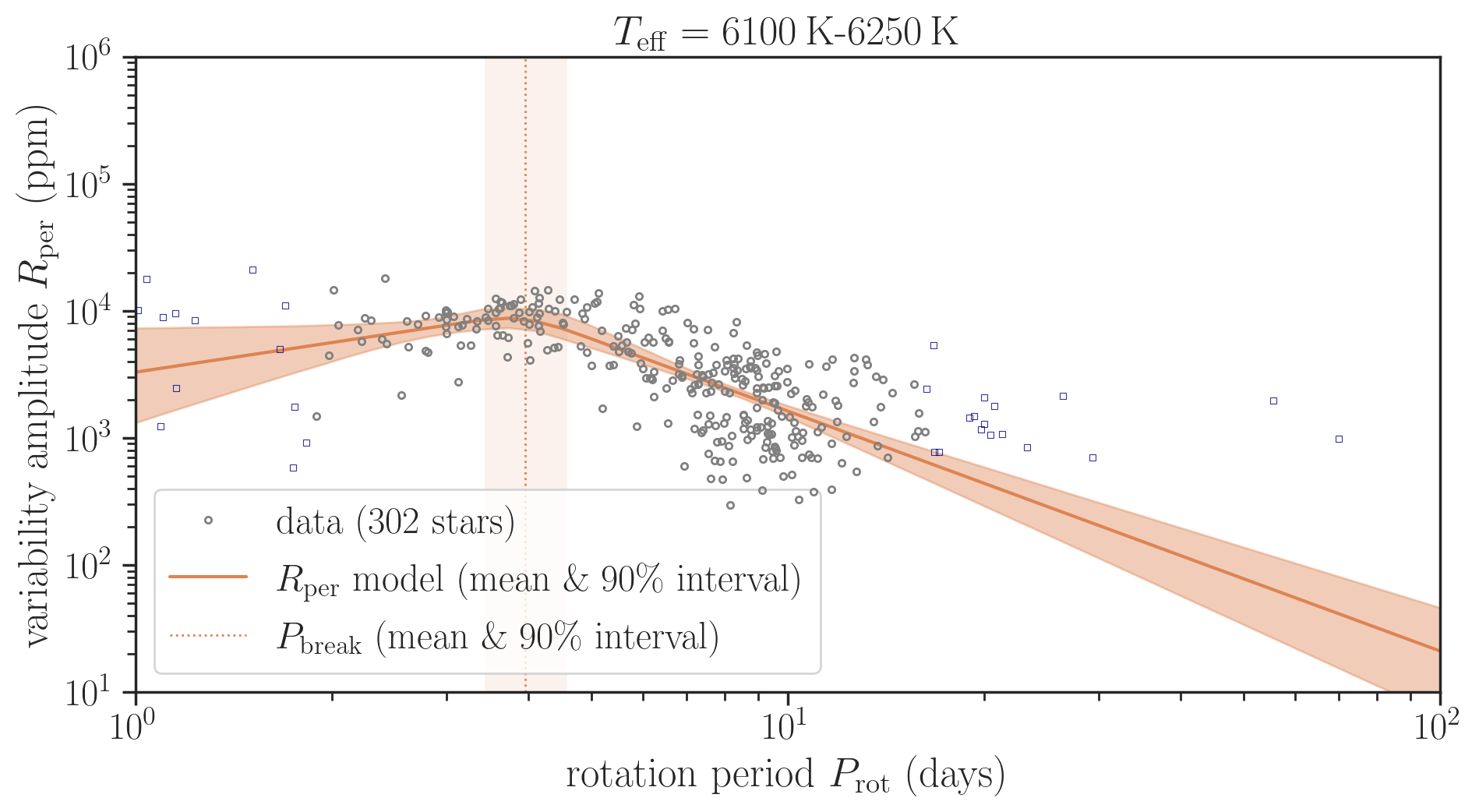}
    \plottwo{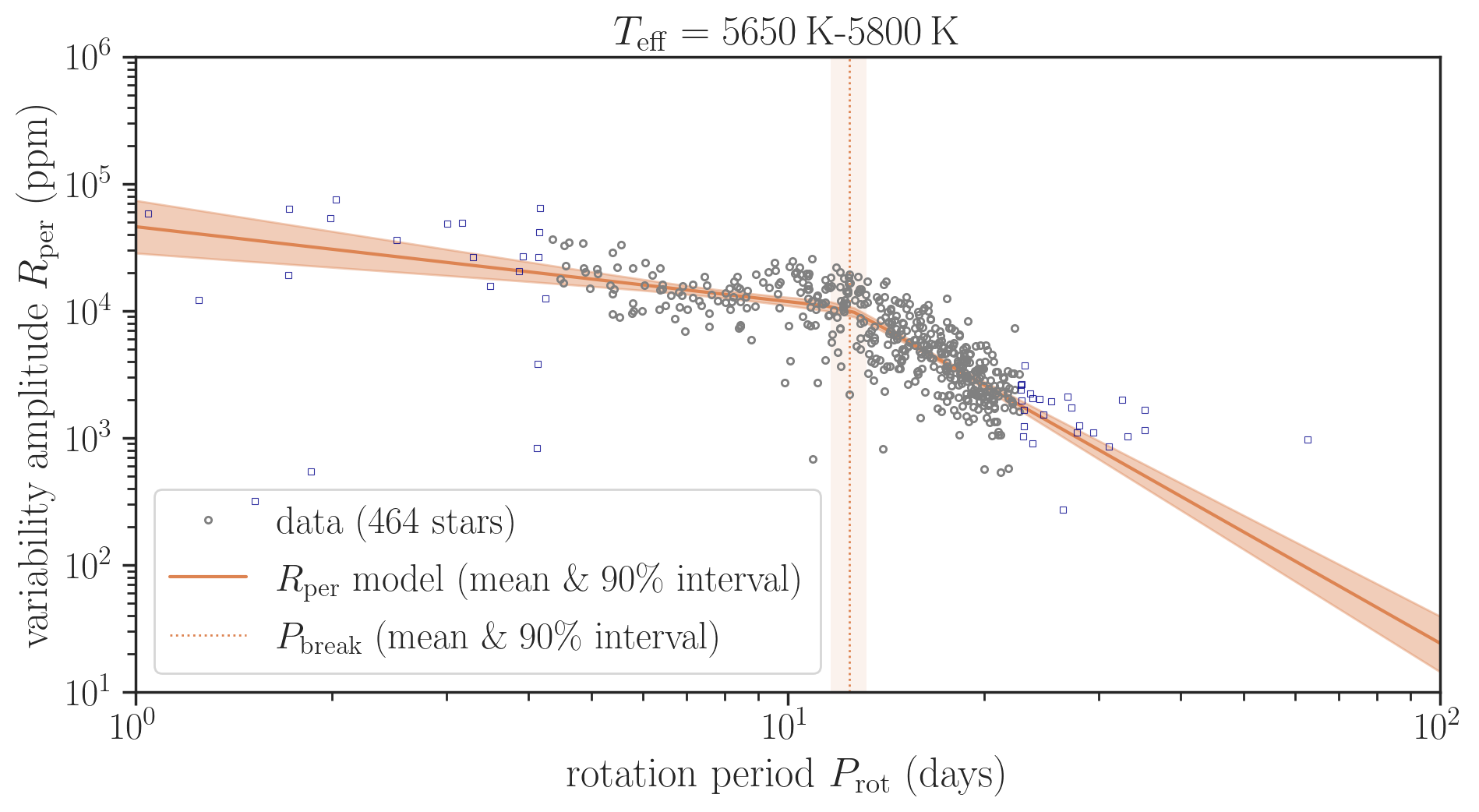}{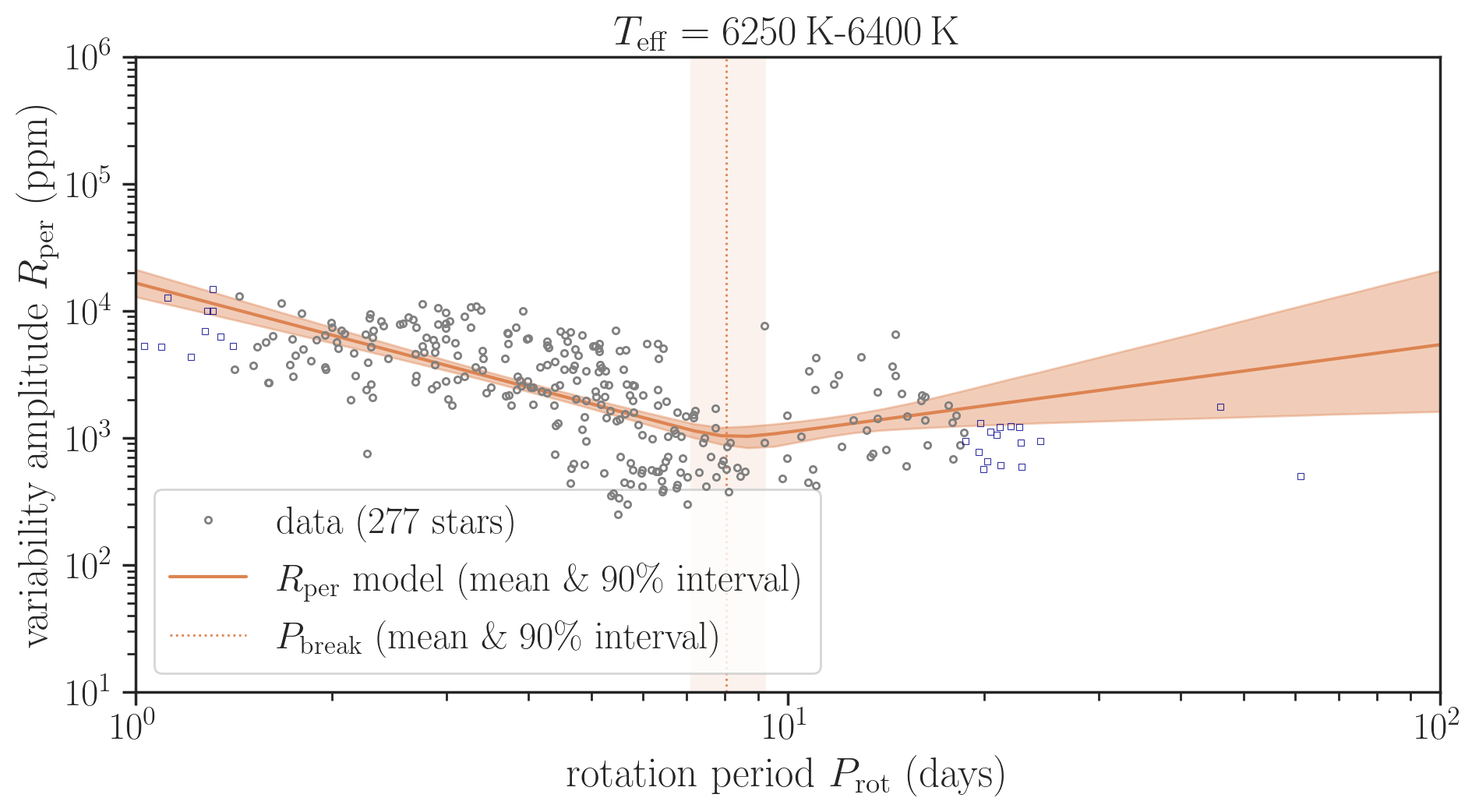}
    \caption{
    (Continued)
    }
\end{figure}


\begin{figure}
    \epsscale{1.15}
    \plottwo{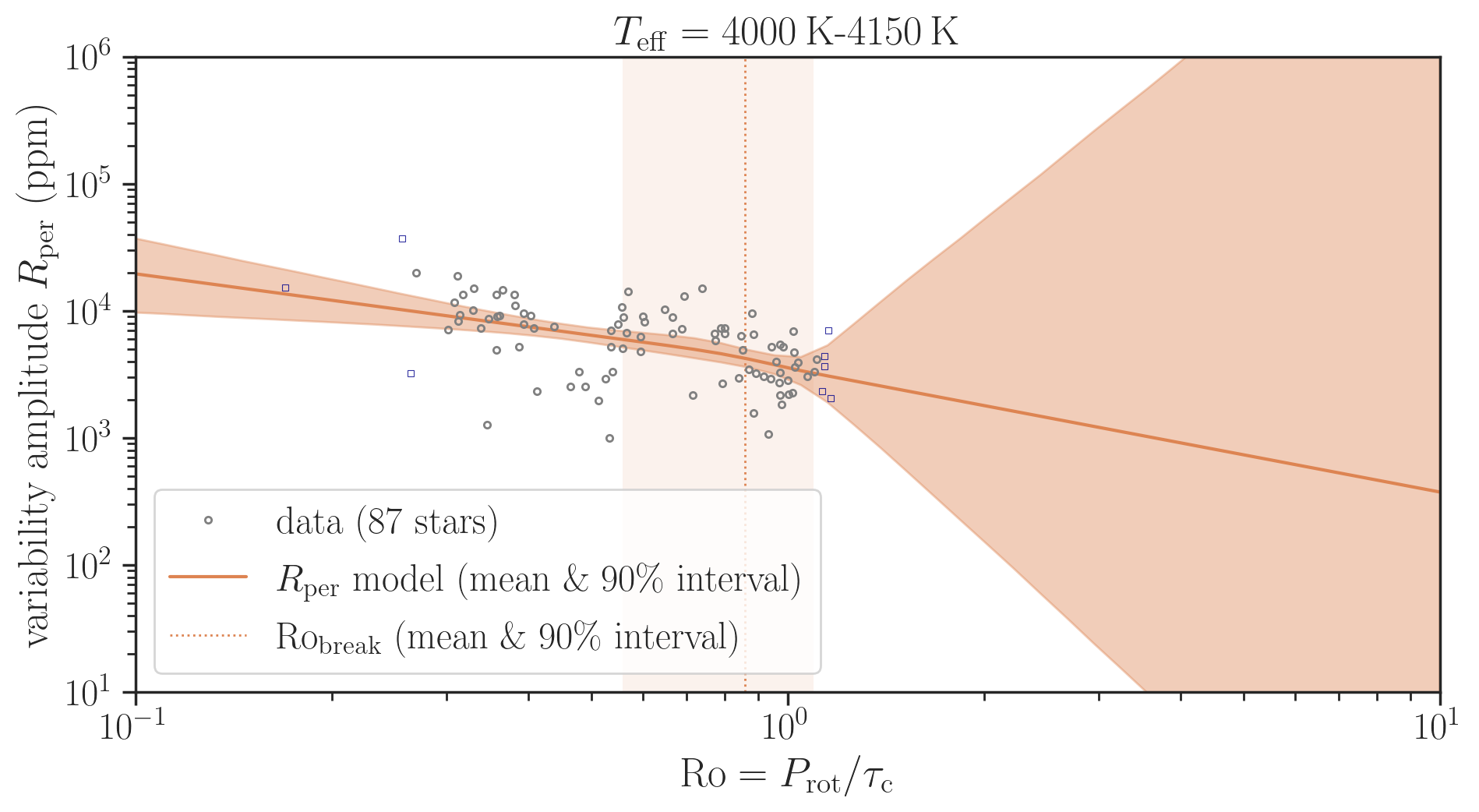}{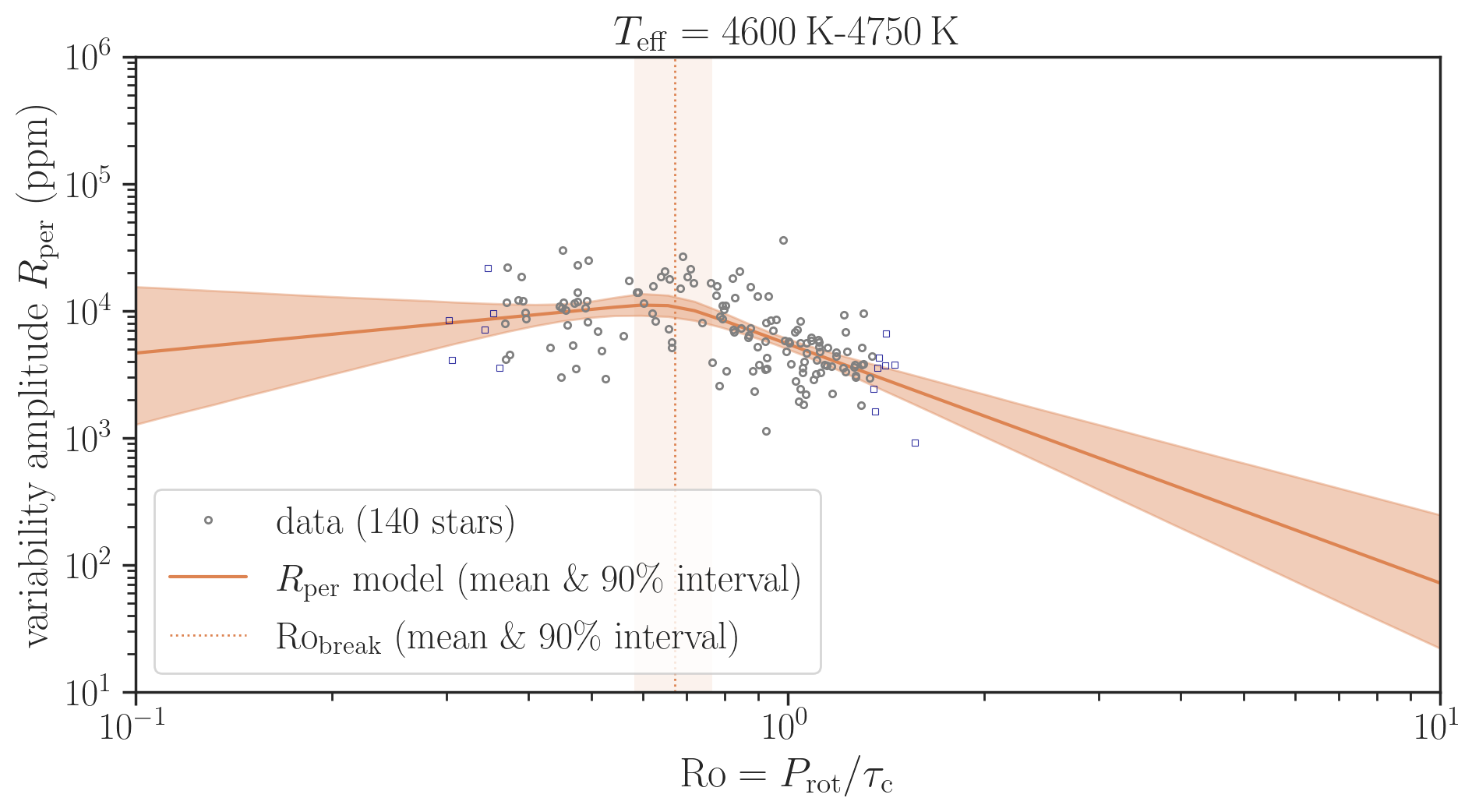}
    \plottwo{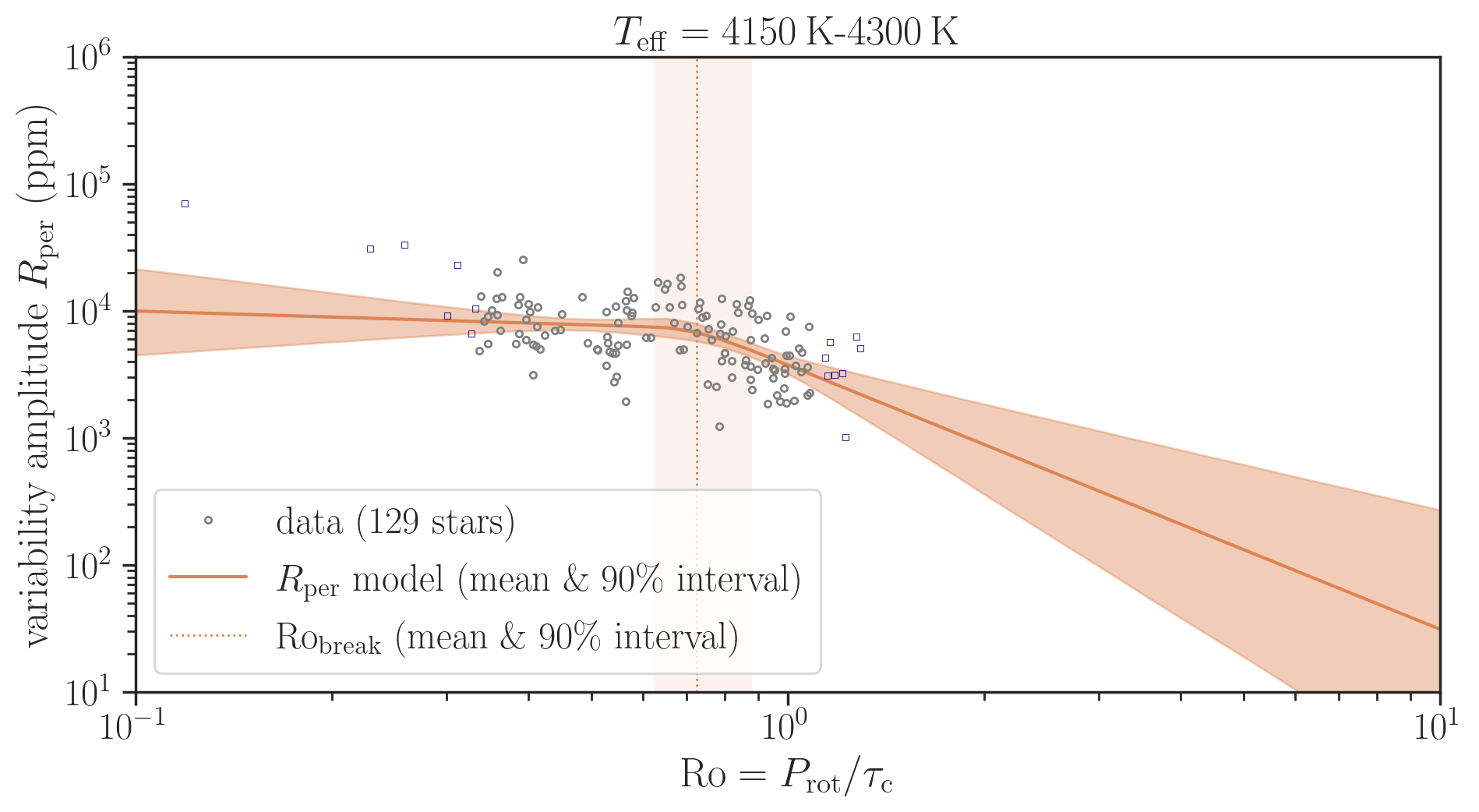}{teff4750-4900_romodel}
    \plottwo{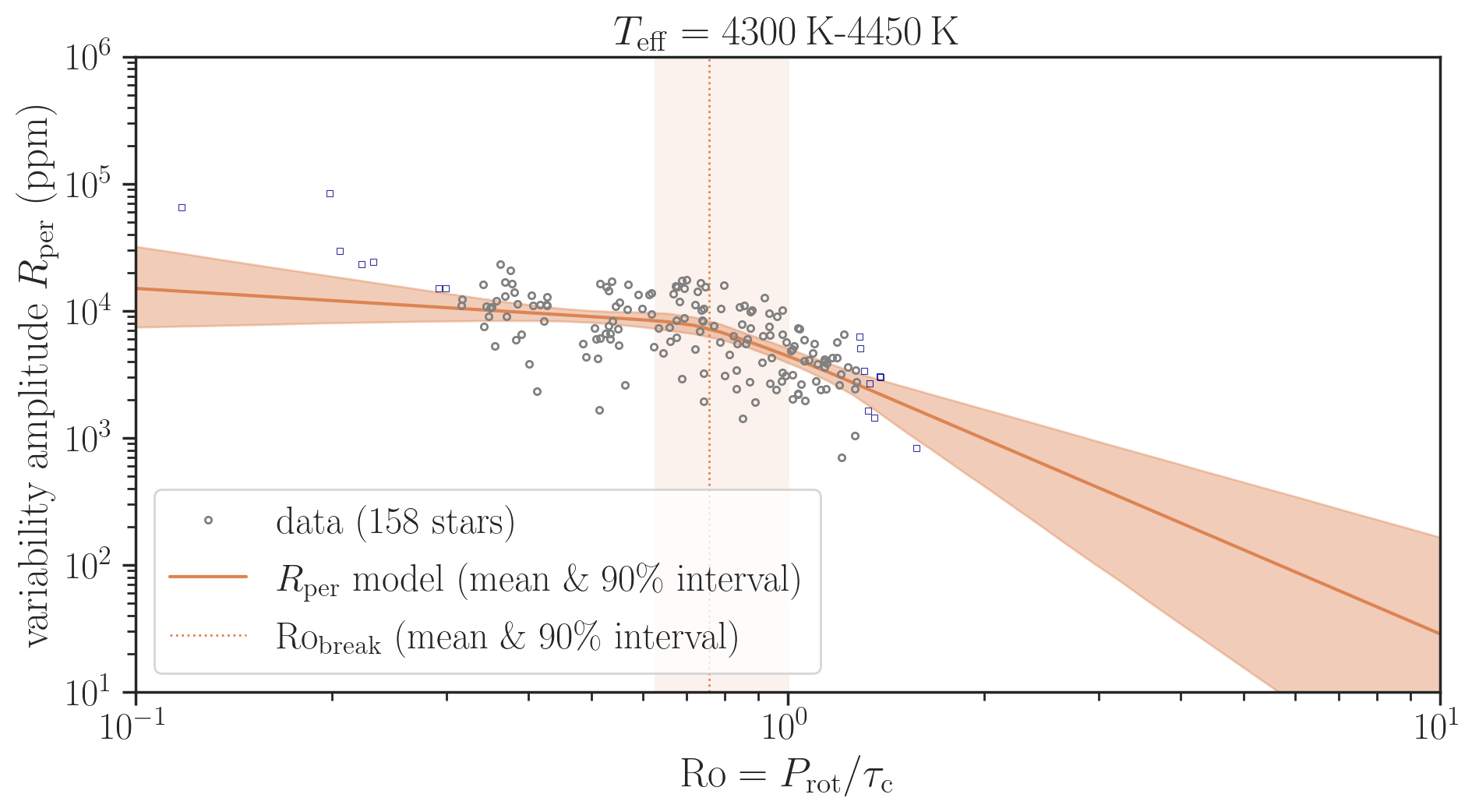}{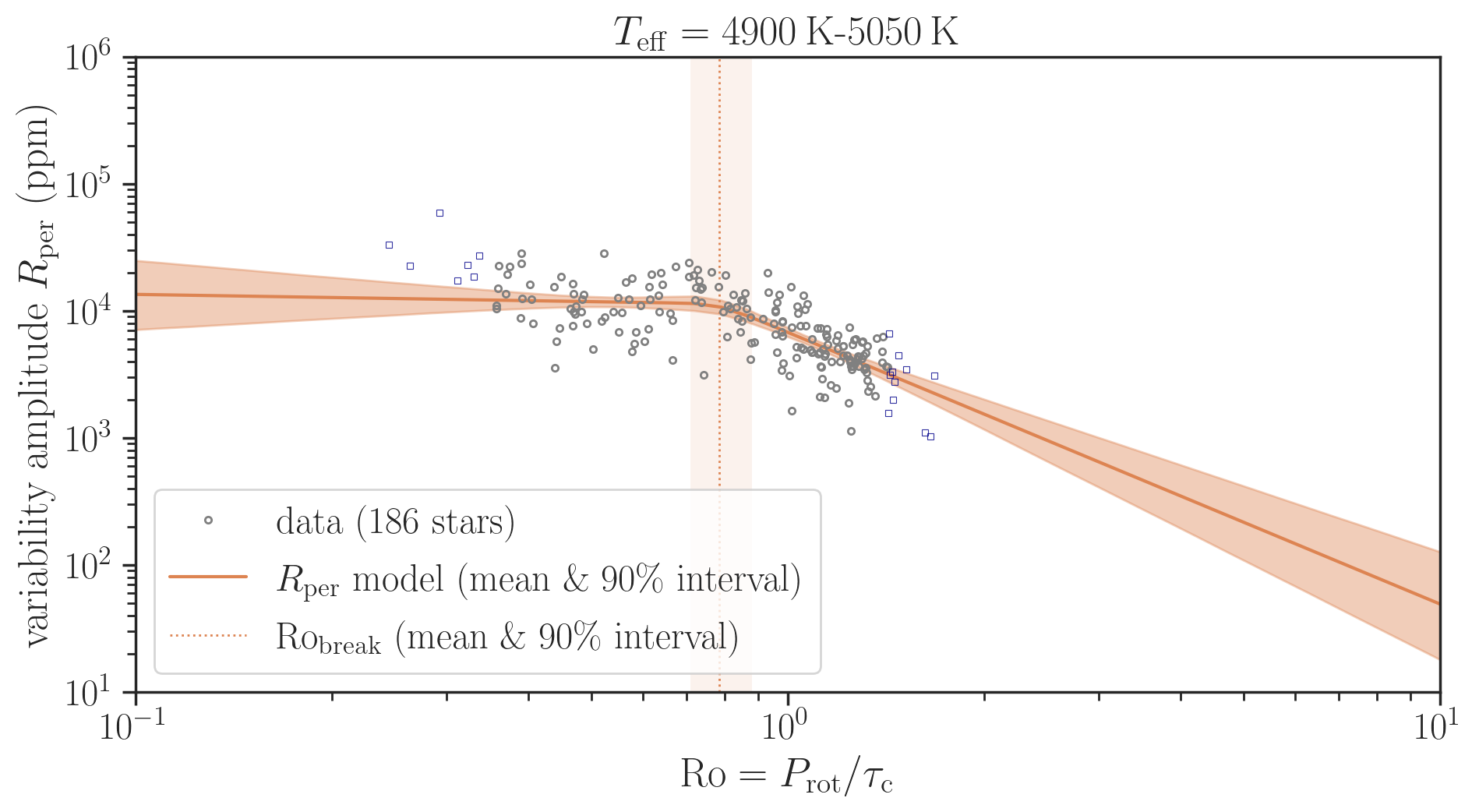}
    \plottwo{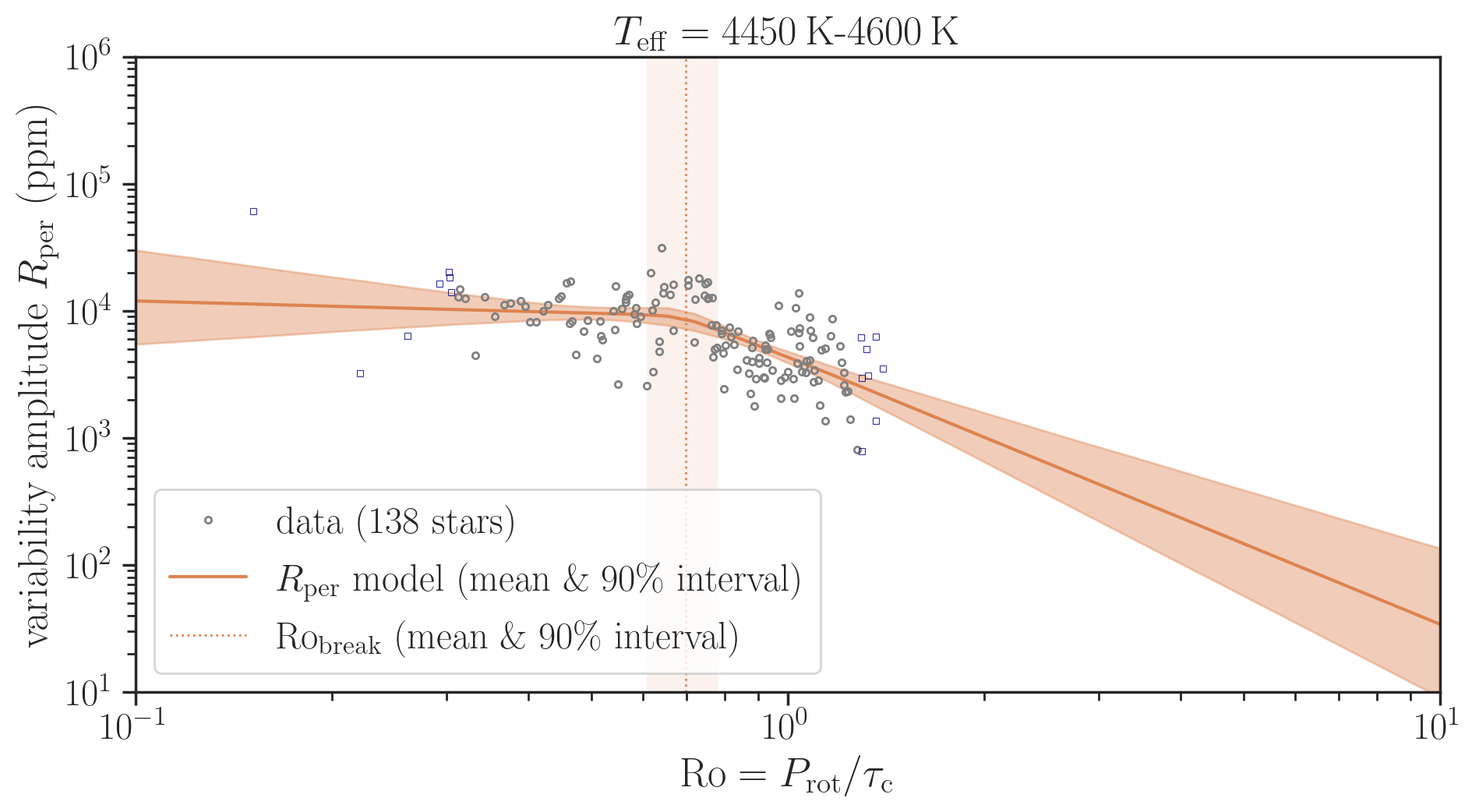}{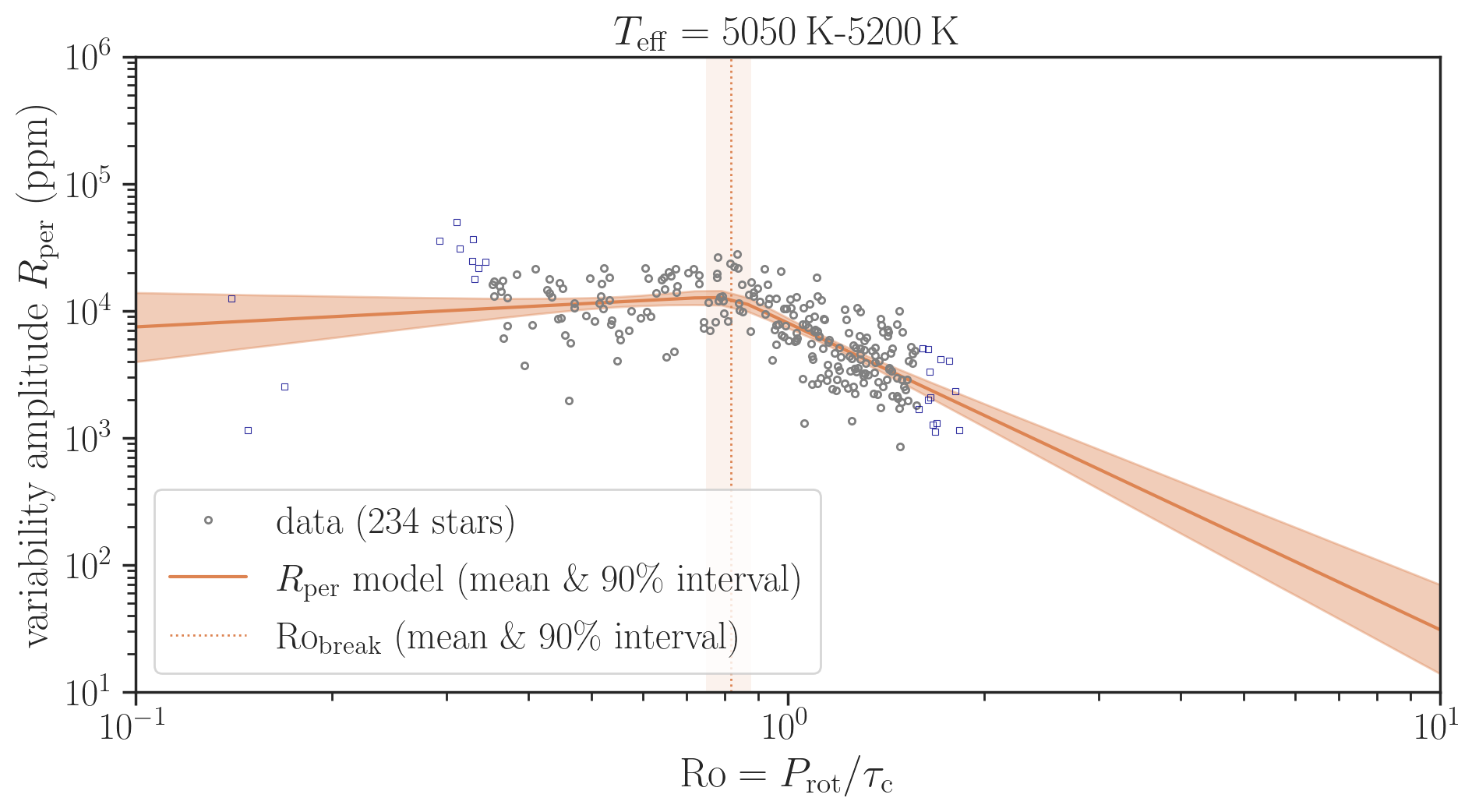}
    \caption{
    Spot-modulation amplitudes $\amp$ and Rossby numbers $\ro$ for stars in each $\teff$ bin.
    {\it Gray circles}:  Data points. Blue open squares show the ones that were not used for modeling.
    {\it Orange solid line and shade}: Broken power-law model. 
    {\it Vertical orange dotted line and shade}: Inferred location of the break, $\robreak$. See Section \ref{ssec:amp_ro} for details.
    }
    \label{fig:r_ro_all}
\end{figure}
\addtocounter{figure}{-1}
\begin{figure}
    \epsscale{1.15}
    \plottwo{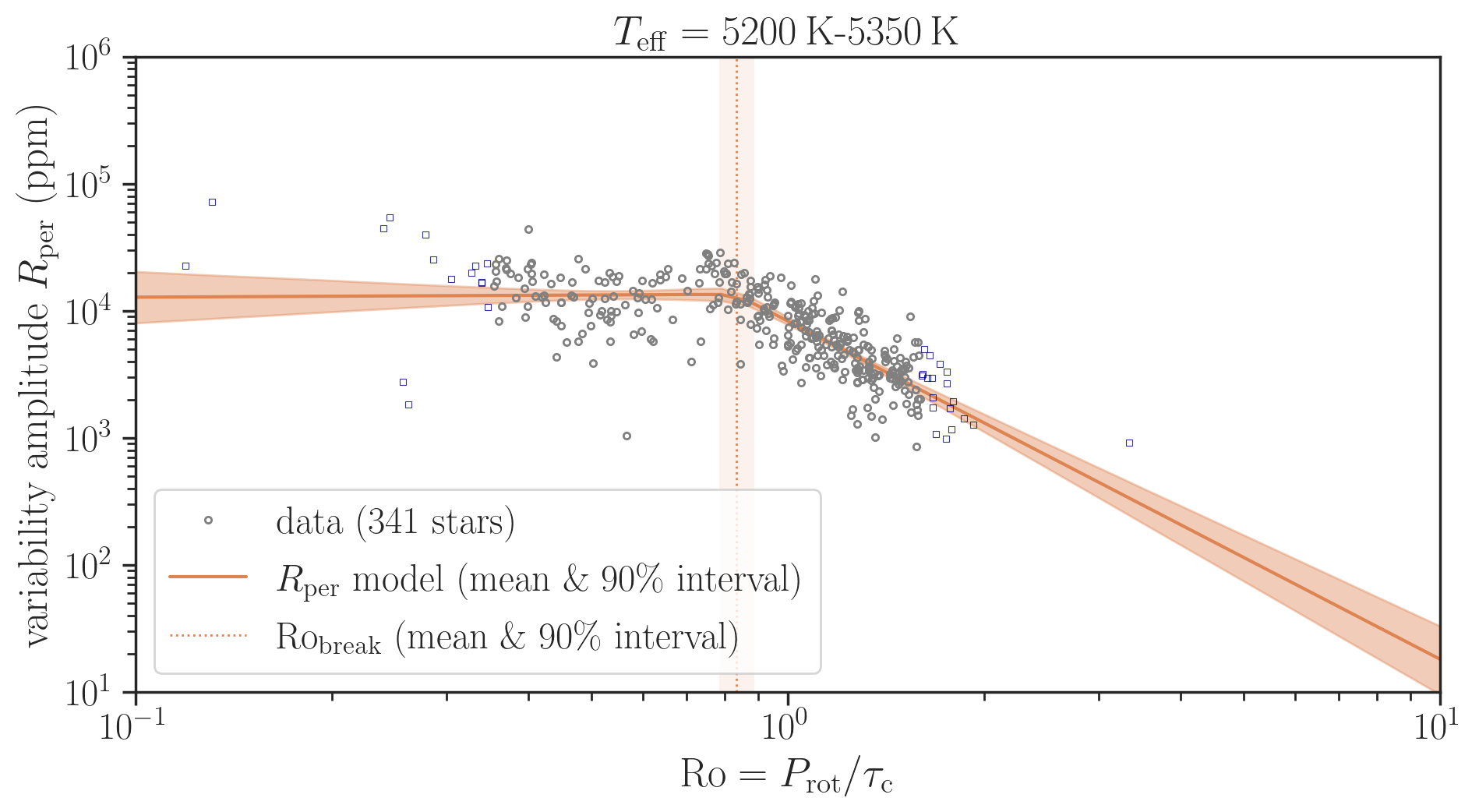}{teff5800-5950_romodel}
    \plottwo{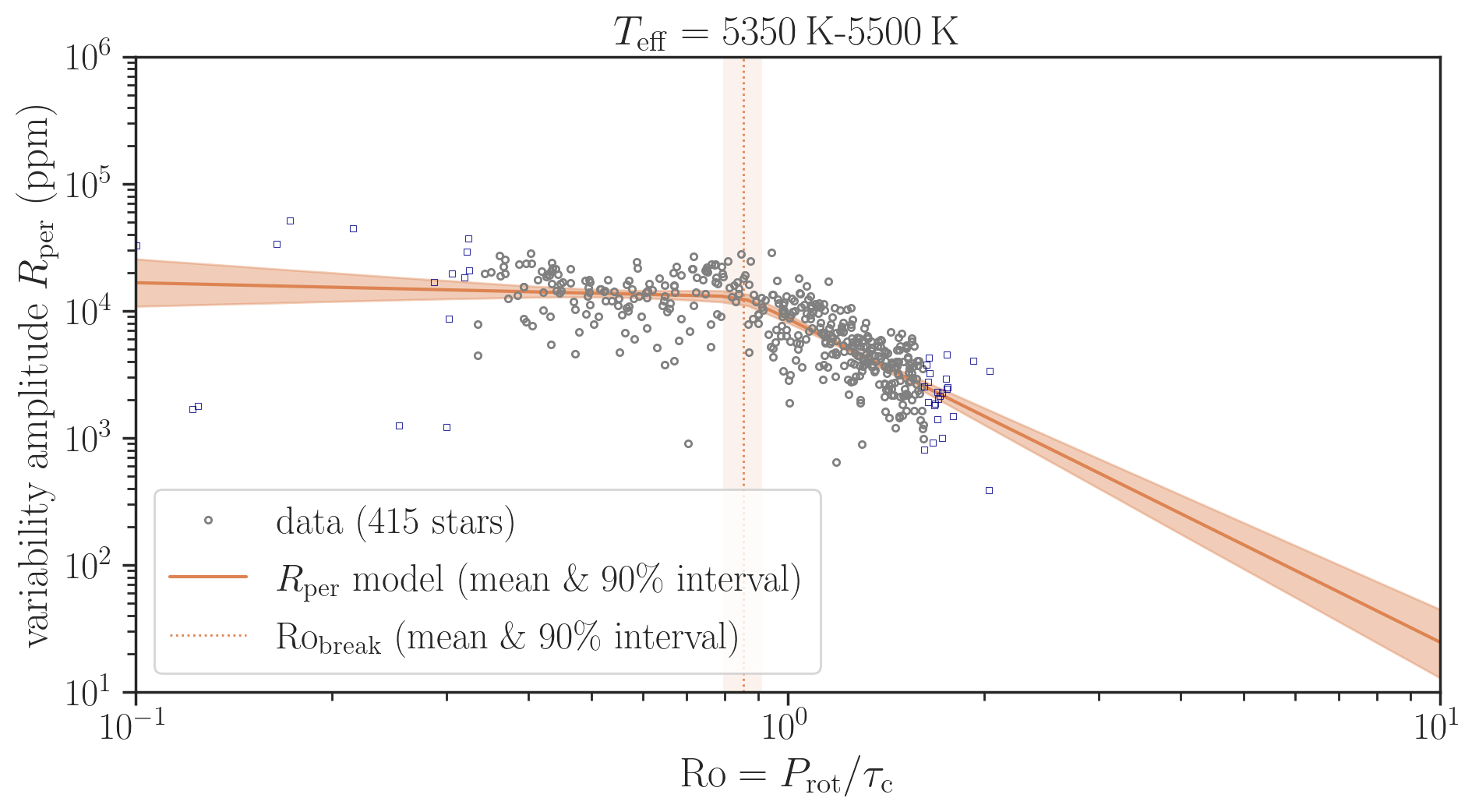}{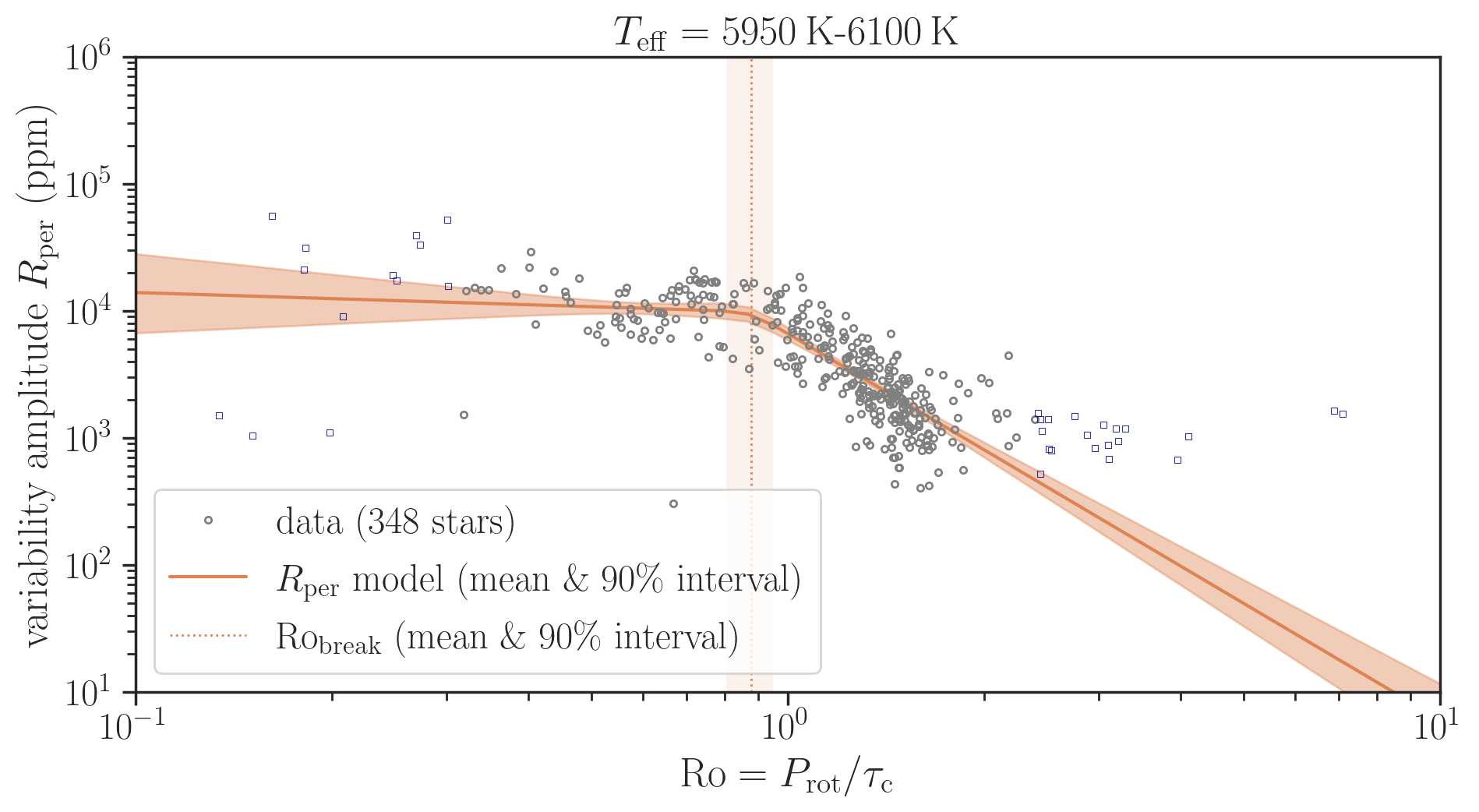}
    \plottwo{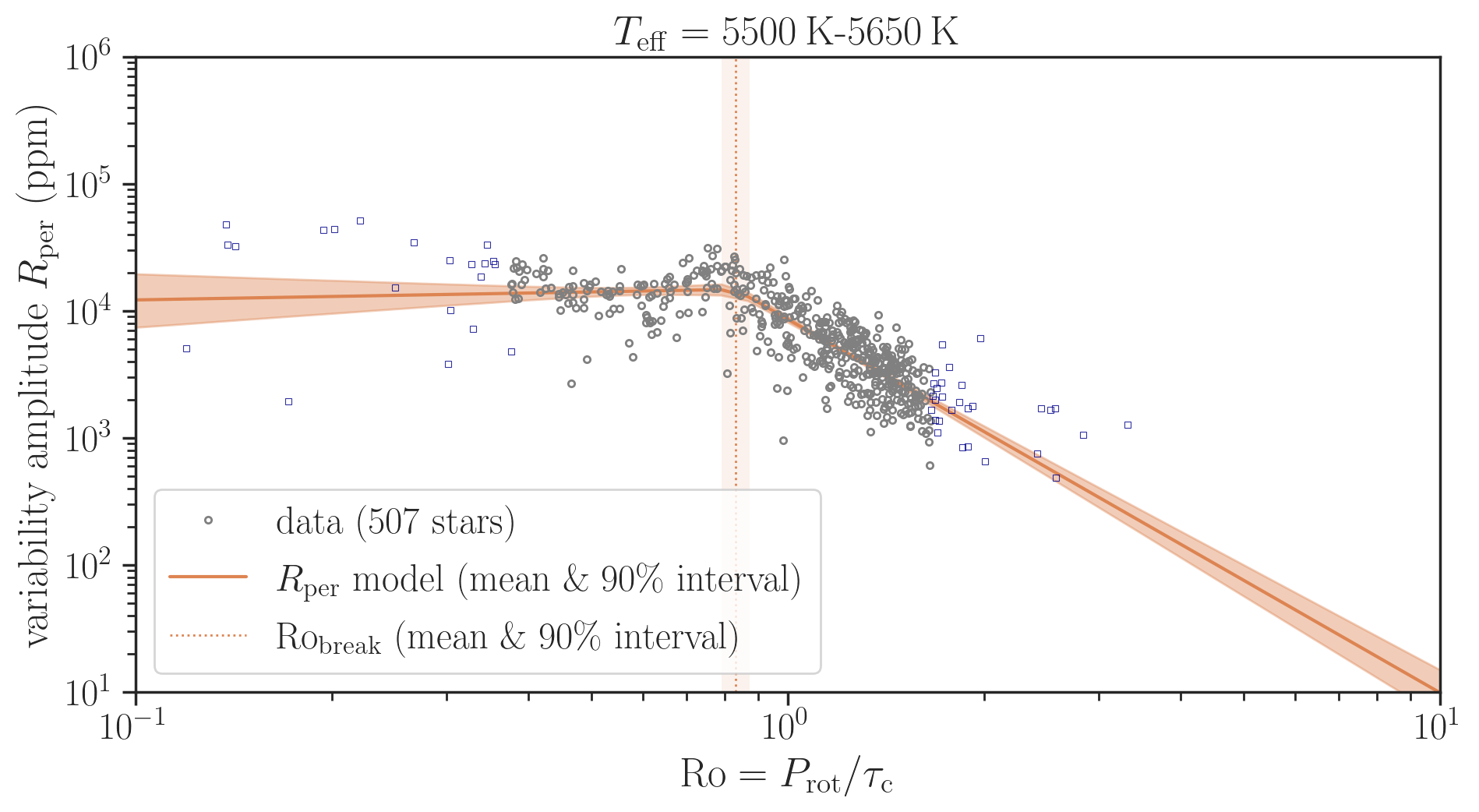}{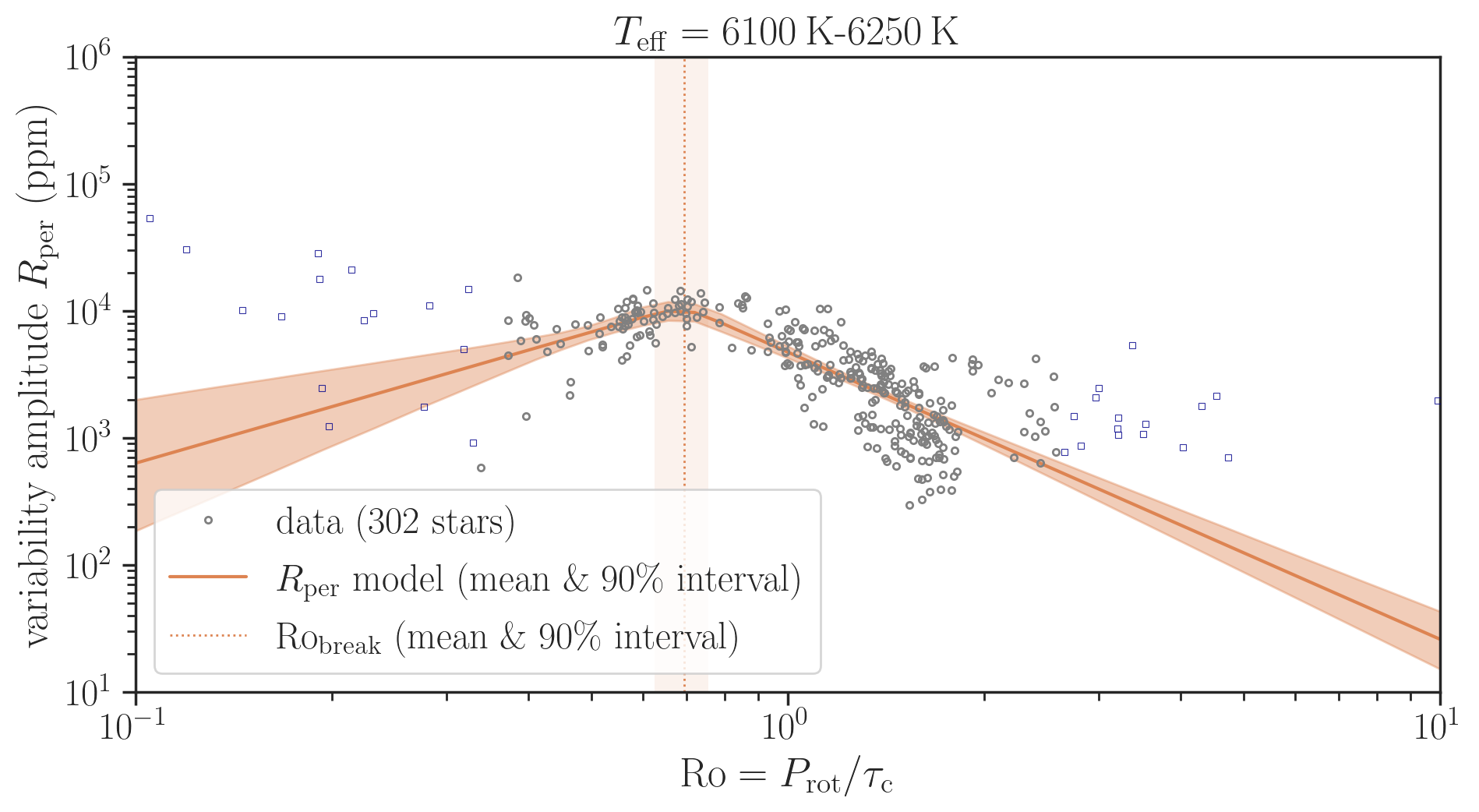}
    \plottwo{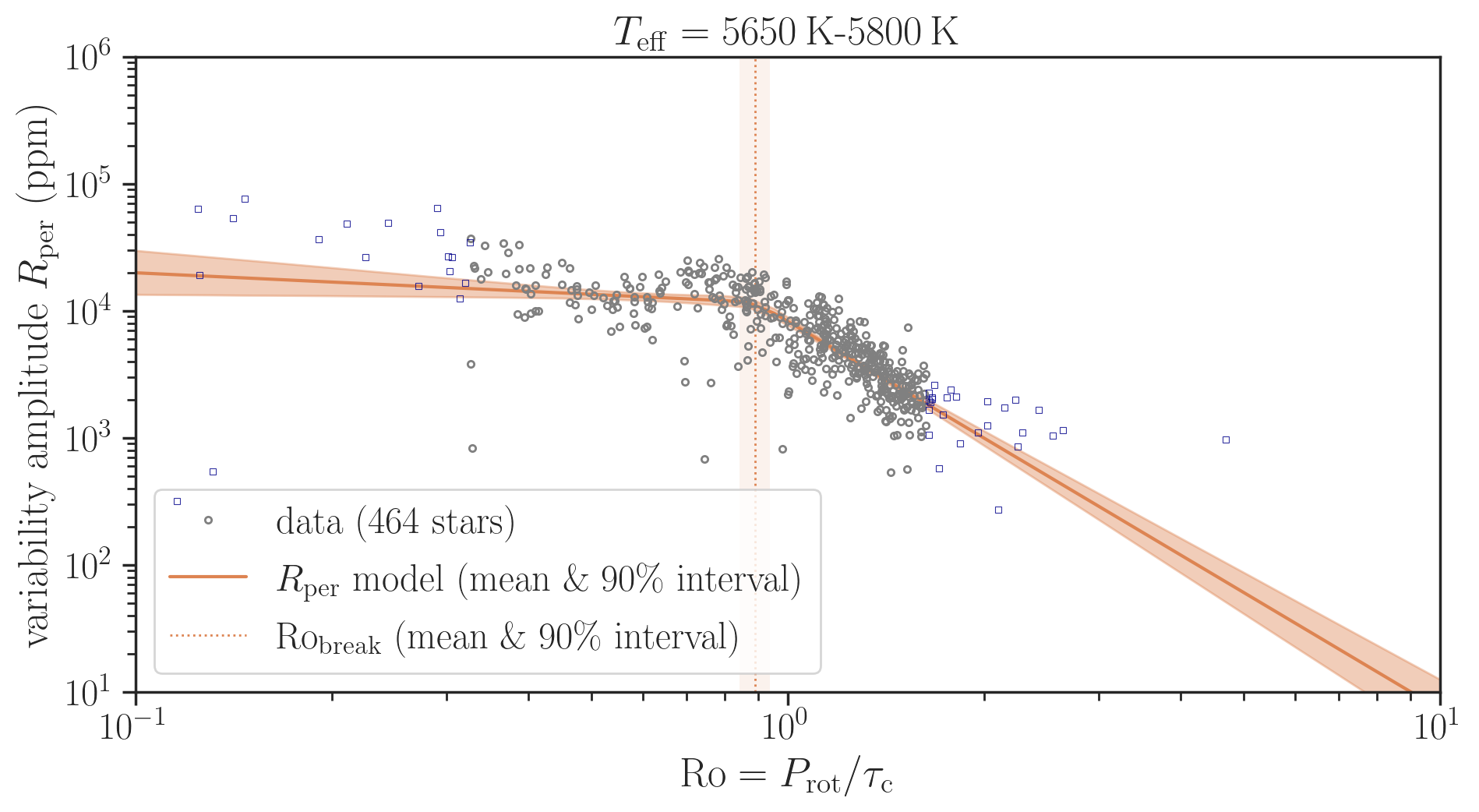}{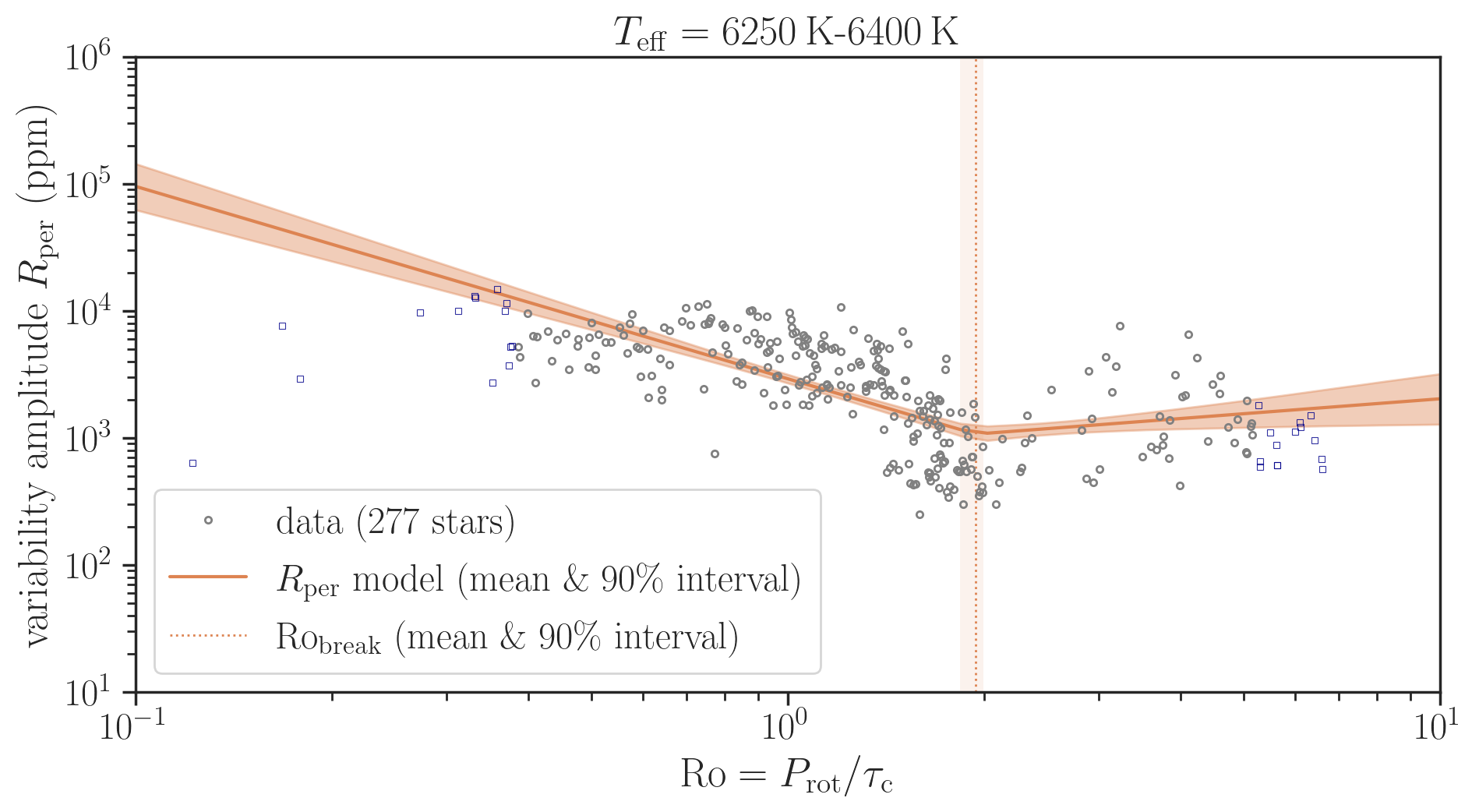}
    \caption{
    (Continued)
    }
\end{figure}

\section{Analyses Adopting Different $\tauc$ Prescriptions}\label{sec:tauc}

\subsection{Relation Between $\amp$ and $\ro$}\label{ssec:tauc_amp}

We have adopted the $\tauc$ prescription by \citet{2011ApJ...741...54C} in the main text (see Section~\ref{ssec:amp_ro}).
For nearly solar-mass stars, the difference from the more recent scales is almost multiplicative and only shifts the $\ro$ axis by a factor of a few \citep[see Figure 9 of][]{2021ApJ...912..127S, 2021ApJ...910..110L}. 
However, the difference becomes larger for stars much cooler or hotter than the Sun in a manner that depends on $\teff$.

\begin{figure*}
    \epsscale{1.05}
    \plotone{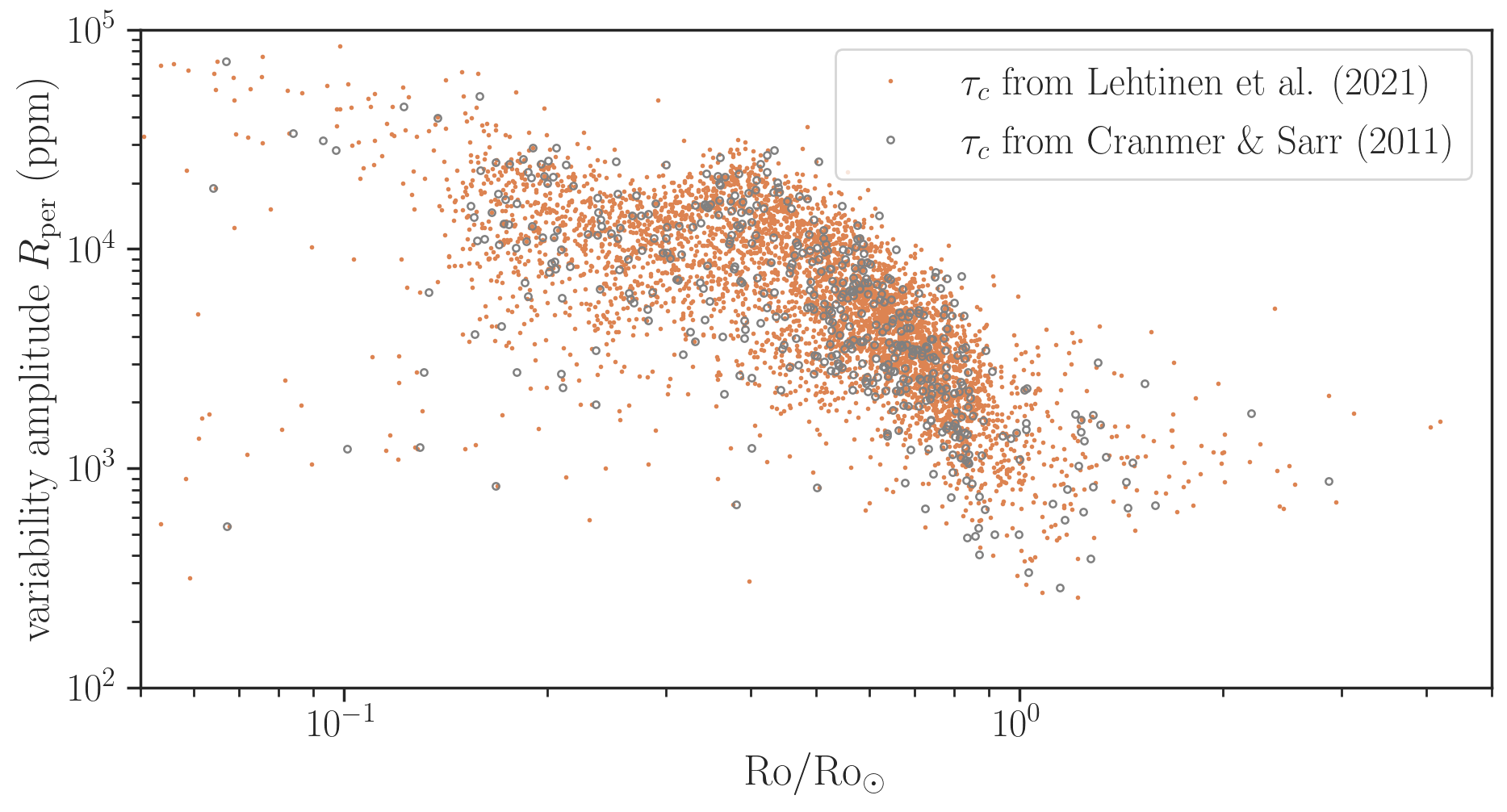}
    \plotone{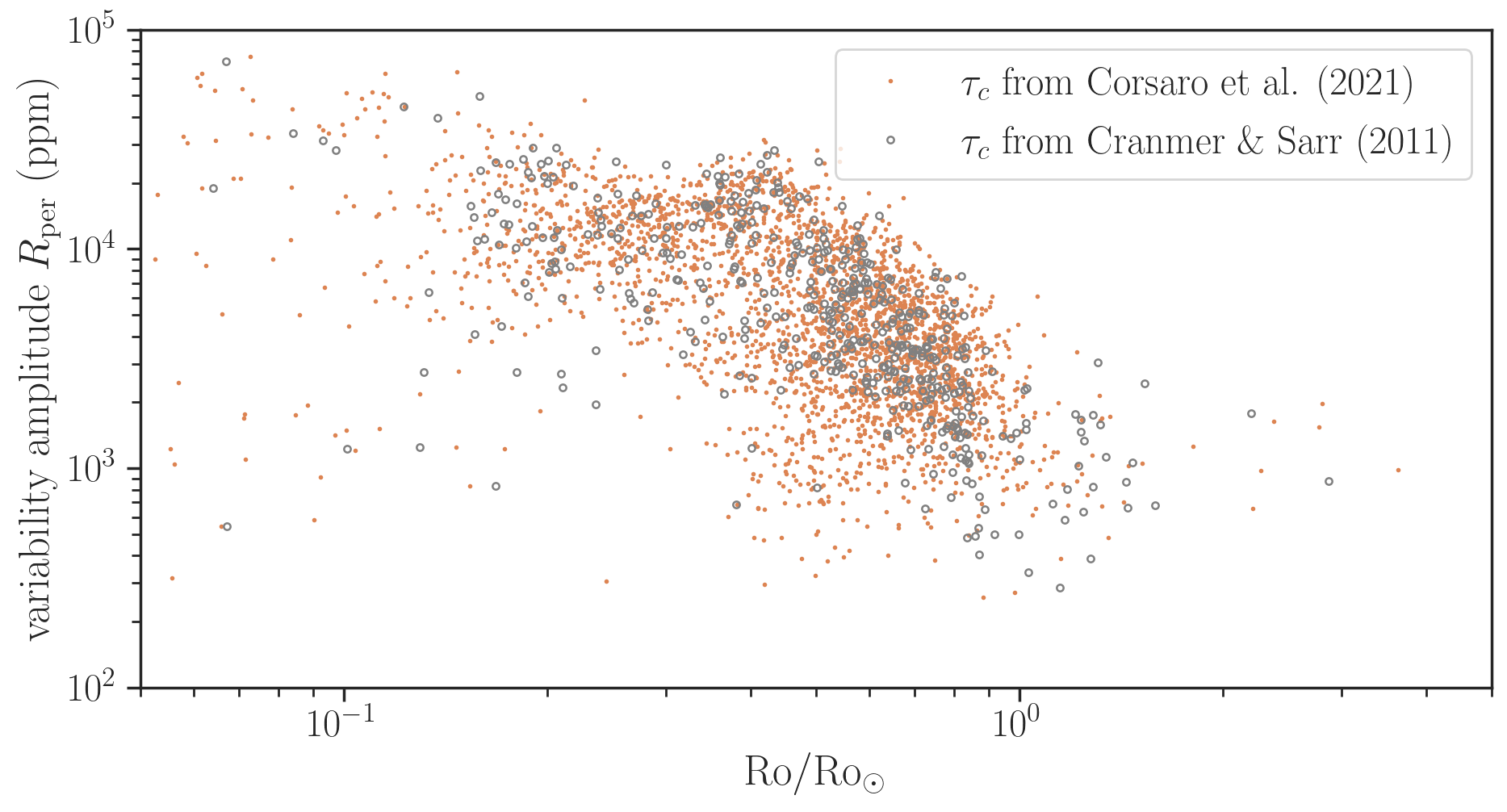}
    \caption{
    {\it Orange dots}: Spot-modulation amplitudes $\amp$ and Rossby number (scaled by the Solar value) computed using the $\tauc$ prescriptions in \citet[][{\it top}]{2021ApJ...910..110L} and in \citet[][{\it bottom}]{2021A&A...652L...2C}. Only the stars with $\teff<6,250\,\mathrm{K}$ are shown. In the bottom panel, stars with $G_\mathrm{BP}-G_\mathrm{RP}>1$ are also excluded because the \citet{2021A&A...652L...2C} prescription cannot be applied to those lower-mass stars.
    {\it Gray open circles}: $\amp$ and $\ro/\ro_\odot$ computed using the $\tauc$ formula in \citet{2011ApJ...741...54C}. This is the sample used for the main analysis, although here it is thinned by a factor or 10 to improve the visibility of the plots.
    }
    \label{fig:tauc}
\end{figure*}

To check on how sensitive our conclusions might be on the adopted $\tauc$ relation, we computed $\tauc$ for our sample stars using the prescriptions in \citet{2021ApJ...910..110L} and \citet{2021A&A...652L...2C} for which the necessary information is readily available. For the former, we first converted $\teff$ to $B-V$, used it to compute $\tauc$ in the Noyes scale, and converted it to $\tauc$ from YaPSI using equation~1 of  \citet{2021ApJ...910..110L}. For the latter, we used equation~11 of \citet{2021A&A...652L...2C}; here we omitted stars with $G_\mathrm{BP}-G_\mathrm{RP}>1$ for which asteroseicmic calibration was not performed. The resulting $\amp$--$\ro$ relations are compared to what we adopted in the main text in Figure~\ref{fig:tauc}. Here we scaled $\ro$ by the solar value $\ro_\odot$ computed for each prescription so that the shape of the $\amp$--$\ro$ relations can be compared between different prescriptions.

Overall, the results show that $\amp$--$\ro$ relations based on these prescriptions are similar to the one we found using the formula in \citet{2011ApJ...741...54C}. More points exist below our standard $\amp$--$\ro$ relation for $\tauc$ from \citet{2021A&A...652L...2C} (bottom panel), which also appear to exist in their Figure 2. The presence of these points, however, does not affect our conclusion on the impact of detection bias: if some stars indeed fall below the $\amp$--$\ro$ relation we assumed, the bias simply becomes even more significant.

The difference in the scatter of $\amp$ at a given $\ro$, however, does indicate that the $\teff$ (in)dependence of the $\amp$--$\ro$ relation is sensitive to the adopted $\tauc$ prescription. In Section~\ref{sec:analysis} we found that its shape is not very sensitive to $\teff$, but this is not guaranteed in other prescriptions. This fact --- in addition to the $\teff$-dependent detection bias discussed in Section~\ref{ssec:r_ro_bias} --- makes it even more difficult to study possible $\teff$ dependence.
Again, this subtlety does not alter the discussion in Section~\ref{sec:detection} because it mainly relies on an empirical fact that the typical $\amp$ is well predicted by $\ro$ within a certain dispersion, but may need to be taken into account in quantitative analyses of the observed population that explicitly model the detection function.

\subsection{Comparison Between Different Activity Indicators}\label{ssec:tauc_comp}

\begin{figure*}
    \epsscale{1.1}
    \plotone{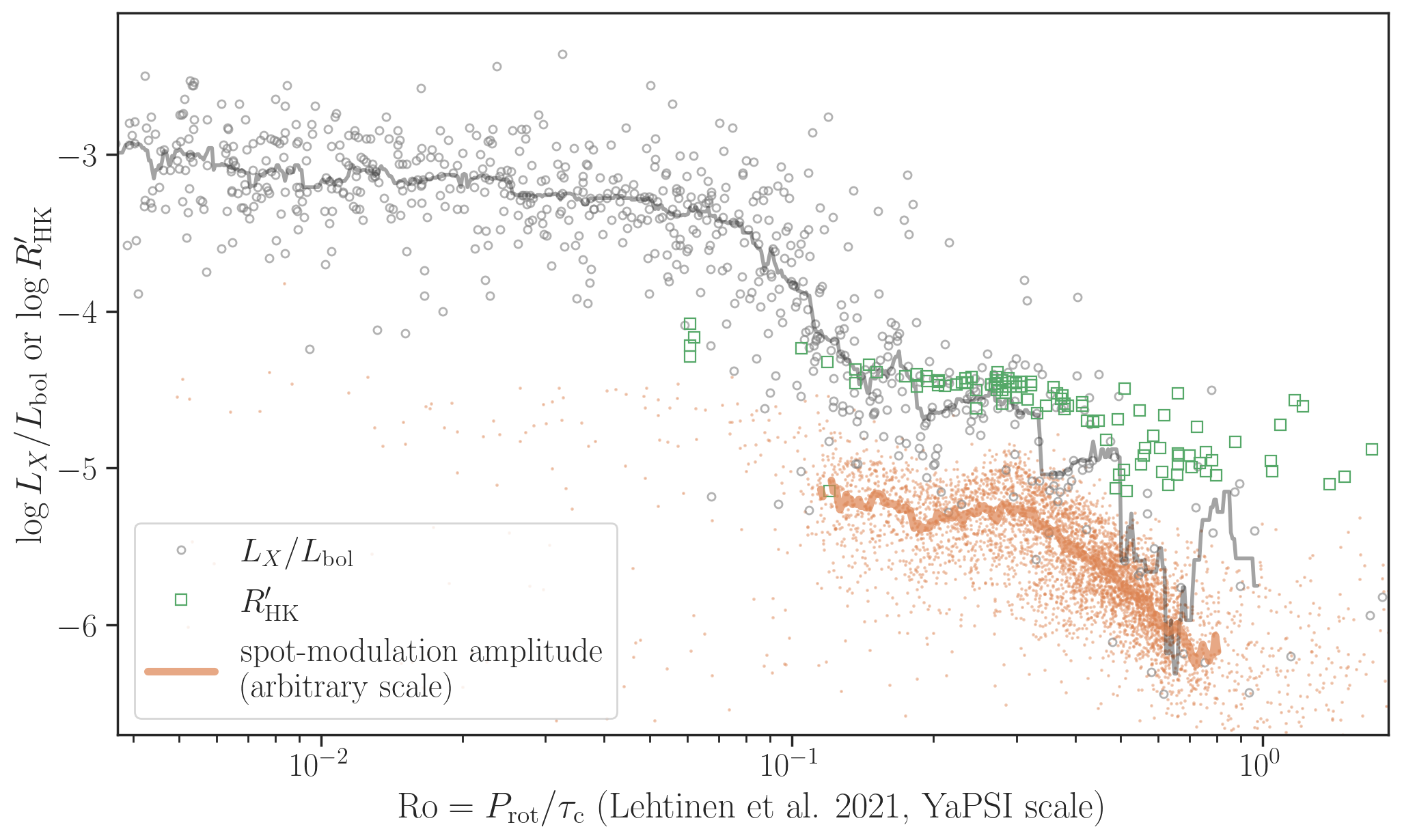}
    \caption{
    Same as Figure~\ref{fig:comparison}, except that $\tauc$ is computed for all the data sets following \citet{2021ApJ...910..110L}.
    }
    \label{fig:comparison_l21}
\end{figure*}

In Figure~\ref{fig:comparison_l21}, we reproduced \ref{fig:comparison} using the $\tauc$ prescription in \citet{2021ApJ...910..110L} as computed in Appendix~\ref{ssec:tauc_amp}. We see transitions in all indicators around $\ro \sim 0.3$, which is $\sim 0.4\,\ro_\odot$ in this prescription that gives $\ro_\odot\approx 0.8$. So this location agrees with what we found in the main text. We did not do the same for the \citet{2021A&A...652L...2C} prescription, because it is not applicable to most of the X-ray sample stars that are less massive than the Sun.

\bibliography{references_masuda}
\bibliographystyle{aasjournal}



\end{document}